\documentclass[%
 reprint,
 superscriptaddress,
 nofootinbib,
 amsmath,amssymb,
 aps,
 prd,
 nolongbibliography
]{revtex4-2}

\usepackage{graphicx}% Include figure files
\usepackage{dcolumn}% Align table columns on decimal point
\usepackage{bm}% bold math
\usepackage{xcolor}
\usepackage{units}
\usepackage{verbatim}
\usepackage{subcaption} % Allows use of subfigures
\usepackage{hyperref}% add hypertext capabilities
\usepackage{placeins}  % for FloatBarrier
\usepackage{stackengine}  % for parentheses around superscripts
\usepackage{multirow}
\usepackage{fixme}
\fxsetup{status=draft,inline=true,margin=false,theme=colorsig}
\FXRegisterAuthor{jw}{jaw}{JW}
\usepackage{xspace}  % used below

\newcommand{\dcp}{\ensuremath{\delta_{\textrm{CP}}}\xspace}
\newcommand{\ssth}[1]{\ensuremath{\sin^2 \theta_{#1}}\xspace}
\newcommand{\sstth}[1]{\ensuremath{\sin^2 2\theta_{#1}}\xspace}
\newcommand{\dmsq}[1]{\ensuremath{\Delta m^2_{#1}}}

\newcommand{\thsol}{\ensuremath{\theta_{12}}\xspace}
\newcommand{\threac}{\ensuremath{\theta_{13}}\xspace}
\newcommand{\thatm}{\ensuremath{\theta_{23}}\xspace}

\newcommand{\dmlg}{\dmsq{32}\xspace}
\newcommand{\sdcp}{\ensuremath{\sin\delta_{\textrm{CP}}}\xspace}

\newcommand{\Ehad}{\ensuremath{E_{\mathrm{had}}}}
\newcommand{\hadfrac}{\ensuremath{f_{\mathrm{had}}}}

\newcommand{\nue}{\ensuremath{\nu_e}\xspace}

\newcommand{\nuebar}{\ensuremath{\bar{\nu}_e}\xspace}
\newcommand{\numu}{\ensuremath{\nu_{\mu}}\xspace}
\newcommand{\numubar}{\ensuremath{\bar{\nu}_{\mu}}\xspace}
\newcommand{\nutau}{\ensuremath{\nu_{\tau}}\xspace}
\newcommand{\nutaubar}{\ensuremath{\bar{\nu}_{\tau}}\xspace}
	
\newcommand{\nuone}{\ensuremath{\nu_{1}}\xspace}
\newcommand{\nuonebar}{\ensuremath{\bar{\nu}_{1}}\xspace}
\newcommand{\nutwo}{\ensuremath{\nu_{2}}\xspace}
\newcommand{\nutwobar}{\ensuremath{\bar{\nu}_{2}}\xspace}
\newcommand{\nuthree}{\ensuremath{\nu_{3}}\xspace}
\newcommand{\nuthreebar}{\ensuremath{\bar{\nu}_{3}}\xspace}

\newcommand{\numutonue}{\ensuremath{\nu_{\mu} \rightarrow \nu_e}\xspace}
\newcommand{\numutonuebar}{\ensuremath{\bar{\nu}_{\mu} \rightarrow \bar{\nu}_e}\xspace}

\newcommand{\pnuenuebar}{\ensuremath{\mathcal{P}(\bar{\nu}_e \rightarrow \bar{\nu}_e)}\xspace}

\newcommand{\pnumunue}{\ensuremath{\mathcal{P}(\nu_{\mu} \rightarrow \nu_e)}\xspace}
\newcommand{\pnumunuebar}{\ensuremath{\mathcal{P}(\bar{\nu}_{\mu} \rightarrow \bar{\nu}_e)}\xspace}

\newcommand{\pnumunumu}{\ensuremath{\mathcal{P}(\nu_{\mu} \rightarrow \nu_{\mu})}\xspace}
\newcommand{\pnumunumubar}{\ensuremath{\mathcal{P}(\bar{\nu}_{\mu} \rightarrow \bar{\nu}_{\mu})}\xspace}

\newcommand{\J}{\ensuremath{J}}

\newcommand{\mrrtt}{\ensuremath{\textrm{M} \textrm{R}^2 \textrm{T}^2}}
\newcommand{\hmcmc}{\mbox{HMCMC}}
\newcommand{\ppfx}{\textsc{PPFX}}
\newcommand{\geant}{\textsc{Geant4}\xspace}
\newcommand{\genie}{\textsc{GENIE}\xspace}
\newcommand{\valencia}{Val\`{e}ncia}
\begin{document}

\preprint{APS/123-QED}

\title{Expanding neutrino oscillation parameter measurements in NOvA using a Bayesian approach}

\newcommand{\ANL}{Argonne National Laboratory, Argonne, Illinois 60439, 
USA}
\newcommand{\Bandirma}{Bandirma Onyedi Eyl\"ul University, Faculty of 
Engineering and Natural Sciences, Engineering Sciences Department, 
10200, Bandirma, Balıkesir, Turkey}
\newcommand{\ICS}{Institute of Computer Science, The Czech 
Academy of Sciences, 
182 07 Prague, Czech Republic}
\newcommand{\IOP}{Institute of Physics, The Czech 
Academy of Sciences, 
182 21 Prague, Czech Republic}
\newcommand{\Atlantico}{Universidad del Atlantico,
Carrera 30 No.\ 8-49, Puerto Colombia, Atlantico, Colombia}
\newcommand{\BHU}{Department of Physics, Institute of Science, Banaras 
Hindu University, Varanasi, 221 005, India}
\newcommand{\UCLA}{Physics and Astronomy Department, UCLA, Box 951547, Los 
Angeles, California 90095-1547, USA}
\newcommand{\Caltech}{California Institute of 
Technology, Pasadena, California 91125, USA}
\newcommand{\Cochin}{Department of Physics, Cochin University
of Science and Technology, Kochi 682 022, India}
\newcommand{\Charles}
{Charles University, Faculty of Mathematics and Physics,
 Institute of Particle and Nuclear Physics, Prague, Czech Republic}
\newcommand{\Cincinnati}{Department of Physics, University of Cincinnati, 
Cincinnati, Ohio 45221, USA}
\newcommand{\CSU}{Department of Physics, Colorado 
State University, Fort Collins, CO 80523-1875, USA}
\newcommand{\CTU}{Czech Technical University in Prague,
Brehova 7, 115 19 Prague 1, Czech Republic}
\newcommand{\Dallas}{Physics Department, University of Texas at Dallas,
800 W. Campbell Rd. Richardson, Texas 75083-0688, USA}
\newcommand{\DallasU}{University of Dallas, 1845 E 
Northgate Drive, Irving, Texas 75062 USA}
\newcommand{\Delhi}{Department of Physics and Astrophysics, University of 
Delhi, Delhi 110007, India}
\newcommand{\JINR}{Joint Institute for Nuclear Research,  
Dubna, Moscow region 141980, Russia}
\newcommand{\Erciyes}{
Department of Physics, Erciyes University, Kayseri 38030, Turkey}
\newcommand{\FNAL}{Fermi National Accelerator Laboratory, Batavia, 
Illinois 60510, USA}
\newcommand{\FSU}{Florida State University, Tallahassee, Florida 32306, USA}
\newcommand{\UFG}{Instituto de F\'{i}sica, Universidade Federal de 
Goi\'{a}s, Goi\^{a}nia, Goi\'{a}s, 74690-900, Brazil}
\newcommand{\Guwahati}{Department of Physics, IIT Guwahati, Guwahati, 781 
039, India}
\newcommand{\Harvard}{Department of Physics, Harvard University, 
Cambridge, Massachusetts 02138, USA}
\newcommand{\Houston}{Department of Physics, 
University of Houston, Houston, Texas 77204, USA}
\newcommand{\IHyderabad}{Department of Physics, IIT Hyderabad, Hyderabad, 
502 205, India}
\newcommand{\Hyderabad}{School of Physics, University of Hyderabad, 
Hyderabad, 500 046, India}
\newcommand{\IIT}{Illinois Institute of Technology,
Chicago IL 60616, USA}
\newcommand{\Indiana}{Indiana University, Bloomington, Indiana 47405, 
USA}
\newcommand{\INR}{Institute for Nuclear Research of Russia, Academy of 
Sciences 7a, 60th October Anniversary prospect, Moscow 117312, Russia}
\newcommand{\Iowa}{Department of Physics and Astronomy, Iowa State 
University, Ames, Iowa 50011, USA}
\newcommand{\Irvine}{Department of Physics and Astronomy, 
University of California at Irvine, Irvine, California 92697, USA}
\newcommand{\Jammu}{Department of Physics and Electronics, University of 
Jammu, Jammu Tawi, 180 006, Jammu and Kashmir, India}
\newcommand{\Lebedev}{Nuclear Physics and Astrophysics Division, Lebedev 
Physical 
Institute, Leninsky Prospect 53, 119991 Moscow, Russia}
\newcommand{\Magdalena}{Universidad del Magdalena, Carrera 32 No 22-08 Santa Marta, Colombia}
\newcommand{\MSU}{Department of Physics and Astronomy, Michigan State 
University, East Lansing, Michigan 48824, USA}
\newcommand{\Crookston}{Math, Science and Technology Department, University 
of Minnesota Crookston, Crookston, Minnesota 56716, USA}
\newcommand{\Duluth}{Department of Physics and Astronomy, 
University of Minnesota Duluth, Duluth, Minnesota 55812, USA}
\newcommand{\Minnesota}{School of Physics and Astronomy, University of 
Minnesota Twin Cities, Minneapolis, Minnesota 55455, USA}
\newcommand{\Mississippi}{University of Mississippi, University, Mississippi 38677, USA}
\newcommand{\NISER}{National Institute of Science Education and Research,
Khurda, 752050, Odisha, India}
\newcommand{\Oxford}{Subdepartment of Particle Physics, 
University of Oxford, Oxford OX1 3RH, United Kingdom}
\newcommand{\Panjab}{Department of Physics, Panjab University, 
Chandigarh, 160 014, India}
\newcommand{\Pitt}{Department of Physics, 
University of Pittsburgh, Pittsburgh, Pennsylvania 15260, USA}
\newcommand{\QMU}{Particle Physics Research Centre, 
Department of Physics and Astronomy,
Queen Mary University of London,
London E1 4NS, United Kingdom}
\newcommand{\RAL}{Rutherford Appleton Laboratory, Science 
and 
Technology Facilities Council, Didcot, OX11 0QX, United Kingdom}
\newcommand{\SAlabama}{Department of Physics, University of 
South Alabama, Mobile, Alabama 36688, USA} 
\newcommand{\Carolina}{Department of Physics and Astronomy, University of 
South Carolina, Columbia, South Carolina 29208, USA}
\newcommand{\SDakota}{South Dakota School of Mines and Technology, Rapid 
City, South Dakota 57701, USA}
\newcommand{\SMU}{Department of Physics, Southern Methodist University, 
Dallas, Texas 75275, USA}
\newcommand{\Stanford}{Department of Physics, Stanford University, 
Stanford, California 94305, USA}
\newcommand{\Sussex}{Department of Physics and Astronomy, University of 
Sussex, Falmer, Brighton BN1 9QH, United Kingdom}
\newcommand{\Syracuse}{Department of Physics, Syracuse University,
Syracuse NY 13210, USA}
\newcommand{\Tennessee}{Department of Physics and Astronomy, 
University of Tennessee, Knoxville, Tennessee 37996, USA}
\newcommand{\Texas}{Department of Physics, University of Texas at Austin, 
Austin, Texas 78712, USA}
\newcommand{\Tufts}{Department of Physics and Astronomy, Tufts University, Medford, 
Massachusetts 02155, USA}
\newcommand{\UCL}{Physics and Astronomy Department, University College 
London, 
Gower Street, London WC1E 6BT, United Kingdom}
\newcommand{\Virginia}{Department of Physics, University of Virginia, 
Charlottesville, Virginia 22904, USA}
\newcommand{\WSU}{Department of Mathematics, Statistics, and Physics,
 Wichita State University, 
Wichita, Kansas 67260, USA}
\newcommand{\WandM}{Department of Physics, William \& Mary, 
Williamsburg, Virginia 23187, USA}
\newcommand{\Wisconsin}{Department of Physics, University of 
Wisconsin-Madison, Madison, Wisconsin 53706, USA}
\newcommand{\deceased}{Deceased.}
\affiliation{\ANL}
\affiliation{\Atlantico}
\affiliation{\Bandirma}
\affiliation{\BHU}
\affiliation{\Caltech}
\affiliation{\Charles}
\affiliation{\Cincinnati}
\affiliation{\Cochin}
\affiliation{\CSU}
\affiliation{\CTU}
\affiliation{\Delhi}
\affiliation{\Erciyes}
\affiliation{\FNAL}
\affiliation{\FSU}
\affiliation{\UFG}
\affiliation{\Guwahati}
\affiliation{\Houston}
\affiliation{\Hyderabad}
\affiliation{\IHyderabad}
\affiliation{\IIT}
\affiliation{\Indiana}
\affiliation{\ICS}
\affiliation{\INR}
\affiliation{\IOP}
\affiliation{\Iowa}
\affiliation{\Irvine}
\affiliation{\JINR}
\affiliation{\Magdalena}
\affiliation{\MSU}
\affiliation{\Duluth}
\affiliation{\Minnesota}
\affiliation{\Mississippi}
\affiliation{\NISER}
\affiliation{\Panjab}
\affiliation{\Pitt}
\affiliation{\QMU}
\affiliation{\SAlabama}
\affiliation{\Carolina}
\affiliation{\SMU}
\affiliation{\Sussex}
\affiliation{\Syracuse}
\affiliation{\Texas}
\affiliation{\Tufts}
\affiliation{\UCL}
\affiliation{\Virginia}
\affiliation{\WSU}
\affiliation{\WandM}
\affiliation{\Wisconsin}

\author{M.~A.~Acero}
\affiliation{\Atlantico}

\author{B.~Acharya}
\affiliation{\Mississippi}

\author{P.~Adamson}
\affiliation{\FNAL}

\author{N.~Anfimov}
\affiliation{\JINR}

\author{A.~Antoshkin}
\affiliation{\JINR}

\author{E.~Arrieta-Diaz}
\affiliation{\Magdalena}

\author{L.~Asquith}
\affiliation{\Sussex}

\author{A.~Aurisano}
\affiliation{\Cincinnati}

\author{A.~Back}
\affiliation{\Indiana}

\author{N.~Balashov}
\affiliation{\JINR}

% ...

\collaboration{The NOvA Collaboration}
\noaffiliation

\date{\today}

\begin{abstract}
	NOvA is a long-baseline neutrino oscillation experiment that measures oscillations in charged-current $\numu \rightarrow \numu$ (disappearance) and $\numu \rightarrow \nue$ (appearance) channels, and their antineutrino counterparts, using neutrinos of energies around \unit[2]{GeV} over a distance of \unit[810]{km}.
	In this work we reanalyze the dataset first examined in our previous paper [Phys. Rev. D \textbf{106}, 032004 (2022)] using an alternative statistical approach based on Bayesian Markov Chain Monte Carlo.
	We measure oscillation parameters consistent with the previous results.
	We also extend our inferences to include the first NOvA measurements of the reactor mixing angle \threac{}, where we find $0.071 \leq \sstth{13} \leq 0.107$, and the Jarlskog invariant, where we observe no significant preference for the CP-conserving value $J=0$ over values favoring CP violation.
	We use these results to examine the effects of constraints from short-baseline measurements of \threac{} using antineutrinos from nuclear reactors when making NOvA measurements of \thatm{}.
	Our long-baseline measurement of \threac{} is shown to be consistent with the reactor measurements, supporting the general applicability and robustness of the PMNS framework for neutrino oscillations.
\end{abstract}

\maketitle

\section{Introduction}
In the three-flavor neutrino oscillation paradigm, transitions among the three flavor eigenstates \nue, \numu, \nutau are governed by the matrix elements between these states and the mass eigenstates \nuone, \nutwo, \nuthree.\footnote{The same is true for their antineutrino counterparts \nuebar, \numubar, \nutaubar and \nuonebar, \nutwobar, \nuthreebar.  Throughout this paper the symbol $\nu$ will refer to both neutrinos and antineutrinos unless otherwise specified.}
Initial constraints on the elements of this Pontecorvo--Maki--Nakagawa--Sakata (PMNS) matrix were obtained by numerous experiments using neutrinos with a variety of energy spectra over various baselines~\cite{Fukuda:1998mi,Fukuda:2002pe,Ahmad:2002jz,Eguchi:2002dm,Michael:2006rx,Abe:2011sj,Abe:2011fz,An:2012eh,Ahn:2012nd}.
Decomposing the PMNS matrix yields a set of rotation-like ``mixing angles'' \thsol{}, \threac{}, \thatm{}, and a phase \dcp{}.
Contemporary neutrino oscillation experiments seek to make precision measurements of these parameters, as well as the differences between the squared mass eigenvalues ($\dmsq{ij} \equiv m_{i}^2 - m_{j}^2$).
These results have fundamental implications for models of neutrino mass and lepton flavor~\cite{Mohapatra:2006gs, Nunokawa:2007qh, Altarelli:2010gt, King:2015aea, Petcov:2017ggy}, as well as models of baryogenesis via charge-parity (CP) symmetry violation~\cite{Fukugita:1986hr,Buchmuller:1996pa,Buchmuller:2004nz,Buchmuller:2005eh,Pilaftsis:1997jf}.
Where multiple experiments can access the same parameters using different neutrino flavors or energies, the overall validity of the three-neutrino framework can also be tested.

Long-baseline (LBL) accelerator neutrino oscillation experiments measure oscillations in $\numu \rightarrow \numu$ (disappearance) and $\numu \rightarrow \nue$ (appearance) channels.
These channels constrain the mixing angles \threac{}, \thatm{}, the mass-squared splitting \dmlg{}, and the CP-violating phase \dcp{}.
Current measurements of \thatm{}~\cite{NOvA:2022wnj, T2K:2023mcm, Super-Kamiokande:2019gzr, IceCubeCollaboration:2023wtb} are consistent with maximal mixing (\thatm{} = $\pi/4$), which would suggest a $\mu\text{-}\tau$ symmetry in the flavors' mixing into the \nuthree mass eigenstate; a nonmaximal value such as \thatm{} $< \pi/4$ (lower octant, LO) or \thatm{} $>\pi/4$ (upper octant, UO) would indicate a preferential coupling of \nutau or \numu, respectively, with \nuthree.
Current experimental uncertainties on \thatm{} are the largest among the mixing angles \cite{ParticleDataGroup:2022pth}.
LBL oscillation measurements, where neutrinos traverse significant quantities of matter, are also impacted by the coherent forward scattering of \nue{}s on electrons in the Earth~\cite{Wolfenstein:1977ue}. 
This modifies the oscillation probabilities \pnumunue and \pnumunuebar with opposite signs. 
The direction of the resulting change in rate depends on whether \nuthree is the heaviest neutrino state (Normal Ordering, NO) or the lightest (Inverted Ordering, IO).
Current observations from LBL (and other) experiments~\cite{T2K:2023smv, Super-Kamiokande:2019gzr, NOvA:2022wnj, MINOS:2013xrl} prefer the NO, though the strength of the NO hypothesis depends on which measurements are included \cite{Esteban:2020cvm}.
A similar story holds for \dcp{}, where LBL experiments provide the only constraints~\cite{Esteban:2020cvm}.

NOvA is a long-baseline neutrino oscillation experiment that observes the \numu disappearance and \nue appearance channels using neutrinos of energies around \unit[2]{GeV} over a distance of \unit[810]{km}.
Previously~\cite{NOvA:2016vij,NOvA:2016kwd,NOvA:2017ohq,NOvA:2017abs,NOvA:2018gge,NOvA:2019cyt,NOvA:2021nfi}, we have presented NOvA constraints on \dmsq{32}, \ssth{23}, and \dcp using a classical frequentist approach.
However, the Feldman--Cousins technique that is required in order to obtain correct frequentist confidence regions for these variables~\cite{Feldman:1997qc, NOvA:2022wnj} poses challenges when confronted with highly degenerate sets of parameters.
It also does not allow for \textit{post hoc} transformations of the variables considered in the analysis.
In this work we present a new analysis of the dataset from Ref.~\cite{NOvA:2021nfi}, based on Bayesian Markov Chain Monte Carlo, which enables us to extend our inferences to include \threac{} and the Jarlskog invariant \J, for which $\J \neq 0$ unambiguously indicates CP violation.
We use these results to examine the implications of assuming short-baseline, nuclear-reactor antineutrino constraints on \threac{} when making measurements of other oscillation parameters in NOvA.

We also investigate the consistency of the PMNS framework by comparing the constraint from reactor experiments with our long-baseline measurement of \threac{}.

\section{3-flavor neutrino oscillations in NO\lowercase{v}A}
\label{sec:NOvA}
In this paper, we reanalyze data collected from an exposure of $13.6 \times 10^{20}$ \unit[14]{kton}-equivalent protons on target (POT) in the neutrino-enriched beam mode and $12.5 \times 10^{20}$\,POT in the analogous antineutrino mode. The dataset, simulations, reconstruction, and estimation of systematic uncertainties remain unchanged for this analysis. A brief overview of these components follows, with detailed descriptions available in Ref.~\cite{NOvA:2021nfi}. Extensive discussion of the new analysis method, its implementations, and the resulting inferences are presented in Sec. \ref{sec:MCMC details}.

\subsection{The NOvA experiment}
NOvA observes oscillations using neutrinos from the Neutrinos at the Main Injector (NuMI) beamline~\cite{NuMI:Adamson2016} at Fermilab using two functionally identical tracking calorimeter detectors that differ primarily in size.
The  \unit[0.3]{kton} near detector (ND), the smaller of the detectors, is located \unit[1]{km} from the neutrino production target, \unit[100]{m} underground.
The far detector (FD), by contrast, is \unit[14]{kton} and is located on the surface at Ash River, Minnesota, \unit[810]{km} from the target.
Both detectors are built from rectangular cells, made from polyvinyl chloride and of $\unit[3.9\times 6.6]{cm^2}$ cross-sectional area, with \unit[3.9]{m} (ND) or \unit[15.5]{m} (FD) length.
These are arranged in alternating horizontal and vertical planes and filled with a mineral-oil-based liquid scintillator.
A stack of alternating active planes and steel plates is placed downstream of the remainder of the ND to range out muons while a small rock overburden is placed above the FD to aid in rejecting the cosmic-ray background. 
The detectors are placed \unit[14.6]{mrad} from the central axis of the neutrino beam to receive a narrow-band neutrino flux predominantly between 1~and~\unit[3]{GeV}.

\subsection{Simulation and selection}
\label{subsec:sel and sim}
We use \geant(v10.4)~\cite{Agostinelli:2002hh, Geant:2017ats} to simulate the production of hadrons from interactions of the primary proton beam with the target as well as their transport through the beam optics. 
These simulations are reweighted using the Package to Predict the FluX (\ppfx)~\cite{Aliaga:2016oaz} to include constraints from external hadron production data~\cite{Paley:2014rpb,Alt:2006fr,Abgrall:2011ae,Barton:1982dg,Seun:2007zz,Tinti:2010zz,Lebedev:2007zz,Baatar:2012fua,Skubic:1978fi,Denisov:1973zv,Carroll:1978hc,Abe:2012av,Gaisser:1975et,Cronin:1957zz,Allaby:1969de,Longo:1962zz,Bobchenko:1979hp,Fedorov:1977an,Abrams:1969jm}. 
Simulated interactions of neutrinos that arise from decays of those hadrons are generated using \genie 3.0.6~\cite{Andreopoulos:2009rq,Andreopoulos:2015wxa} and modified by corrections we derive from NOvA ND and external data. 
In particular, NOvA ND data is used to produce a NOvA-specific tune of the IFIC \valencia{} 2p2h model~\cite{Nieves:2011pp,Gran:2013kda} that describes charged-current neutrino scattering from correlated pairs of nucleons.
The NOvA tune also modifies final-state interactions (simulated with the \genie hN full intranuclear cascade model) using pion--nucleus scattering data~\cite{Allardyce:1973ce,Saunders:1996ic,Meirav:1988pn,Levenson:1983xu,Ashery:1981tq,Ashery:1984ne,PinzonGuerra:2016uae}. 
The outgoing final-state particles are propagated through the detector using \geant and a custom NOvA readout simulation~\cite{Aurisano:2015oxj}.

We use groups of spatially and temporally proximate cells with activity above threshold to apply basic data quality and containment selection cuts to events in both data and simulation.
Within these groups we assign vertices and reconstruct likely particle trajectories.
Ultimately we divide the events into \numu charged-current (CC), \nue CC, neutral-current (NC), or cosmogenic background categories using NOvA's convolutional neural network (CNN)-based classifier~\cite{Aurisano:2016jvx}. 
Boosted decision trees (BDTs) are used to further reject cosmic backgrounds in the FD samples. 
Both sets of tags are utilized together to create \numu{} CC and \nue{} CC candidate samples.
Fully contained \nue{} candidates at the FD are further divided into high and low purity samples based on the CNN score in order to enhance the signal-to-background rejection capability of the fit. 
To improve the statistical strength of the fit, we recover an additional sample of ``peripheral'' events that fail the containment or cosmic rejection BDT but pass stricter particle ID requirements. 
We estimate neutrino energy for \numu{} CC events using the muon track length and the total deposited calorimetric energy of the hadronic system.
The energy for \nue{} CC events is estimated using a function of calorimetric energy that takes as input the energy of the event's reconstructed trajectories divided into electromagnetic and hadronic components, as identified by a separate CNN-based classifier~\cite{Psihas:2019ksa}.

\subsection{Near-to-far extrapolation and systematics}
\label{subsec: extrap and systs}
Predictions for the neutrino event rates at the FD are constrained using the high-statistics neutrino interactions measured in the ND. 
These measurements are used to devise corrections to the ND prediction, which we propagate to the FD by adjusting for the differing efficiency and flux between the detectors using simulations. 
We call this process ``extrapolation.''  
We apply oscillations to this data-driven prediction when comparing to FD data during inference (Sec.~\ref{subsec:osc inf} below).  

Corrections to the signal spectra for both \numu{} disappearance and \nue{} appearance channels arise from the \numu{} spectra at the ND.  
The near-to-far extrapolation for the \numu disappearance samples at the FD is performed in quartiles of hadronic energy fraction ($\hadfrac = \Ehad/E_\nu$, with $E_\nu$ being the reconstructed neutrino energy and \Ehad{} the reconstructed hadronic energy). % comments here to prevent an extra line break being added
\begin{figure*}
    \begin{subfigure}[b]{0.45\textwidth}
      \begin{center}
        \includegraphics[width=\columnwidth]{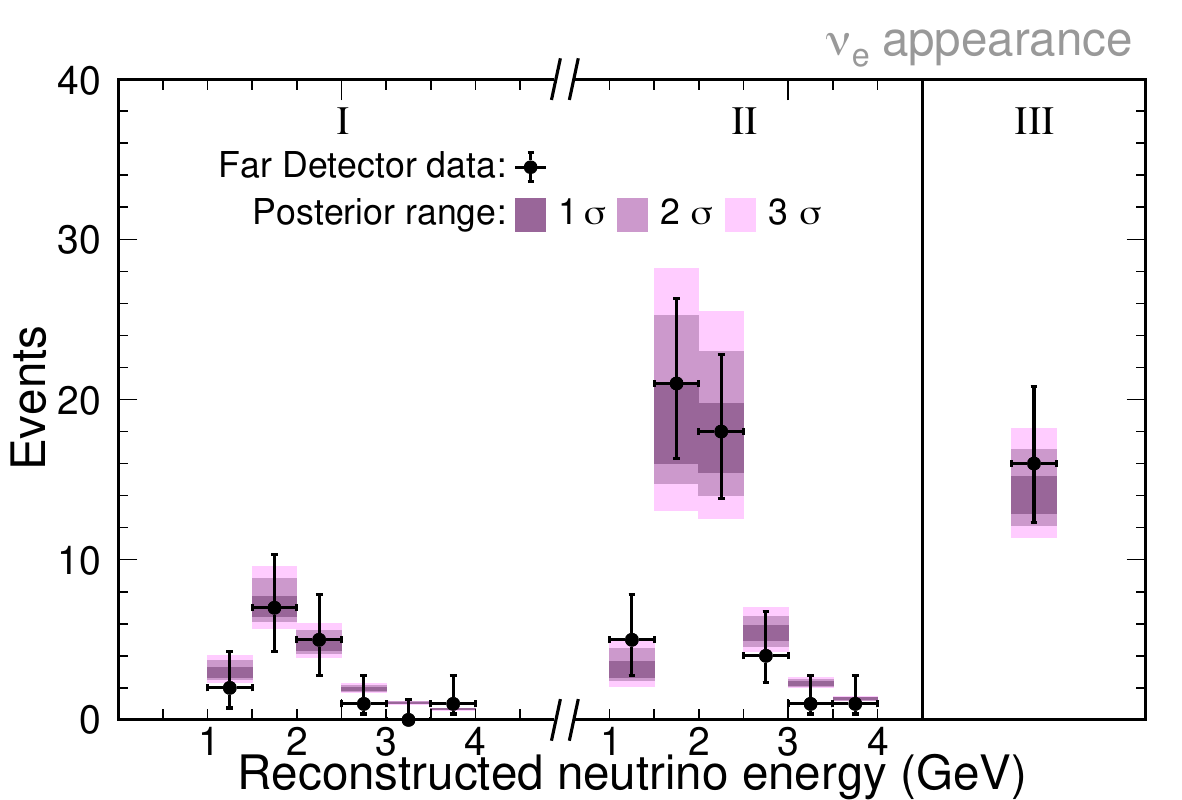}
      \end{center}
    \end{subfigure}
    \begin{subfigure}[b]{0.45\textwidth}
      \begin{center}
        \includegraphics[width=\columnwidth]{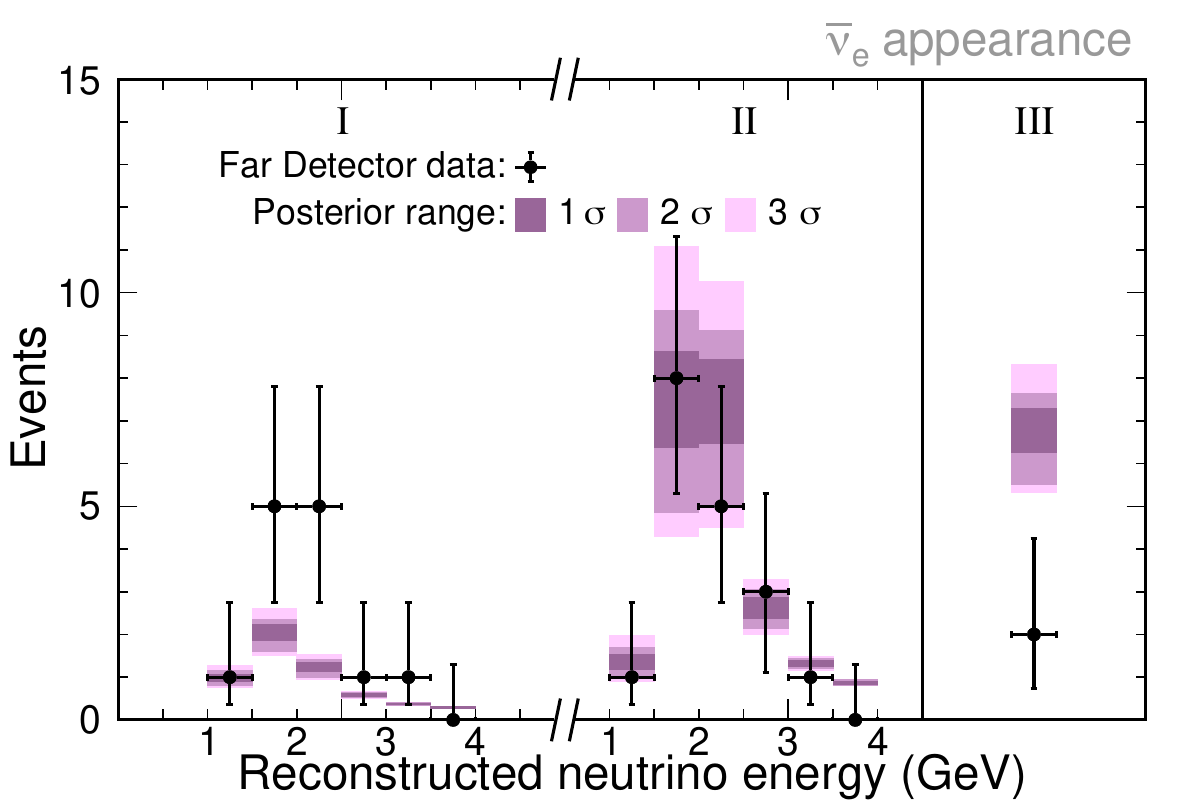}
      \end{center}
    \end{subfigure} 
    \begin{subfigure}[b]{0.45\textwidth}
        \begin{center}
        \includegraphics[width=\columnwidth]{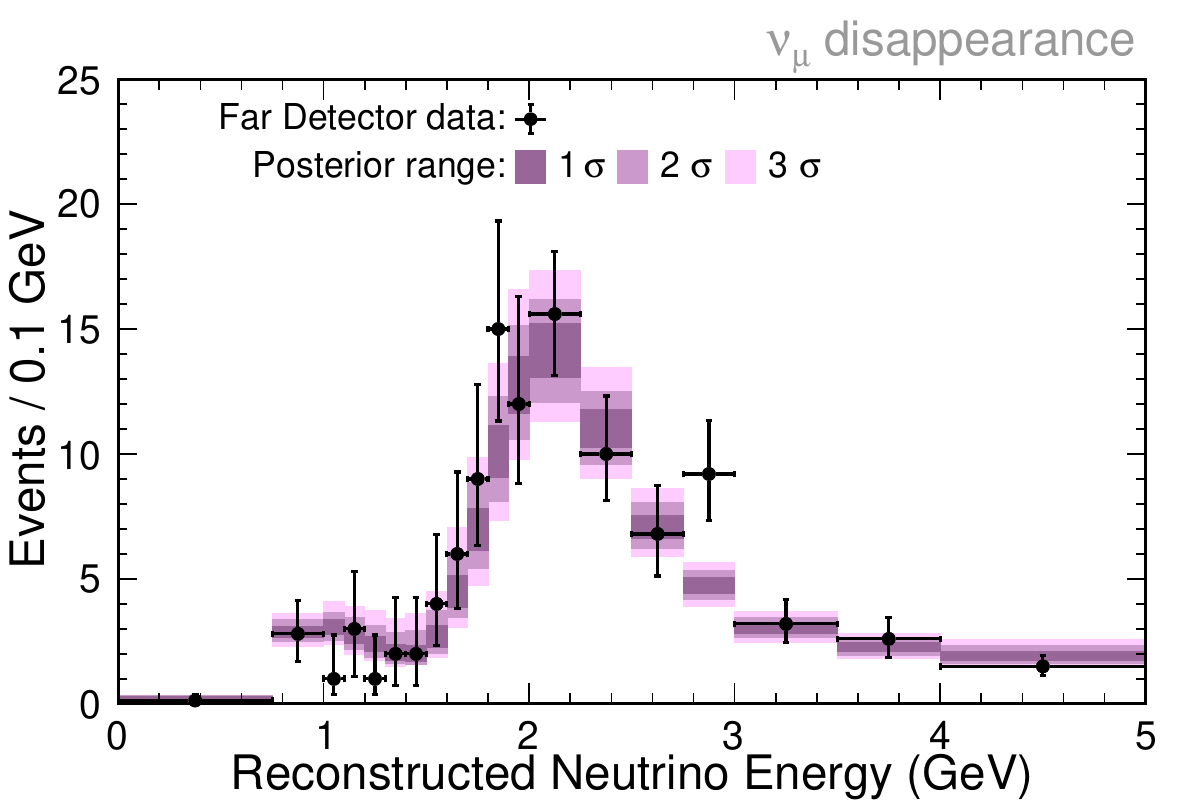}
        \end{center}
    \end{subfigure}
    \begin{subfigure}[b]{0.45\textwidth}
      \begin{center}
        \includegraphics[width=\columnwidth]{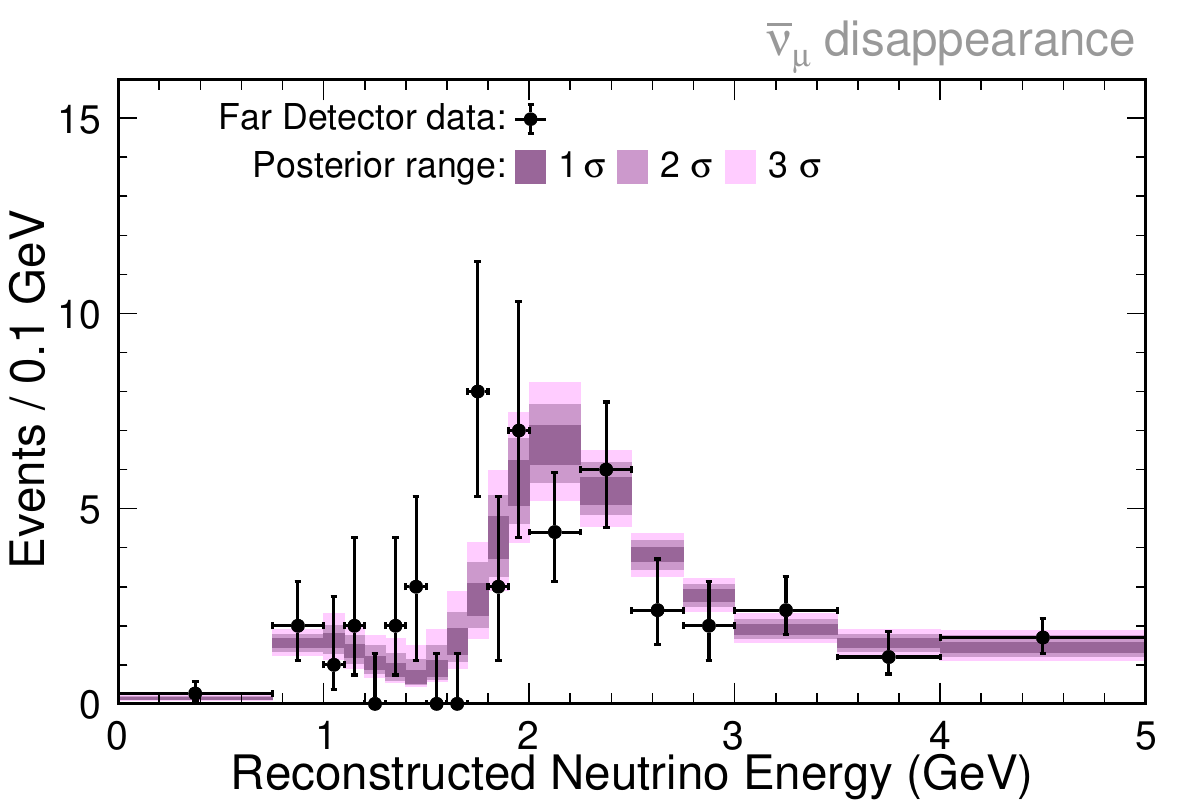}
      \end{center}
    \end{subfigure} 
    \caption{Reconstructed neutrino energy distribution of selected data
             events~(black crosses) in FD \nue CC samples~(top) and FD \numu CC
             samples~(bottom) in neutrino-enriched beam mode~(left) and
             antineutrino-enriched beam mode~(right). 
             The colored bands correspond to the range of 1\,$\sigma$ (darkest), 2\,$\sigma$, and 3\,$\sigma$ (lightest)
             of the extrapolated FD spectra produced using the combinations of the oscillation and systematic
             parameters sampled by our MCMC algorithms, illustrating the posterior distributions resulting from our data fit (described further in Sec.~\ref{subsec:osc inf}). 
             The FD \nue samples are divided into bins of low (I) and high (II) particle ID confidence
             as well as the peripheral (III) sample discussed in
             Sec.~\ref{subsec:sel and sim}. The four $E_{frac}$ \numu
             subsamples have been combined together in each of the lower two plots.}
    \label{fig:pppred}
\end{figure*} %

Extrapolating in the \hadfrac{} bins has the effect of grouping events that share similar hadronic system characteristics so that compatible events are constrained together in the FD sample, despite the detectors' somewhat different acceptances.  
Performing the oscillation fit in these bins enhances sensitivity to the oscillation dip as a function of $E_\nu$ since we achieve a finer energy resolution for bins with smaller \hadfrac{}.   
We further subdivide the near-to-far extrapolation into three bins of reconstructed transverse momentum of the outgoing charged lepton $p_T$ for both appearance and disappearance channels. 
Like the \hadfrac{} subdivision, this allows us to better match the constraints from the much smaller ND to the FD, in this case adjusting for the differing containment of events with leptons that emerge at large angles relative to the beam direction (i.e., large $p_T$).
The predictions in $p_T$ bins are summed prior to the oscillation fit.
We constrain the small beam backgrounds in the \nue appearance channels with a similar procedure based on ND \nue candidates after first decomposing them into NC, \numu~CC, and intrinsic \nue categories using data-driven constraints.
(\nuebar beam backgrounds are all constrained together rather than being decomposed this way.)
Cosmic backgrounds for all FD samples are determined from dedicated FD cosmic data samples.
The remaining minor backgrounds are estimated from simulation.
More detail on these procedures can be found in our previous paper \cite{NOvA:2021nfi}. 

We evaluate the impact of systematic uncertainties on the analysis by repredicting the sample spectra described above with altered parameters in the simulation.
Uncertainties in the neutrino flux and interaction model are treated using event reweighting. 
Uncertainties in the detector calibration and custom modeling of light in the detectors, on the other hand, must be fully resimulated.
After suppressing those that result in negligible changes to our spectra, 67 sets of systematically shifted simulations remain from these techniques, one set for each uncertainty.
Each variation in each set is extrapolated to the FD using ND data via the same process as above, which constrains the impact of the uncertainties on the predicted spectra.  
We use the extrapolated variations to obtain 67 parameterized interpolations of the uncertainties' constrained effect on the observables, and 67 free parameters corresponding to the latter are what are marginalized during the oscillation inference in Sec.~\ref{sec:MCMC details}. 

Figure~\ref{fig:pppred} shows the reconstructed energy spectra of the data observed at the FD using the \nue and \numu CC selections described in Sec.~\ref{subsec:sel and sim}, separated by the neutrino-enriched and antineutrino-enriched beam modes. 
Overlaid on these plots are bands of FD predictions produced according to the extrapolation procedure above.  
These illustrate the spectra predicted using the 68.3\%, 95.4\%, and 99.7\% highest probability values of the systematics just described and of the relevant oscillation parameters, determined using the MCMC algorithms that will be detailed in the next section.
For economy, here and in the following these ranges are labeled with the conventional shorthand in terms of Gaussian standard deviations $\sigma$, using the corresponding $z$-scores of 1\,$\sigma$, 2\,$\sigma$, and 3\,$\sigma$, respectively.

A Poisson likelihood \cite{ParticleDataGroup:2022pth} computed over the bins between the FD data and ND-constrained predictions such as those shown here forms the likelihood component of the posterior computed in Bayes' theorem below.

\section{Oscillation parameter inferences using Markov Chain Monte Carlo}
\label{sec:MCMC details}

We derive posterior probability density distributions for relevant oscillation parameters using Bayes' theorem \cite{Bayes:1763}.
Marginalizing away the nuisance parameters of our model, which include the tens of systematic uncertainties described in Sec.~\ref{subsec: extrap and systs}, is a challenging problem because it requires an integral over many dimensions.
We therefore turn to a Monte Carlo method for computing the posterior: Markov Chain Monte Carlo (MCMC).
In MCMC, we draw a collection of sequential samples from the posterior with a frequency proportional to the posterior probability density.
Histograms that approximate the posterior shape (with accuracy governed by the sample count) may be computed in any variable(s) of interest using these samples.
In so doing, any dimensions not explicitly summed are implicitly marginalized.\footnote{For an accessible introduction to MCMC, the reader is referred to Ref.~\cite{Speagle:2020}.  An exhaustive treatment may be found at Ref.~\cite{Brooks:2011a}.}
In our implementations, we obtain MCMC samples that draw values from the oscillation parameter space $(\dmlg, \ssth{13}, \ssth{23}, \dcp)$ and the 67 aforementioned parameters corresponding to our systematic uncertainties, for a total of 71 degrees of freedom.

Numerous algorithms for obtaining MCMC samples exist; we have implemented two for this analysis.
The conclusions obtained from them agree with one another.
Though descriptions of both methods are readily found in the literature, there are certain implementation choices that must be made for each, which we discuss in Sec.~\ref{subsec:MCMC NOvA impl}.
Sec.~\ref{subsec:osc inf} lays out our resulting inferences on the parameters.

\subsection{NOvA MCMC implementations}
\label{subsec:MCMC NOvA impl}

\subsubsection{``ARIA:'' \mrrtt algorithm}
\label{subsubsec: ARIA}

The traditional MCMC algorithm, the Metropolis--Rosenbluth--Rosenbluth--Teller--Teller (\mrrtt{}) method\footnote{Historically this method was known as the ``Metropolis'' (or ``Metropolis--Hastings'') method, referring to the first author of the seminal papers~\cite{Metropolis:1953am, Hastings:1970aa}.  We follow more recent convention and hereafter refer to it by all of the authors' names.  See also Ref.~\cite{Gubernatis:2005zz}.} is straightforward.
We call our implementation ARIA, in honor of Arianna Rosenbluth, who first implemented the method in machine code~\cite{Rosenbluth:2003}.
Proceeding from an initial seed in the parameter space, subsequent samples are selected by proposing a jump to a new set of coordinates, and accepting or rejecting that proposal according to an acceptance rule \cite{ParticleDataGroup:2022pth}.
This process is repeated until a sufficient number of samples have been collected.
There is no explicit stopping criterion.
The method also does not specify the distribution to be used in the proposal algorithm.
In our implementation, we use a common choice, which is of a multivariate Gaussian.
We determine the characteristic length scales and correlations of the Gaussian empirically in order to optimize sampling efficiency (see App.~\ref{app:step sizes}).
Our ARIA results below have $10^5$ effective samples (see App.~\ref{app:warmup and thinning}).

\subsubsection{``Stan:'' Hamiltonian MCMC}

Though the \mrrtt{} method proposes samples quickly, they are typically highly autocorrelated.
Its sampling proposals can also be inefficient if the posterior has both sharply concentrated and broader regions.
Other MCMC methods have been developed to address these shortcomings, including one called ``Hamiltonian'' MCMC inference (\hmcmc).
We implemented a C++ interface to the Stan modeling platform~\cite{Stan-manual} to obtain \hmcmc{} samples.

The main difference between \hmcmc{} and \mrrtt{} is how proposals are generated.
Rather than proposing randomly, \hmcmc{} views the posterior surface as a topographical one that can be explored by a fictitious particle.
Samples correspond to this particle's trajectories under the influence of a gravitational potential whose gradient aligns with the direction of higher posterior density.
Endowing the particle with an initial momentum that counterbalances the centripetal force from gravitation produces stable trajectories that traverse the highest density region of posterior space \cite{Betancourt:2017}.
\hmcmc{} does this by numerically integrating Hamilton's equations for the fictitious particle system with its position $\vec{q}$ (which correspond to the parameters of interest) and momentum $\vec{p}$ coordinates, and a Hamiltonian $H = -\log (\text{posterior})$.
This approach produces samples that are nearly uncorrelated at the expense of additional computing cycles to compute the gradient of the posterior.
The $7 \times 10^{7}$ Stan samples we obtained may be compared to the $10^{5}$ effective samples mentioned above for ARIA.
We find Stan's default choices of the sampling distribution for the pseudoparticle kinetic energies and the integration stopping condition to be sufficient for our needs (see App.~\ref{app:Stan defaults}).

Its topographical nature means that unlike \mrrtt{}, \hmcmc{} is ill-suited to parameters that assume only one of a discrete set of values, which would manifest as discontinuities in the trajectories considered.
This presents a difficulty in neutrino oscillation parameter inference, where the absolute value of \dmsq{32} is known with relatively good precision, but its sign (which determines the neutrino mass ordering) remains an important unknown.
While it is possible to allow \hmcmc{} to explore the entire range of \dmsq{32}, we find in practice that this results in poor exploration, as few trajectories manage to ``jump'' across the wide disfavored region $|\dmsq{32}| \lesssim \unit[2 \times 10^{-3}]{eV^2}$.
Instead, at the end of each trajectory determination, we introduce a separate \mrrtt{}-like step which considers the possibility of changing the sign of \dmsq{32} according to the prior probability chosen for it (50\%, i.e., uniform prior; see Sec.~\ref{subsubsec:priors} below).
If the acceptance ratio between both mass orderings satisfies the \mrrtt{} criteria, the proposed sign is retained; if not, it is reverted to its previous value.\footnote{Though explicitly treating the mass ordering in this way is not required by the ARIA method, since it does not require continuity between samples, we find that it significantly improves the sampler performance.  Thus we use it for ARIA as well.}

\subsubsection{Choices of prior}
\label{subsubsec:priors}

In Bayesian inference, the posterior probabilities are influenced by the choice of prior probability densities, which encode assumptions made about the parameters before the data is examined.
If the data used for a measurement is sufficient, its constraint on the posterior will typically overwhelm the prior, rendering the prior choice unimportant.
However, when data is sparse, or when the prior vanishes in regions of the parameter space, the choice of prior may affect the result.
The priors we choose differ according to the parameters being considered.

\paragraph{Parameters of interest: $|\dmsq{32}|$, $\mathrm{sgn}(\dmsq{32})$, \ssth{23}}
\label{par:LBL osc}
We prefer to use ``uninformed'' priors, which do not favor any particular value, for the physics parameters we intend to measure directly.
In practice, this usually amounts to a prior uniform in the variable in question (which, in the special case of the binary parameter $\mathrm{sgn}(\dmsq{32})$, corresponds to 50\% probability for each of the two options).
However, uniformity is not preserved under a change of variable: for instance, a prior uniform in \thatm{} is not uniform in the measured variable \ssth{23}.
In our results below, we have studied the impact of priors uniform both in a particular variable and relevant functions of it, and we report when the prior choice significantly affects the results.

\begin{figure}[h]
	\includegraphics[width=\linewidth]{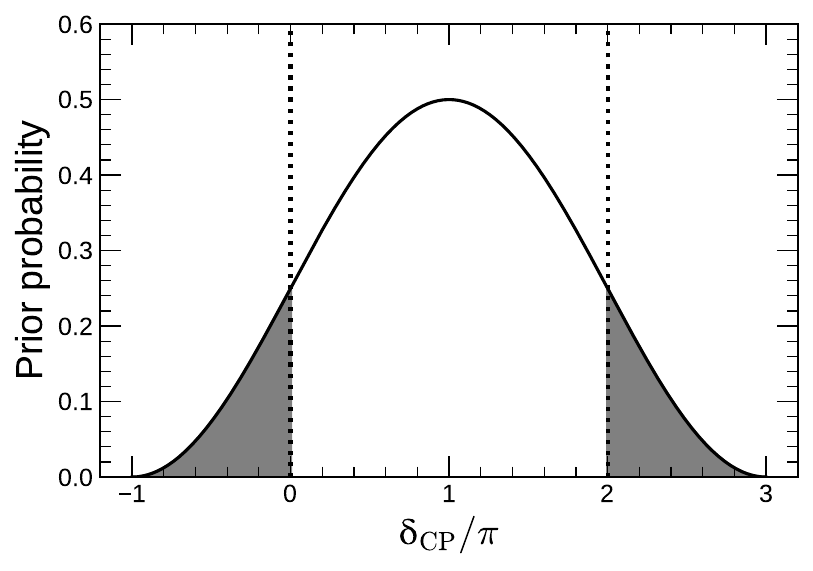}
	\caption{Special prior distribution from Eq.~\ref{eq:dcp prior} used for \dcp.  The sum of the prior for any point within the region between the dotted lines with the corresponding point in the gray shaded areas that differs from it by $\pm 2\pi$ is always $\frac{1}{2}$.}
	\label{fig:dcp tophat prior}
\end{figure}

\paragraph{Parameter of interest: \dcp}
\label{par:dcp}
While \dcp{} is intended to have a uniform prior as well, it receives additional special treatment.
Its cyclical nature as the phase of a complex number results in an infinite set of values having identical consistency with the data.
This can cause MCMC samplers never to converge on a single value for \dcp{} and consequently to sample much more slowly.
To combat this problem, we developed a novel special prior over \dcp:
\begin{equation}
	\label{eq:dcp prior}
	\Pi(\dcp) = 
		\begin{cases}
			\frac{1}{2} \sin^2\left( \frac{1}{4}(\dcp + \pi)\right), & -1 \leq \dcp/\pi \leq 3 \\
			0,                                                       & \text{otherwise}.
		\end{cases}
\end{equation}
This function is illustrated in Fig.~\ref{fig:dcp tophat prior}.
This prior forces the samplers to remain near a single phase of \dcp, \mbox{$0 \leq \dcp/\pi \leq 2$}, as the prior vanishes outside of [-1, 3].
Because it makes the transition to the vanishing regions in a differentiable manner, this prior is suitable for use with \hmcmc{}.
Moreover, the sum of the prior's value at every point $0 \leq \dcp/\pi \leq 2$ with that of all the points outside that interval that share the same phase is a value that does not depend on \dcp:
\begin{equation}
	\sum_{n=-\infty}^{\infty} \Pi(\dcp + 2\pi n) = \frac{1}{2}.
\end{equation}
This implies that the prior behaves identically to a uniform prior when used in conjunction with the oscillation probability.
For the rest of this paper, whenever we refer to a ``uniform'' prior in \dcp{}, we mean this prior.

When in the subsequent sections we study the effect of choosing a prior uniform in \dcp{}\ vs.\ $\sin(\dcp)$, we reweight the MCMC samples obtained with the prior above, taking the Jacobian factor $\partial(\sin(\dcp)) = \cos(\dcp)$ as the weight.

\paragraph{Parameter of interest: \sstth{13}}
\label{par:ssthreac}
As discussed further in Sec.~\ref{subsec:osc inf} below, we consider two separate cases for \sstth{13}: one where its prior is treated identically to that of \ssth{23} (see~\ref{par:LBL osc}), and one where measurements from short-baseline reactor antineutrino oscillations are applied as a constraint.
In the latter case, we impose a Gaussian prior with standard deviation obtained from the 2019 world average of reactor measurements~\cite{ParticleDataGroup:2018ovx}, analogous to the treatment in our most recent frequentist result~\cite{NOvA:2021nfi}: $\sstth{13} = 0.085 \pm 0.003$.

\paragraph{Systematic uncertainties}
Our implementation varies systematic uncertainty parameters in units of their standard deviation from the model's nominal value. 
The response of the prediction as a function of standard deviation from nominal for each systematic uncertainty is the same as in the frequentist analysis~\cite{Nayak:2021eal}.
The model being fitted uses predictions constructed using the extrapolation described in sec.~\ref{subsec: extrap and systs}, which has the effect of implicitly applying the relevant constraints from the ND.
As this method of employing the ND information does not rely on fitting model parameters to the ND data---rather, it effectively reduces the systematic uncertainty variations to consist of only the components not shared by the ND and FD---the systematic uncertainty parameters remain \textit{a priori} uncorrelated in our MCMC sampling.
Thus, for all systematic uncertainties we use a unit Gaussian distribution, uncorrelated from all other parameters, as their prior.

\subsection{Oscillation inferences: results and discussion}
\label{subsec:osc inf}

In this section we describe our inferences regarding the PMNS neutrino oscillation
parameters, given the data, model, and the Bayesian methodology described
above. 
Two separate sets of results are obtained.
The first uses a Gaussian prior on \sstth{13}, imposing a constraint from reactor antineutrino experiments (Sec.~\ref{ssub:pmns_params}-\ref{subsubsec:CPV and J}).
The second uses a prior uniform in \sstth{13}, yielding results constrained only by the NOvA data (Sec.~\ref{subsubsec: NOvA-only}). 
Samples obtained from ARIA and Stan produce essentially identical distributions in all the variables considered (App.~\ref{app:ARIA-Stan equivalence}).

\subsubsection{Goodness of fit}
\label{subsubsec:goodness of fit}

To evaluate the goodness of the fit for this Bayesian analysis we use posterior predictive $p$-values (PPP)~\cite{Gelman:1996}. 
In a PPP test, the coordinates from each MCMC sample are used to form a prediction for the observed spectra.
A Poisson $\chi^2$ statistic is computed between each prediction and the data, which we denote as $\chi^2_{\textrm{data}}$.
A second prediction is made for each sample by applying Poisson fluctuations to the prediction above, and a second $\chi^2_{\textrm{pseudodata}}$ calculated between this pseudodata and the original prediction.
Because the data spectra are unchanged in this process, $\chi^2_{\textrm{data}}$ incorporates only variations in the oscillation parameters and systematic uncertainties.
Conversely, since the pseudodata distributions have Poisson fluctuations applied but use the same oscillation parameters and systematic uncertainty pulls as the base model for each MCMC sample, $\chi^2_{\textrm{pseudodata}}$ treats only statistical uncertainties.

\begin{figure}[h!]
	\centering
	\includegraphics[width=0.9\columnwidth]{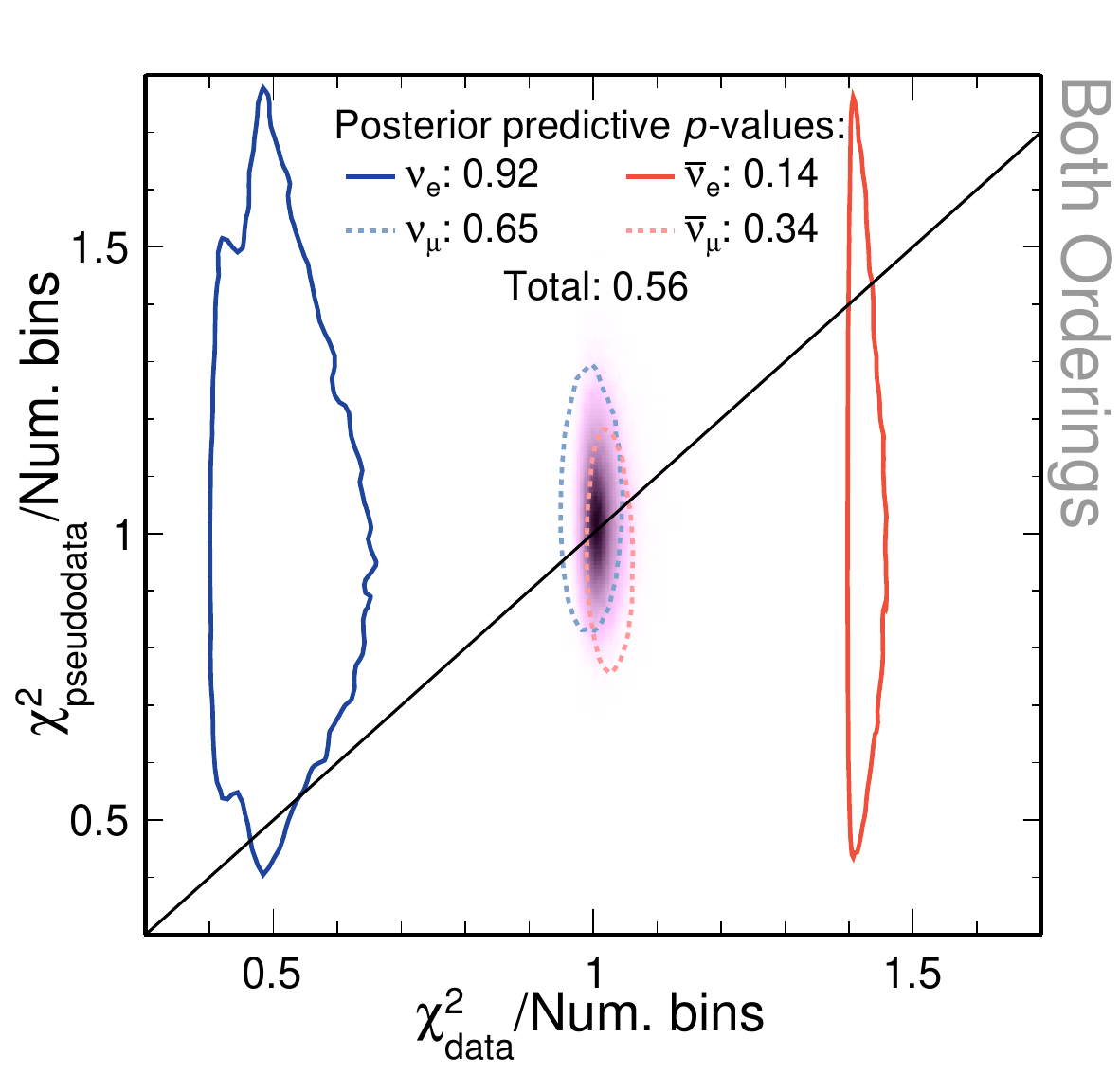}
	\caption{Posterior predictive $p$-values from real data MCMC samples,
		with \threac constraint from reactor experiments applied. The purple distribution is a 
		scatterplot of the binned $\chi^2$ computed between the model and real data spectra ($x$-axis), 
		against a similar $\chi^2$ between the model and pseudodata spectra ($y$-axis), both
		divided by the number of degrees of freedom (DOF) in the fit, computed for each MCMC sample. 
		%		(See the text for more on how the pseudodata spectra are constructed.)
		The dashed and solid contours show 1\,$\sigma$ intervals from the same
		posterior-predictive distributions calculated only for $\nu_e$ (dark blue, left
		solid), $\bar{\nu}_e$ (red, right solid), $\nu_{\mu}$ (light blue, left dashed) and
		$\bar{\nu}_{\mu}$ (light red, right dashed) data samples. The posterior
		predictive $p$-values shown in the legend are the fraction of each distribution that lies above
		the diagonal $\chi^2_{\textrm{data}} = \chi^2_{\textrm{pseudodata}}$ line
		(black).  
		%	The fact that the purple distribution is centered at $(1, 1)$ and is evenly distributed above and below the the diagonal, with the associated $p$-value of 0.56, indicates a good fit.
	}
	\label{fig:ppvwrc}
\end{figure}

The distribution of these $(\chi^2_{\textrm{data}}, \chi^2_{\textrm{pseudodata}})$ pairs for the entire ensemble of spectra considered in Fig.~\ref{fig:pppred} is shown as the purple shading in Fig.~\ref{fig:ppvwrc}.
The PPP then consists of the fraction of points in this ensemble that lie above the $\chi^2_{\textrm{data}} = \chi^2_{\textrm{pseudodata}}$ line. 
In the limit of infinite MCMC samples, a model that perfectly describes the data apart from statistical variations will produce a PPP of 0.5.
We observe that the shaded distribution in Fig.~\ref{fig:ppvwrc} is distributed evenly around 1 unit of $\chi^2$ per bin in both axes, and the PPP we obtain is 0.56.
Both of these imply that the model is a good representation of the data.

We also find good $p$-values for the \numu{} and \numubar{} samples considered independently, whose 1\,$\sigma$ credible regions (those enclosing 68.3\% of the posterior, as with the ranges for the spectra in Fig.~\ref{fig:pppred}) are indicated by the dashed light blue and light red colored contours in Fig.~\ref{fig:ppvwrc}, respectively.
By contrast, the effect of fluctuations in the smaller-statistics samples can be seen more readily in the analogous \nue{} (solid dark blue) and \nuebar{} (solid red) contours.
Both contours are much larger than their \numu{} counterparts, particularly along the y-axis, which corresponds to the dimension where statistical uncertainties are considered.
In the \nue{} contour, the offset downwards from unity along the x-axis, and the corresponding shift above the diagonal along the y-axis, together suggest a set of fluctuations that are relatively close to the Asimov (unfluctuated) model prediction.
Each member of the ensemble of pseudodata spectra thus typically has larger $\chi^2$ relative to the Asimov than that of the data.
This relative closeness of the \nue{} prediction to the data can also be seen in the top left panel of Fig.~\ref{fig:pppred}.
The unusual shape of this contour arises because the PPP distribution for \nue{} consists of two modes superimposed upon one another, corresponding to the high-probability regions within the NO that will be discussed below in Fig.~\ref{fig:datawrdcpthatm2d}.
On the other hand, the situation is reversed for the \nuebar{} spectra, where the data fluctuations seen in Fig.~\ref{fig:pppred} result in larger deviations from the Asimov prediction than the bulk of the pseudodata spectra, and the contour consequently shifts downward from the diagonal.
The fact that combining all four subsamples together produces a PPP that is closer to 0.5 than any of them are individually is good evidence that the deviations from PPP of 0.5 in the various subsamples are predominantly driven by statistical, rather than systematic, effects.\footnote{The reader may find further information on interpretations of posterior predictive $p$-values in Refs.~\cite{Gelman:2002,Gelman:2013}.}
Moreover, studies where Poisson fluctuations were applied to the model to produce fake data spectra, which were then subjected to the PPP computation process, indicated that both large and small PPP values such as those we observe here do naturally occur in the subsamples.
We conclude therefore that our set of MCMC samples reflect PMNS parameters with a good description of the physics exhibited in our data.

\subsubsection{PMNS parameter measurements}
\label{ssub:pmns_params}

\begin{figure}[htbp]
	\centering
	\includegraphics[width=0.95\columnwidth]{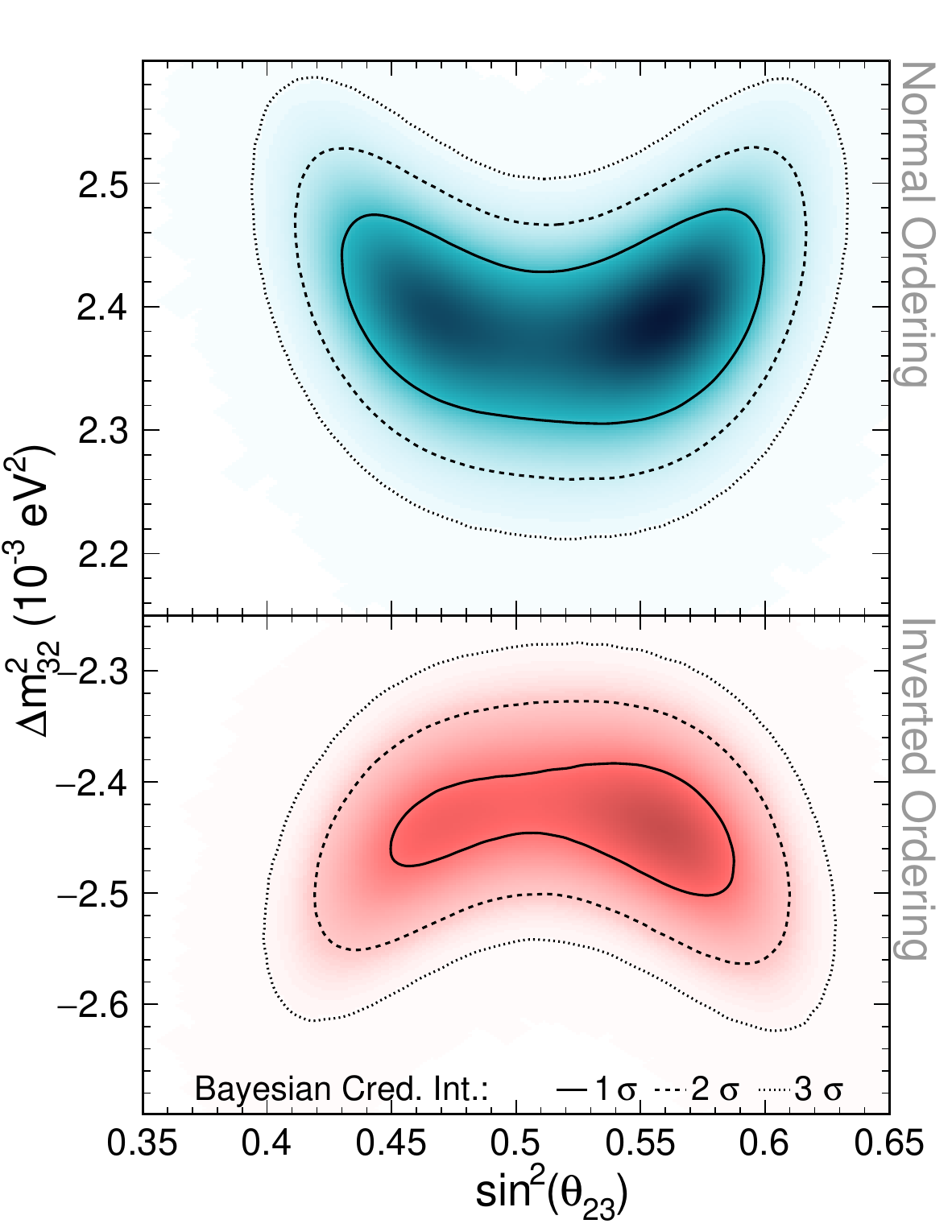}
	\caption{Binned posterior probability densities (shaded) for
		\ssth{23}--\dmlg, marginalized over both mass orderings and plotted
		separately for the normal (top) and the inverted (bottom) mass orderings
		(marginalization over the mass orderings is explained at the beginning of
		Sec.~\ref{ssub:pmns_params}). Contours indicate regions enclosing
		1\,$\sigma$, 2\,$\sigma$ and 3\,$\sigma$ of the posterior probability.}
	\label{fig:datawrcdmthatm2d}
\end{figure}

\begin{figure}[htpb]
	\centering
	\includegraphics[width=\columnwidth,trim={0 0 0 1cm},clip]{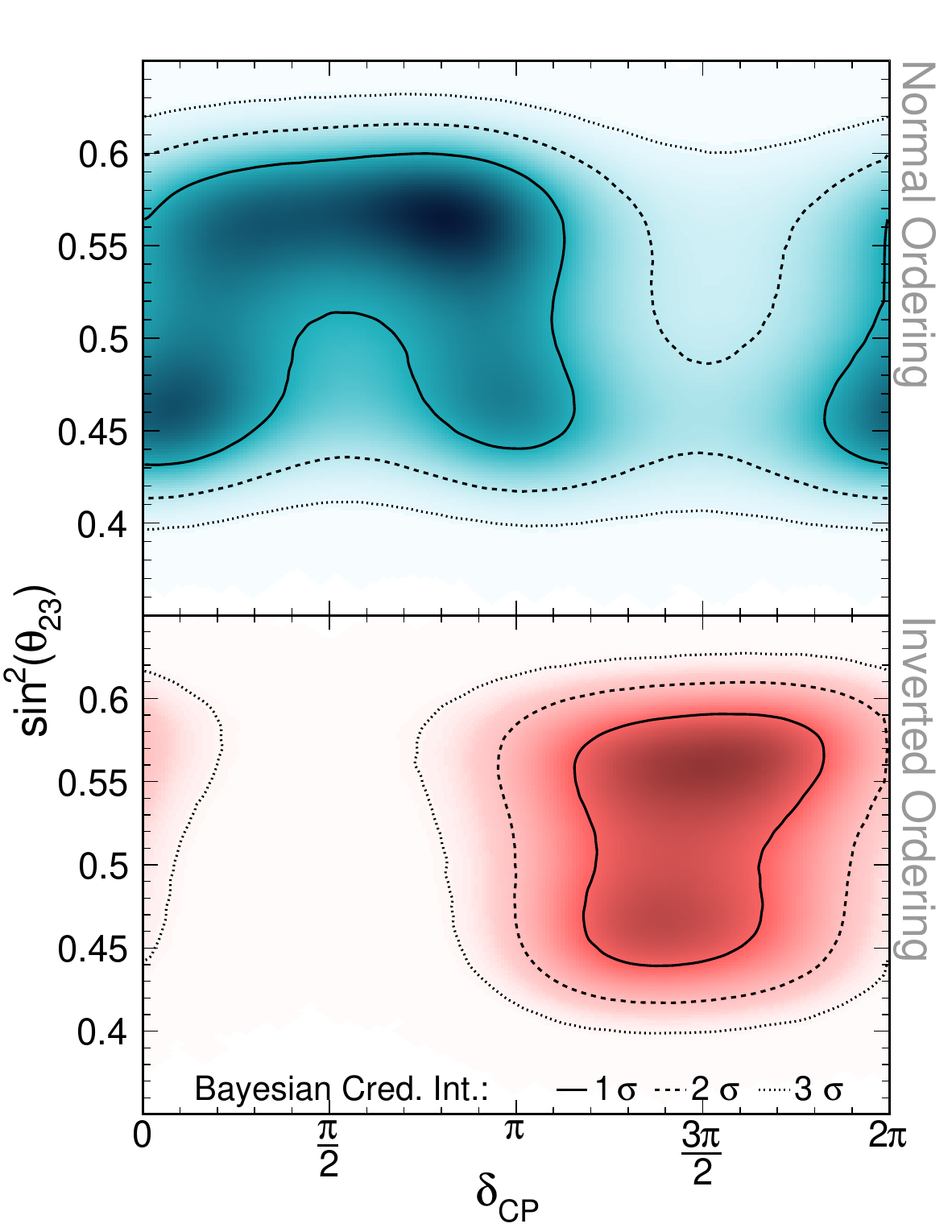}
  \caption{Binned posterior probability density (shaded) with 1, 2, and
           3\,$\sigma$ credible intervals for \dcp--\ssth{23}, marginalized
           over both mass orderings for the normal mass ordering (top, blue)
           and the inverted mass ordering (bottom, red). The external constraint on \threac{} from reactor experiments was applied.}
	\label{fig:datawrdcpthatm2d}
\end{figure}

We produced Bayesian credible regions for the PMNS neutrino oscillation parameters using the data spectra and the MCMC samplers described in sections~\ref{subsec:MCMC NOvA impl} and~\ref{subsec: extrap and systs}, respectively.
These credible regions together with the posterior probability distributions
are shown in Figs.~\ref{fig:datawrcdmthatm2d}~and~\ref{fig:datawrdcpthatm2d}.
In both cases the constraint on \threac from the reactors used is the 2019 PDG's combination
of extant measurements~\cite{ParticleDataGroup:2018ovx} and is applied in the form of
a Gaussian prior on \sstth{13}, as explained in section~\ref{par:ssthreac}.
In the figures in this section, credible intervals that show the normal and inverted mass orderings separately are created by first making one shared credible interval that spans both.  
The separate panels then display the relevant regions from this shared interval that apply to the specified ordering.
By constructing them in this manner, we ensure the NO and IO intervals share a highest-posterior density point and posterior probability distribution, preserving any NOvA preference towards one of the mass orderings in the credible region limits.
Similarly, the distributions and intervals labeled ``both orderings'' are created from MCMC samples over all values for \dmsq{32}, summing together the normal and inverted ordering posteriors before extracting the credible intervals.
Versions of these distributions with an alternate marginalization scheme that considers each ordering independently may be found in the Supplemental Material~\cite{supp}.

\begin{table}[htbp]
	\centering
	
	\caption{Highest posterior density points (HPD) together with the
		1\,$\sigma$ Bayesian credible interval limits for the PMNS
		parameters of interest, marginalized over all the mass ordering (MO)
		hypotheses. Marginalization over the mass orderings is explained at
		the beginning of Sec.~\ref{ssub:pmns_params}. In these results a Gaussian 
		prior corresponding to the reactor constraint on \sstth{13} 
		(see sec. \ref{par:ssthreac}) is applied.}
	
	\begin{tabular}{c  c  c  c}
		\hline
		\hline\\[-2.3ex]
		& MO & HPD & $1\,\sigma$ \\
		\hline\\[-2.3ex]
		
		\multirow{1}{*}{\dcp} & Both & $0.91\pi$ & $[0.02\pi, 0.31\pi]\cup[0.68\pi, 1.67\pi]$ \\
		(Prior uniform  & Normal & $0.89\pi$ &  $[0.54\pi, 1.07\pi]\cup[1.99\pi, 0.48\pi]$  \\
		in \dcp)                & Inverted & $1.44\pi$ &  $[1.26\pi, 1.65\pi]$ \\
		\hline\\[-2.3ex]
		
		\multirow{3}{*}{\ssth{23}} & Both & $0.56$ &  $[0.45, 0.49]\cup[0.52, 0.59]$ \\
		& Normal & $0.56$ &  $[0.44, 0.59]$ \\
		& Inverted & $0.56$ &  $[0.55, 0.57]$ \\
		\hline\\[-2.3ex]
		
		\dmsq{32}& Normal & $2.39$ &  $[2.32, 2.46]$ \\
		($\times10^{-3}$ eV$^{2}$) & Inverted & $-2.44$ &  $[-2.47, -2.41]$ \\
		\hline
		\hline
	\end{tabular}
	\label{tab:pmnswrc}
\end{table}
Table~\ref{tab:pmnswrc} shows the highest posterior probability density (HPD) points together with the 1\,$\sigma$ credible intervals.
These points are given for all the PMNS oscillation parameters of interest, split into both, normal, and inverted mass orderings (using the same methodology regarding the mass ordering as for the figures).
For some of the parameters the 1\,$\sigma$ region spans disjoint areas; we denote this with a union symbol $\cup$. 
These high posterior probability regions are in generally good agreement with the frequentist
analysis of the same dataset~\cite{NOvA:2021nfi}.

\begin{figure}[htpb]
			\includegraphics[width=0.95\columnwidth,trim={0 0 0 1cm},clip]{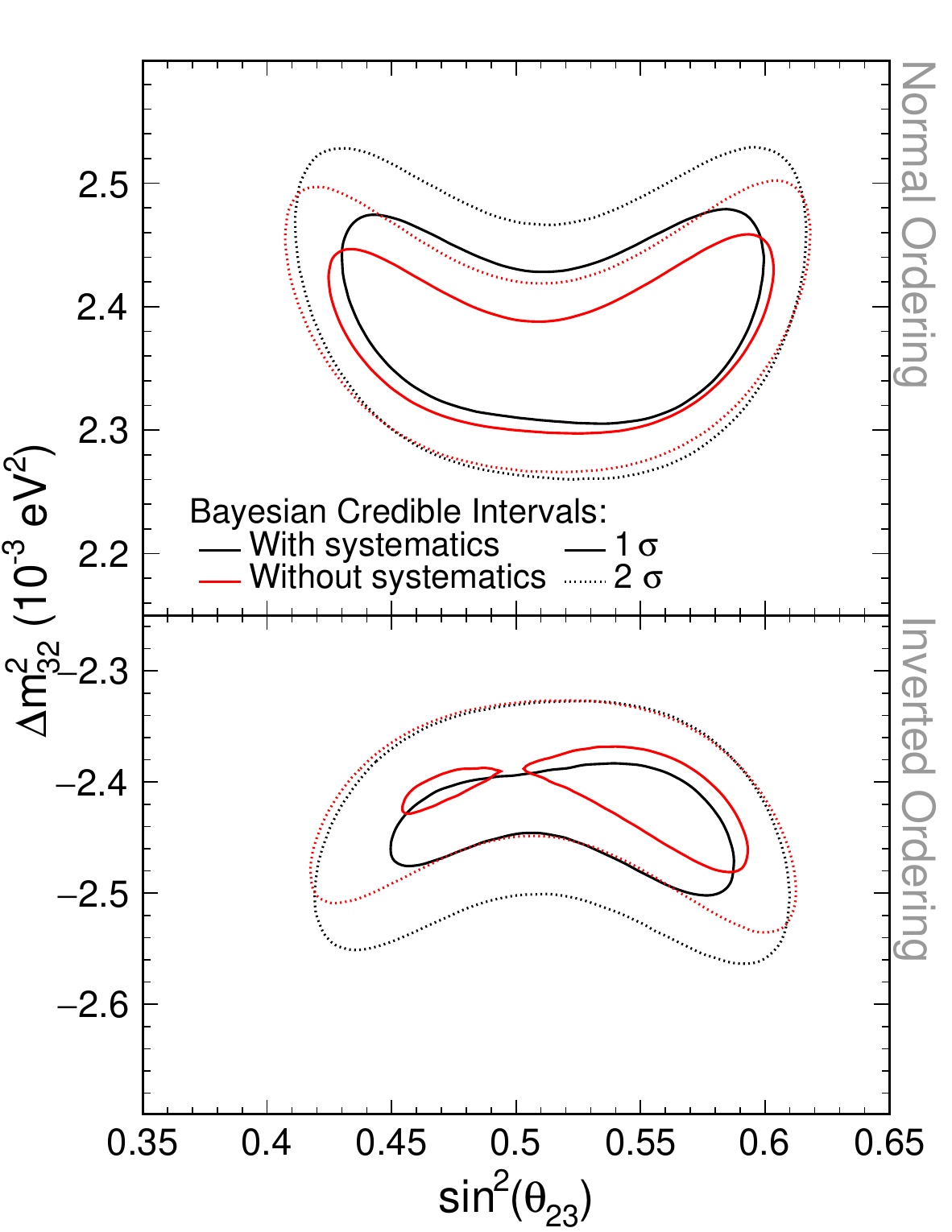}
  \caption{Credible interval comparisons for \mbox{\ssth{23}--\dmlg} when sampling with only 
           statistical uncertainties (red) and with all the NOvA systematic
           parameters (black). The external constraint on $\theta_{13}$ from the
           reactor experiments was applied in these fits.}	
	\label{fig:datawrcstatsvssystdisapp}
\end{figure}

\begin{figure}[htpb]
	\begin{center}
		\includegraphics[width=\columnwidth]{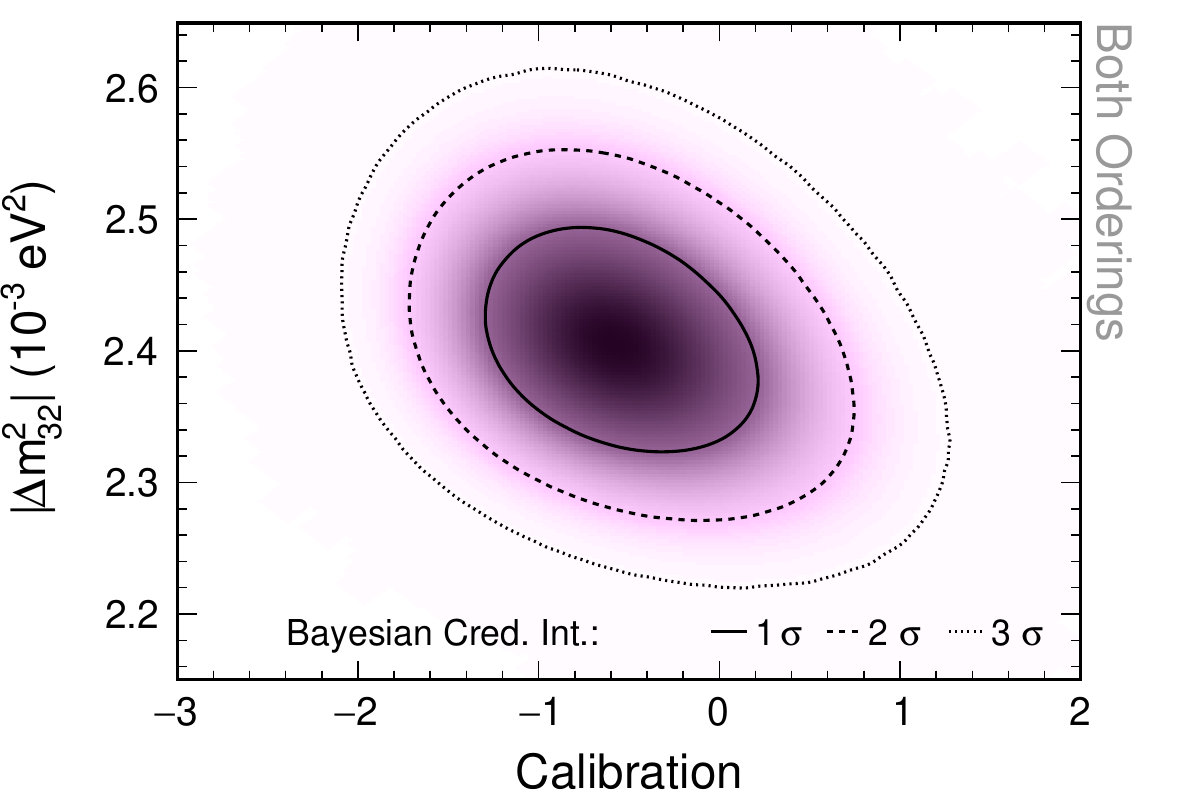}
	\end{center}
  \caption{Binned posterior probability density (shaded) with 1, 2, and
  	       3\,$\sigma$ credible intervals in $|\dmlg|$ and the NOvA absolute calibration
           systematic uncertainty, where the horizontal axis is measured in units of
           standard deviations from the nominal value. The external constraint
           on $\theta_{13}$ from reactor experiments was applied in this
           fit.}	
	\label{fig:datawrcdmcalib2d}
\end{figure}

Figure~\ref{fig:datawrcdmthatm2d} shows the \ssth{23}--\dmlg plane, 
where the denser MCMC samples (darker color) and larger credible
regions in the upper panel relative to the lower indicate a mild preference for
the normal ordering. This conclusion holds in the presence of the entire
systematic uncertainty model discussed in Sec.~\ref{subsec: extrap and systs}.
However, the preferred regions in each mass ordering depend in a nontrivial way
on the systematic uncertainties, as illustrated in Fig.~\ref{fig:datawrcstatsvssystdisapp}.
The most significant effect is in \dmlg, where the systematic effects not only
broaden the preferred region, but also shift the most probable value to larger
absolute magnitudes. The most important of the uncertainties contributing to
this movement is in the absolute calibration of the calorimetric energy scale.
Because this directly affects reconstructed neutrino energies, it shifts the
expected number of events in the trough of the \numu and \numubar disappearance
spectra and is thus anticorrelated with \dmlg{} (correlation coefficient $-0.29$), as can be seen in
Fig.~\ref{fig:datawrcdmcalib2d}. 
Shifting \dmlg{} in this way also moves \ssth{23} closer to maximal disappearance ($\sin^{2}{\theta_{23}^{\mathrm{MD}}} \approx 0.51$; the value depends slightly on \ssth{13}).
Because the \numu{} disappearance spectra are essentially identical for values reflected across the maximal disappearance line, the credible regions are nearly symmetric around it; the degeneracy is broken only by the \nue{} appearance spectra, which have less statistical power.
Thus, the credible regions for smaller $|\ssth{23} - \sin^{2}{\theta_{23}^{\mathrm{MD}}}|$ appear narrower, even though the sensitivity is unchanged.
This effect is responsible for the apparently slightly tighter constraint on \ssth{23} observed in Fig.~\ref{fig:datawrcstatsvssystdisapp} under the effect of systematic uncertainies.

%\FloatBarrier


\begin{figure}[htpb]
			\includegraphics[width=\columnwidth,trim={0 0 0 1cm},clip]{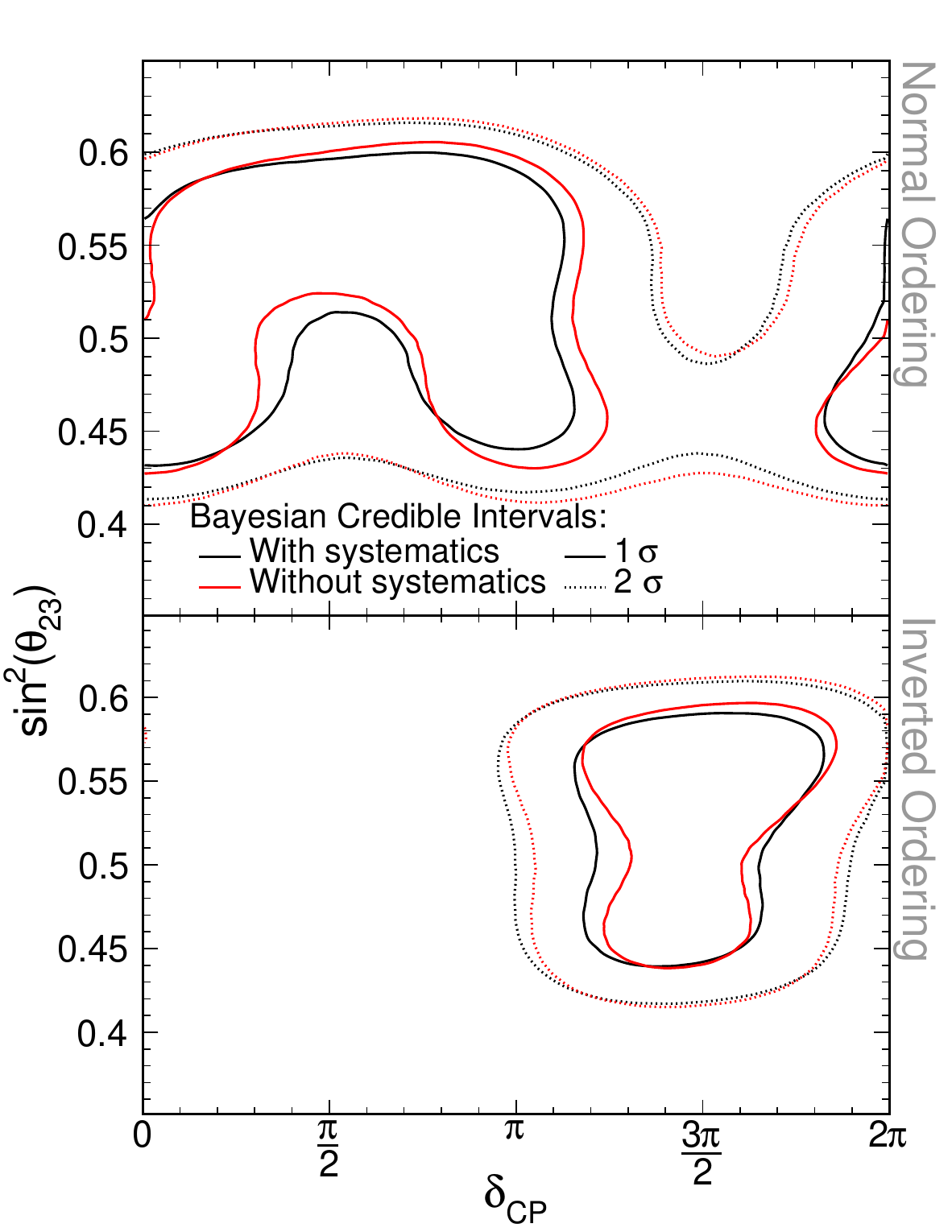}
  \caption{Credible interval comparisons for \mbox{\dcp--\ssth{23}} between
           statistical-only fits and fits including all the NOvA systematic
           parameters, with external constraint on \threac from reactor experiments.}
	\label{fig:datawrcstatsvssystapp}
\end{figure}

The situation is different in the \dcp{}--\ssth{23} plane, which is shown in
Fig.~\ref{fig:datawrdcpthatm2d}. Here, we observe preferences for
CP-nonconserving values of \dcp{} (i.e., nonintegral values of $\dcp/\pi$) in
both normal and inverted orderings, and for the upper octant ($\ssth{23} > 0.5$)
of \thatm{} (reflected also in Fig.~\ref{fig:datawrcdmthatm2d}), though the
CP-conserving points are only weakly disfavored in NO. However, in contrast to
the conclusions for \ssth{23}--\dmlg, these inferences are
minimally affected by the presence of systematic uncertainties, as
Fig.~\ref{fig:datawrcstatsvssystapp} makes clear---even if the
resolution does degrade slightly and the credible regions grow. 
Here the minimal impact of systematic uncertainties
owes primarily to the smaller statistics of the \nue{} and especially \nuebar{}
appearance samples that drive the sensitivity to these parameters,
which results in statistical uncertainties dominating the uncertainty budget.


To more quantitatively assess the mass ordering and octant preferences, we give
the posterior probabilities inferred for each combination of hypotheses in
Table~\ref{tab:bayesfactorwrc}. We also express them in a less prior-dependent
way using Bayes factors.
In both cases the evidence for either option is weak.\footnote{The reader unfamiliar with the interpretation of Bayes factors is referred to the standard treatments of Jeffreys~\cite{Jeffreys:1961} or Kass~\&~Raftery~\cite{KassAndRaftery:1995}.}
The analogous Gaussian $p$-values also correspond to significances of less than
1\,$\sigma$. The
interpretations of the octant and the mass ordering hypothesis preferences are
in good agreement with the 2020 frequentist analysis of the same dataset, which
used profiling with Feldman--Cousins corrections instead of
marginalization~\cite{NOvA:2021nfi, NOvA:2022wnj}. This general agreement is
also true for the parameters' intervals, with small differences expected given
the differing statistical methods used. To examine the CP-conservation
situation more comprehensively, taking the whole PMNS matrix into
consideration, we will use the Jarlskog invariant measure, explored in the next section.

\begin{table}[htpb]
  \centering
  \caption{Bayes factors (posterior probabilities) in all the \thatm{} octant
           and mass ordering hypotheses combinations. A weak preference towards
           the normal mass ordering and upper octant is observed. Numbers
           extracted from a fit with the external \threac constraint.
           Probabilities summed across rows or columns may differ slightly 
           from the totals due to rounding.  In these results 
           the reactor constraint on \sstth{13} is applied.}
  \label{tab:bayesfactorwrc}
  \begin{tabular}{c c c c}
    \hline
    \hline\\[-2.3ex]
    & Normal     & Inverted    & \multirow{2}{*}{Total}\\
    & Ordering   & Ordering    &  \\
    \hline\\[-2.3ex]
    Upper Octant & 0.71 (0.42) & 0.26 (0.21) & 1.67 (0.63) \\
    Lower Octant & 0.35 (0.26) & 0.13 (0.12) & 0.60 (0.38) \\
    \hline\\[-2.3ex]
    Total & 2.08 (0.68) & 0.48 (0.33) & (1.0)\\
    \hline
    \hline
  \end{tabular}
\end{table}

\subsubsection{CP violation -- Jarlskog invariant}
\label{subsubsec:CPV and J}
The Jarlskog invariant is a measure of the strength of charge-parity violation that is independent of how the mixing matrix is parameterized. 
The neutrino-mixing form of the Jarlskog~\cite{Denton:2019} parallels its development in the quark sector~\cite{Jarlskog:1985}. Under the three-neutrino-flavor assumption, its definition is:
\begin{align}
	\label{eq:jarlskoginvariant}
	\J \equiv & \cos(\theta_{12})\cos^2(\theta_{13})\cos(\theta_{23})\sin(\theta_{12}) \nonumber \\
              & \times \sin(\theta_{13})\sin(\theta_{23})\sin(\delta_{\mathrm{CP}}),
\end{align}
\noindent where if any of the factors is zero, the invariant vanishes and the
neutrino mixing matrix is CP-conserving. Nonzero values of $J$, on the other hand,
indicate CP violation. We produce this measurement by taking the MCMC chain
with all the oscillation parameters' values and calculating the
Jarlskog invariant at each MCMC step.
As NOvA is not sensitive to the value of $\theta_{12}$, for \J{} we use a Gaussian prior analogous to the one used for the reactor constraint (sec. \ref{par:ssthreac}) but whose range consists of the PDG's 2019 average of solar and LBL reactor neutrino measurements: $\ssth{12} = 0.307 \pm 0.013$~\cite{ParticleDataGroup:2018ovx}, the same value used in our previous results.

As described in section~\ref{subsubsec:priors}, we use uninformed priors for
the oscillation parameters where possible, meaning that the priors are uniform
in the variable that is being shown. 
The trigonometric dependence of \J{} on the oscillation parameters shown in eq.~\ref{eq:jarlskoginvariant} makes clear that constructing of a prior uniform in \J{} would result in nonuniform distributions over its constituent variables.
We studied numerous priors in the oscillation parameters not constrained by external data and found that for any reasonable choice of prior for \thatm or \dmlg---that is, one that does not vanish or diverge across the range of values allowed by other contemporary experiments---the posterior in \J{} is essentially unchanged, due to the strong constraints afforded by the NOvA data.
Thus the only degree of freedom where the prior is of major concern is \dcp.
Since the Jarlskog invariant is written in
terms of \sdcp, the natural formulation is in terms of a prior uniform in
\sdcp. At the same time, there are theoretical considerations that suggest a
prior uniform in \dcp may be more appropriate~\cite{Denton:2020igp}. We
therefore consider both a prior uniform in \sdcp and one uniform in \dcp.

\begin{figure}[htb]
	\begin{subfigure}[b]{\columnwidth}
		\begin{center}
			\includegraphics[width=0.9\textwidth,trim={0 3.5cm 0 0},clip]{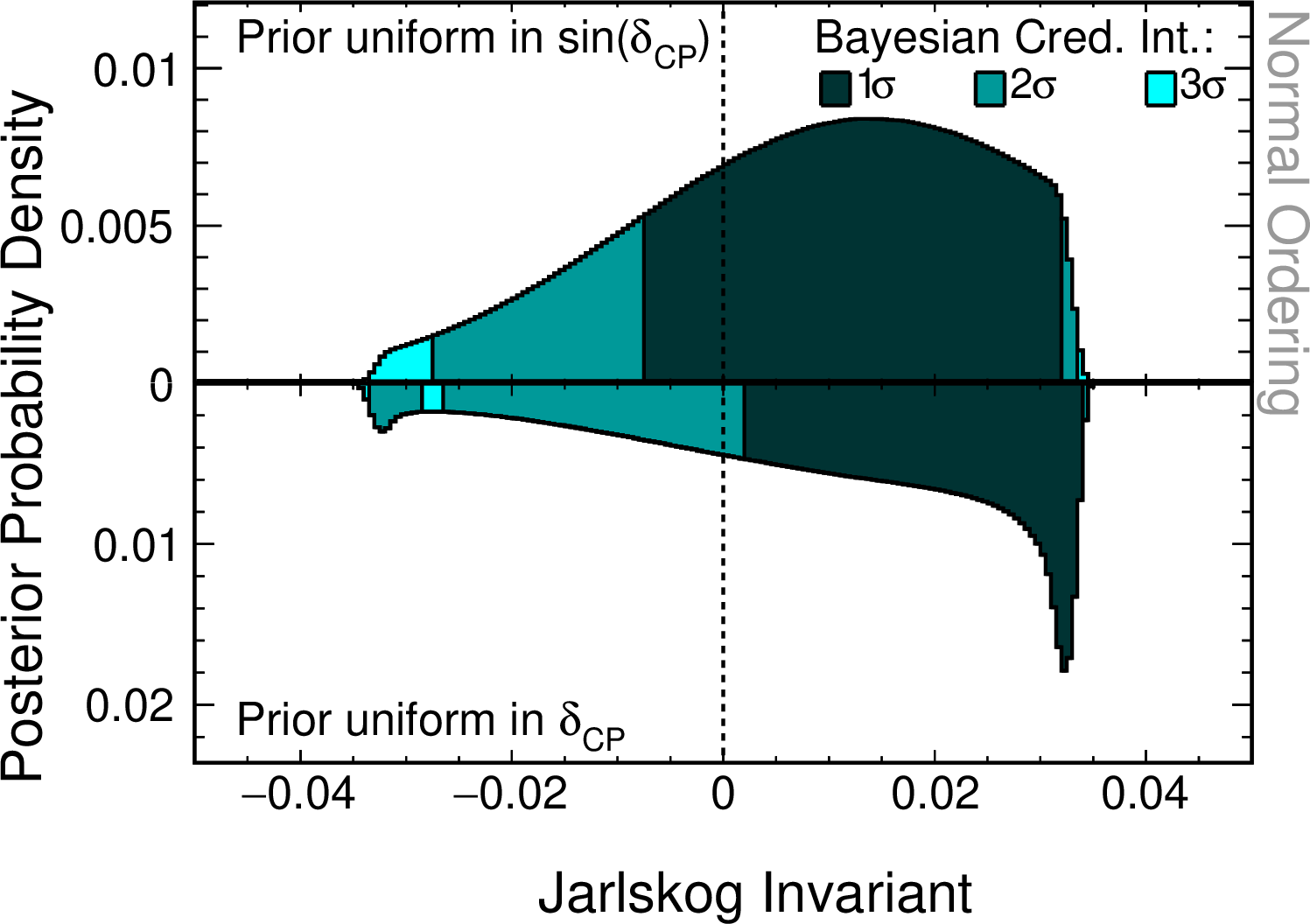}
		\end{center}
	\end{subfigure}
	\begin{subfigure}[b]{\columnwidth}
		\begin{center}
			\includegraphics[width=0.9\textwidth]{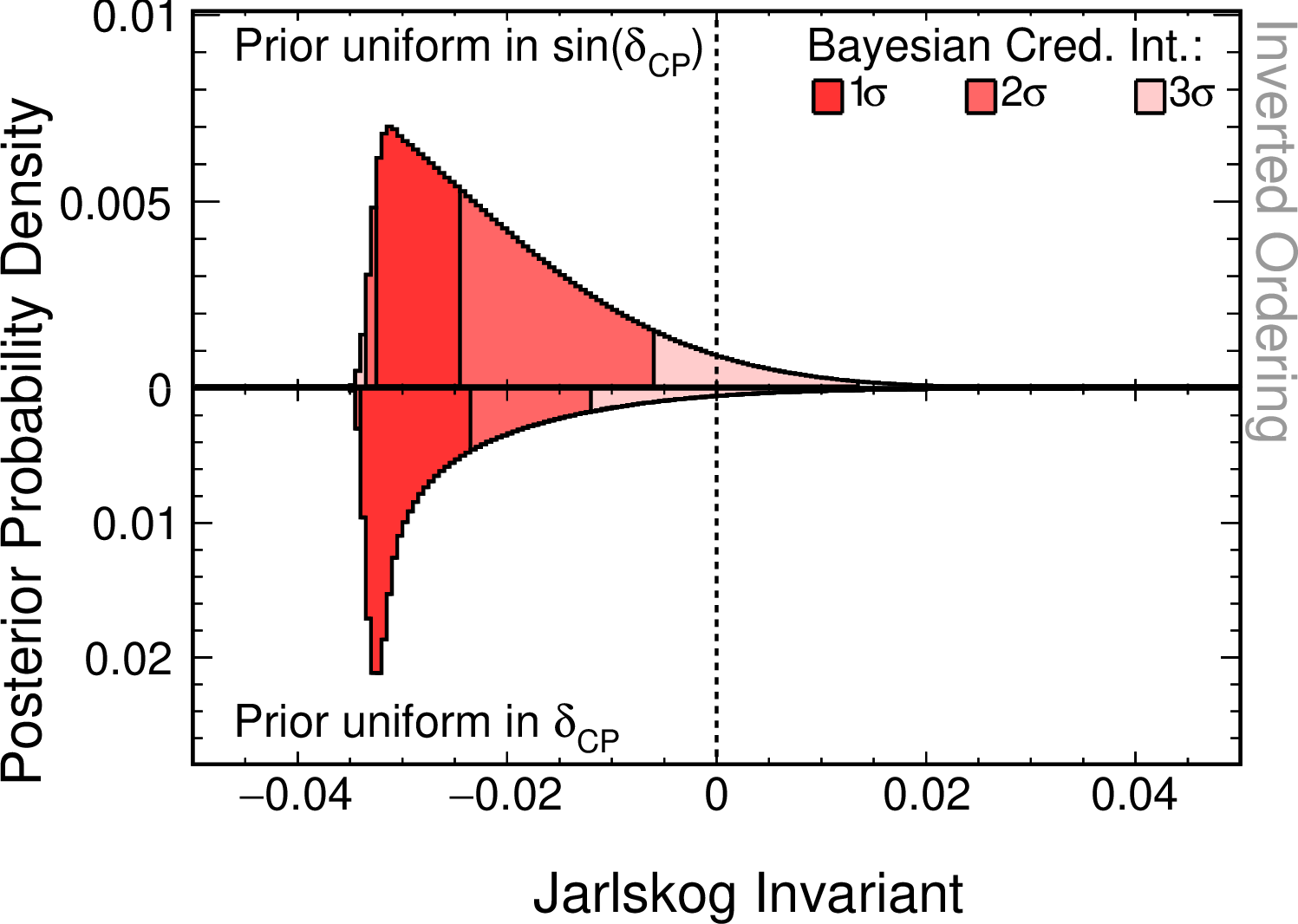}
		\end{center}
	\end{subfigure}
	\caption{Posterior probability density for the Jarlskog invariant,
		marginalized over both mass orderings and plotted separately for the
		NO (top plot) and IO (bottom plot). The top half of each panel shows
		the posterior with a prior uniform in \sdcp, while the bottom half
		uses a prior uniform in \dcp. A CP-conserving line is drawn at
		$\J=0$. The  external constraint in \threac from the reactor
		experiments is used. %
	}
	\label{fig:datawrcjarlskog}
\end{figure}

Figure~\ref{fig:datawrcjarlskog} shows the inferred values of the Jarlskog
invariant extracted from a fit to the NOvA data with the external constraint on
\threac from the reactor experiments, marginalized over the normal (top) and
inverted (bottom) mass orderings. The regions most favored by the data tend
towards CP violation in the NO, although the 1\,$\sigma$ intervals
do include CP conservation when the prior uniform in \sdcp is used.  The IO's preferred
regions are more markedly distant from CP conservation, particularly for the
prior uniform in \dcp{}.  As expected, these trends mirror those observed when
considering CP-conserving and CP-violating values of \dcp{} in
Sec.~\ref{ssub:pmns_params}.

A more quantitative way of measuring the NOvA preference for CP conservation with the
Jarlskog invariant is through Bayes factors. These can be calculated
with the Savage--Dickey density ratio method, which computes Bayes factors for
point hypotheses~\cite{Dickey:1971, Mulder:2020}. Using this method we can calculate the
Bayes factor for the CP-conserving value $\J=0$, nested under the unconstrained
hypothesis where \J{} can take any value. Bayes factors always depend formally
on the choice of prior, but in many circumstances (such as those discussed
above), if two hypotheses being compared use the same prior, it will
cancel. This is not the case in the Savage--Dickey method. 
Therefore, we computed the Bayes factor for $\J=0$ under several hundred combinations of priors for \thatm, \dmlg, and \dcp.
The combinations used in the previous discussion (uniform in \thatm, $|\dmlg|$, and $\mathrm{sgn}(\dmlg)$, plus the two variants in \dcp) produced Bayes factors that were among the most conservative that we found (though it is possible to engineer priors with, for example, smaller ranges where they are uniform to produce even more conservative Bayes factors).
We therefore took these as representative Bayes factors and inverted them to obtain the associated Bayes factors for CP violation over CP
conservation, $\J \neq 0$.

\begin{table}[h]
  \centering
  \caption{Bayes factors for preference of
           CP~violation over CP~conservation, extracted using the Savage--Dickey
           method at $\J=0$. Priors uniform in \dcp and \sdcp are both shown.
           The preferences are given for the normal (NO), inverted (IO),
           and both (BO) mass orderings.}
  \label{tab:bayesfactorjarlskog}

  \begin{tabular}{c c c c}
    \hline
    \hline\\[-2.3ex]
    Prior & NO & IO & BO \\
    \hline\\[-2.3ex]
    Uniform \sdcp & 1.2 & 3.4 & 1.5 \\
    Uniform \dcp  & 1.0 & 3.8 & 1.6 \\
    \hline
    \hline
  \end{tabular}
\end{table}

\begin{figure*}[hbpt]
	\centering
	\includegraphics[width=1.\textwidth]{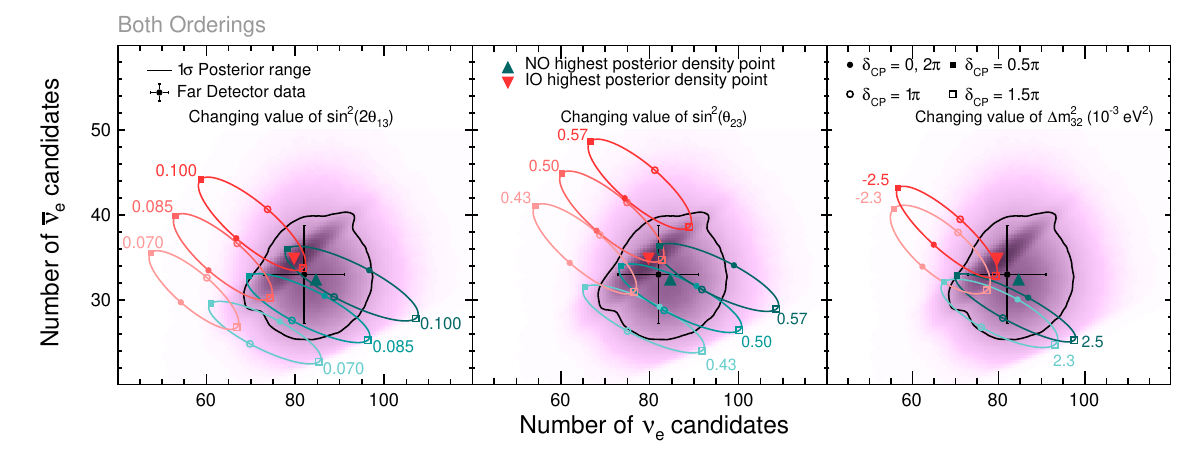}
	\caption{Posterior probability density (purple shading)---that is, the probability that
		a given (\nue, \nuebar) pair of counts is the true underlying value in nature (see text)---compared to the measurement and associated  expected statistical variance (black cross) of the total number of \nuebar{} candidates vs. \nue candidates.
		The black solid contours indicate the regions
		enclosing 68.3\% of the posterior. Each panel overlays a set
		of ellipses corresponding to the predicted event counts over the range of
		possible values of \dcp{}, with the parameter notated nearby held at that
		value (blue for NO, red for IO, with varying shading depending on the
		parameter value) and other parameter values as given in
		Table~\ref{tab:pmnsworc}. The triangular markers show the highest posterior
		density in the \nue--\nuebar{} space for when restricting to NO (blue) and IO (red) hypotheses.
		The posteriors shown here employ a uniform prior over \sstth{13}.}
	\label{fig:ppp bievent}
\end{figure*}

Table~\ref{tab:bayesfactorjarlskog} shows the Bayes factors for CP violation
against CP conservation for normal, inverted, and both mass orderings, calculated
for priors uniform in \sdcp or \dcp. All of these probabilities point
towards a preference for CP nonconservation, although the preferences are
minimal regardless of the \dcp prior or the mass ordering (and an analogous frequentist
$p$-value would indicate significances less than 1\,$\sigma$, apart from the
inverted ordering, where they range $1.1-1.2$\,$\sigma$\footnote{We emphasize that these significances cannot be read off of Fig.~\ref{fig:datawrcjarlskog}.  The probability density shown there is marginalized independently at each value of \J.  However, the binary hypothesis test $\J=0$ vs.~$\J \neq 0$ requires a simultaneous marginalization across the whole $\J \neq 0$ space.  For the latter, a point-hypothesis treatment, such as the the Savage--Dickey formalism we use here, is necessary.}).
% calculation: erfinv(1-(1-4.4/5.4)/2) = 1.189
However, we reiterate that the inherent prior-dependence of the Savage--Dickey method means these values can only be treated as representative of a class of possible interpretations of the evidence.
Future neutrino oscillation experiments, in which the evidence for the ordering is expected to be much stronger, will likely wish to study this problem further.

\subsubsection{Using only NOvA constraints on \threac}
\label{subsubsec: NOvA-only}

\begin{figure*}[htbp]
	\centering
	\begin{subfigure}[b]{0.45\textwidth}
		\begin{center}
			\includegraphics[width=\textwidth]{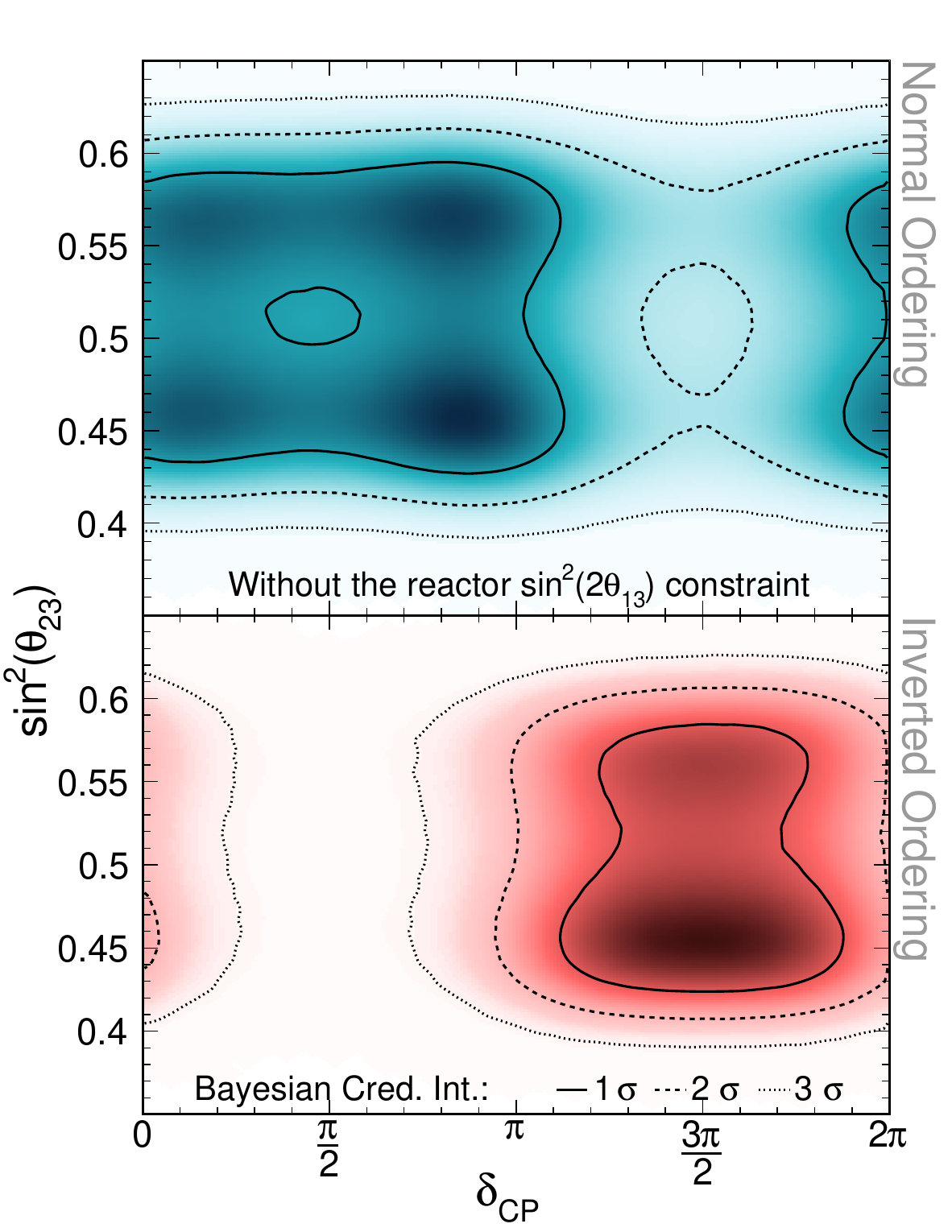}
		\end{center}
		\caption{\dcp--\ssth{23}}
	\end{subfigure}
	\begin{subfigure}[b]{0.45\textwidth}
		\begin{center}
			\includegraphics[width=\textwidth]{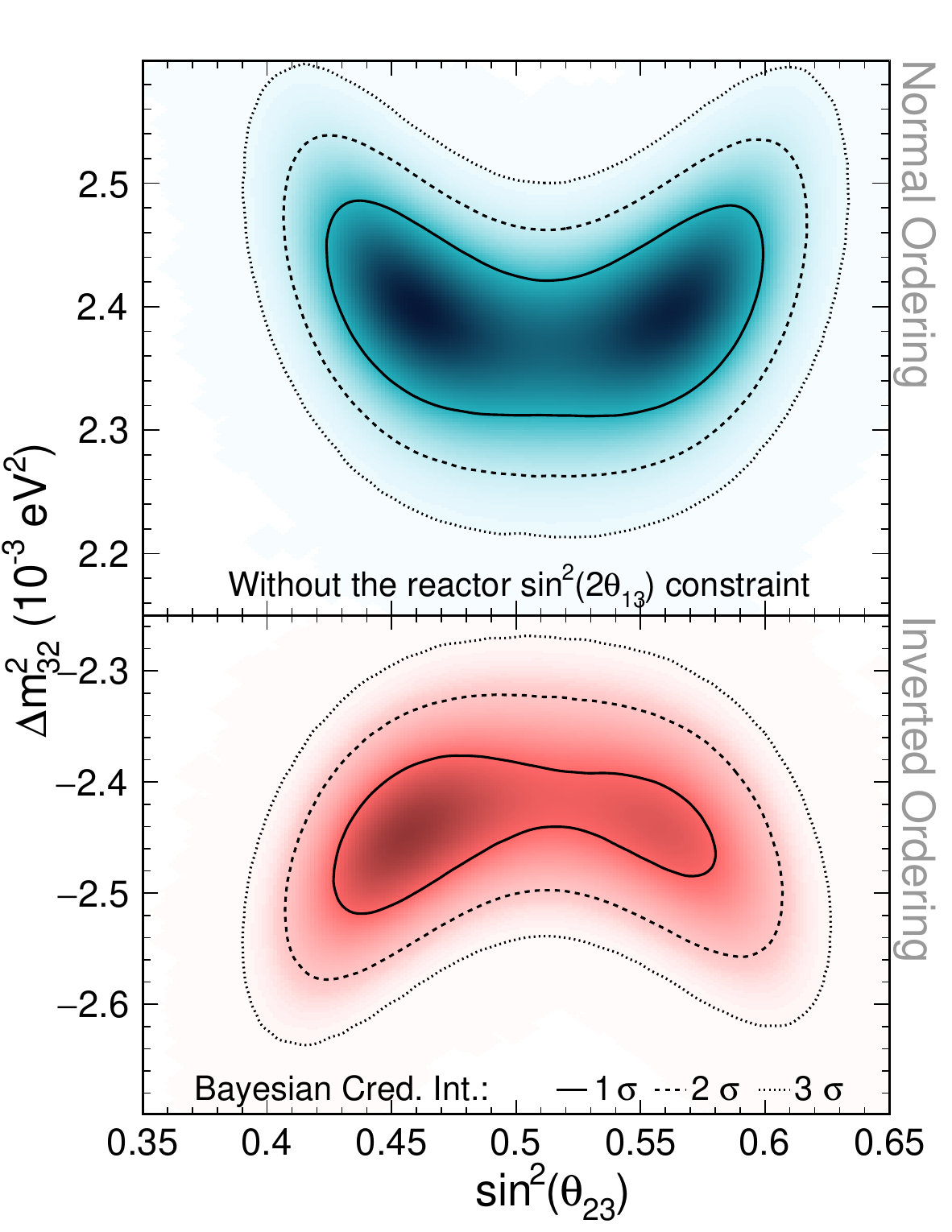}
		\end{center}
		\caption{\ssth{23}--\dmlg}
	\end{subfigure}
	\caption{Posterior probability density (shaded) with 1, 2, and
		3\,$\sigma$ credible intervals for \dcp--\ssth{23} (left) and
		\ssth{23}--\dmlg (right), marginalized over both mass orderings, plotted
		separately for NO (top, blue) and IO (bottom, red). Contours extracted from
		a fit with prior uniform in
		\sstth{13}.}
	\label{fig:dataworcappdiss}
\end{figure*}

NOvA's oscillation measurements simultaneously constrain a combination of mixing angles (\thatm{}, \threac{}), the mass-squared splitting \dmlg, the CP phase \dcp, and the neutrino mass ordering. 
Therefore, applying a strong external 1D constraint on \threac{} from reactor antineutrino experiments---and thereby reducing the available solution space---increases the sensitivity to other parameters.
This is especially true of the octant of \thatm{}, as we will show below.
However, using an uninformed (uniform) prior in \sstth{13} enables us to directly compare a NOvA-only measurement against that of the reactor experiments.
In so doing we can examine the robustness of the PMNS description across short-baseline reactor antineutrino experiments and long-baseline accelerator-based oscillation experiments.
We also may study whether our preferences change in the presence of an external constraint on the data.
In the following, we use recomputed posteriors that assumed a uniform prior over \sstth{13}.
The PPP goodness-of-fit metric from sec. \ref{subsubsec:goodness of fit}, when recomputed for the associated posterior spectra, was found not to change more than 0.01 for any of the samples shown in Fig.~\ref{fig:ppvwrc}, suggesting that the data is well modeled without the constraint.

\begin{table}[htpb]
	\centering
	
	\caption{The highest posterior density points (HPD) together with the
		1\,$\sigma$ Bayesian credible interval limits for the PMNS
		parameters of interest marginalized over all the mass ordering (MO)
		hypotheses. Marginalization over the mass orderings is explained at
		the beginning of Sec.~\ref{ssub:pmns_params}. Extracted from a fit
		with prior uniform in \sstth{13}.}
	
	\begin{tabular}{c  c  c  c}
		\hline
		\hline\\[-2.3ex]
		& MO & HPD & $1\,\sigma$ \\
		\hline\\[-2.3ex]
		
		\multirow{1}{*}{\dcp} & Both & $1.53\pi$ & $[0.67\pi, 1.88\pi]\cup[0.06\pi, 0.12\pi]$ \\
		(Prior uniform      & Normal & $0.85\pi$ &  $[0.48\pi, 1.04\pi]\cup[1.99\pi, 0.41\pi]$  \\
		in \dcp)        & Inverted & $1.52\pi$ &  $[1.25\pi, 1.75\pi]$ \\
		\hline\\[-2.3ex]
		
		\multirow{3}{*}{\sstth{13}} & Both & $0.087$ &  $[0.071, 0.107]$ \\
		& Normal & $0.084$ &  $[0.065, 0.108]$ \\
		& Inverted & $0.094$ &  $[0.083, 0.106]$ \\
		\hline\\[-2.3ex]
		
		\multirow{3}{*}{\ssth{23}} & Both & $0.46$ &  $[0.43, 0.50]\cup[0.53, 0.58]$ \\
		& Normal & $0.46$ &  $[0.43, 0.59]$ \\
		& Inverted & $0.46$ &  $[0.44, 0.48]$ \\
		\hline\\[-2.3ex]
		
		\dmsq{32}& Normal & $2.39$ &  $[2.33, 2.46]$ \\
		($\times10^{-3}$ eV$^{2}$) & Inverted & $-2.44$ &  $[-2.48, -2.40]$ \\
		\hline
		\hline
	\end{tabular}
	\label{tab:pmnsworc}
\end{table}

Without an external constraint on \sstth{13}, the relevant degrees of freedom present a complex space with many intercorrelations among the parameters and numerous possible solutions.
We can use the type of posterior distributions shown in Fig.~\ref{fig:pppred} to explore these possibilities: for example, by computing the total number of predicted \nue{} and \nuebar{} candidates for the parameters of each MCMC sample and comparing the distribution of these predictions in (\nue, \nuebar) candidate space to what we observe in the data.
This is shown in Fig.~\ref{fig:ppp bievent}.
To guide our intuition, we overlay ellipses corresponding to the ranges of predicted number of events for the possible values of \dcp{} at fixed values of the other parameters, subject to constraint from our data in that the highest posterior density (HPD) value in each ordering is used for parameters not explicitly varied.
Though not all dimensions of the way our observed \nue{} and \numu{} candidate energy spectra interact with the oscillation parameters can be represented in this presentation, we can still note several important features.
First, the highest-density (highest-probability) region lies essentially equally along the overlap region between NO and IO ellipses in all panels, meaning we will find good solutions for either, given an appropriate value of \dcp{}.
However, the NO ellipses subtend more of the posterior region, which will result in a slight overall preference for NO.
Second, it is clear from the left panel that the value of \sstth{13} preferred by the reactor average, $\sstth{13} = 0.085$, is compatible with the highest-probability region for NOvA.
Third, comparing the left and center panels, there is obvious degeneracy
between \sstth{13} and \ssth{23}, since independently varying them produces
similar effects on the predictions. 
%It is worth mentioning that \numu and
%\numubar data samples provide a strong constraint on \sstth{23} (corresponding
%to two values in \ssth{23} space), with \nue and \nuebar providing a further
%(but weaker) constraint on \ssth{23}.
And finally, the right panel demonstrates that the value of \dmsq{32} plays a minor (though not negligible) role in the appearance channel; its primary constraints arise from the \numu{} disappearance measurement.
These features will be examined more quantitatively in the distributions that follow.
Table~\ref{tab:pmnsworc} shows the highest posterior probability points
together with the 1\,$\sigma$ credible intervals for all the PMNS oscillation
parameters of interest, extracted from the fit with a prior uniform in
\sstth{13}. Similarly to Table~\ref{tab:pmnswrc}, the values are split into
both, normal, and inverted mass orderings with any disjoint 1\,$\sigma$ regions
denoted with a union symbol $\cup$. 
As compared to the results of the fit with the external constraint
from reactor experiments, the central values for \dcp{} and
\ssth{23} shown here are the most different.
This is not unexpected since there is also a degeneracy between these two parameters,
and both \dcp{} and \ssth{23} have multiple areas of high probability.

Fig.~\ref{fig:dataworcappdiss} shows our preferred regions in \dcp--\ssth{23} and \ssth{23}--\dmlg spaces without the external \threac constraint. 
Comparing with Figs.~\ref{fig:datawrcdmthatm2d}~and~\ref{fig:datawrdcpthatm2d}, it is evident that removing the constraint diminishes the NOvA sensitivity especially to the octant of \thatm{} (there is now near symmetry across $\ssth{23}=0.5$). 
Although the octant preference is weakened, the central values of the PMNS parameters \ssth{23}, \dmsq{32}, and especially \dcp{} do not otherwise see any significant change.
This is as expected from Fig.~\ref{fig:ppp bievent}.
The results without the reactor constraint are therefore fully compatible with the standard NOvA results with the external constraint on \sstth{13} applied.
We also observe that although the sensitivity is reduced without the constraint, and although that for a given value of \dcp{} there is always a combination of parameters within either mass ordering compatible with the data, certain \dcp{}--\ssth{23}--ordering combinations are still excluded with reasonable confidence, such as $(\dcp=\frac{\pi}{2}, \mathrm{IO})$ and $(\dcp=\frac{3\pi}{2}, \ssth{23}=0.5, \mathrm{NO})$.
Combinations such as these produce strong asymmetries in the (\nue, \nuebar) counts, and as Fig.~\ref{fig:ppp bievent} makes clear, such parameter values lie in regions outside the posterior point cloud, and are thus disfavored.
Table~\ref{tab:bayesfactorworc} examines the posterior probabilities for each of
the octant and mass-ordering hypotheses more quantitatively.
Compared to Table~\ref{tab:bayesfactorwrc}, we note a similar weakened preference for the upper octant of \thatm{} for the fit without the reactor constraint.

\begin{table}[htpb]
  \centering
  \caption{Bayes factors (posterior probabilities) for all the
           \thatm{} octant and mass ordering hypotheses, with a marginal
           preference towards the normal mass ordering and upper octant.
           Numbers extracted from a fit with a uniform prior in \sstth{13}.
           Probabilities summed across rows or columns may differ slightly 
           from the totals due to rounding.}
  \label{tab:bayesfactorworc}

  \begin{tabular}{c c c c}
    \hline
    \hline\\[-2.3ex]
                       & Normal   & Inverted  & \multirow{2}{*}{Total} \\
                       & Ordering & Ordering  &  \\
    \hline\\[-2.3ex]
    Upper Octant & 0.53 (0.35)  & 0.20 (0.17) & 1.05 (0.51) \\
    Lower Octant & 0.39 (0.28)  & 0.42 (0.20) & 0.95 (0.49) \\
    \hline\\[-2.3ex]
    Total & 1.70 (0.63) & 0.59 (0.37) & (1.0)\\
    \hline
    \hline
  \end{tabular}
\end{table}

\begin{figure}[htpb]
	\centering
	\includegraphics[width=\columnwidth]{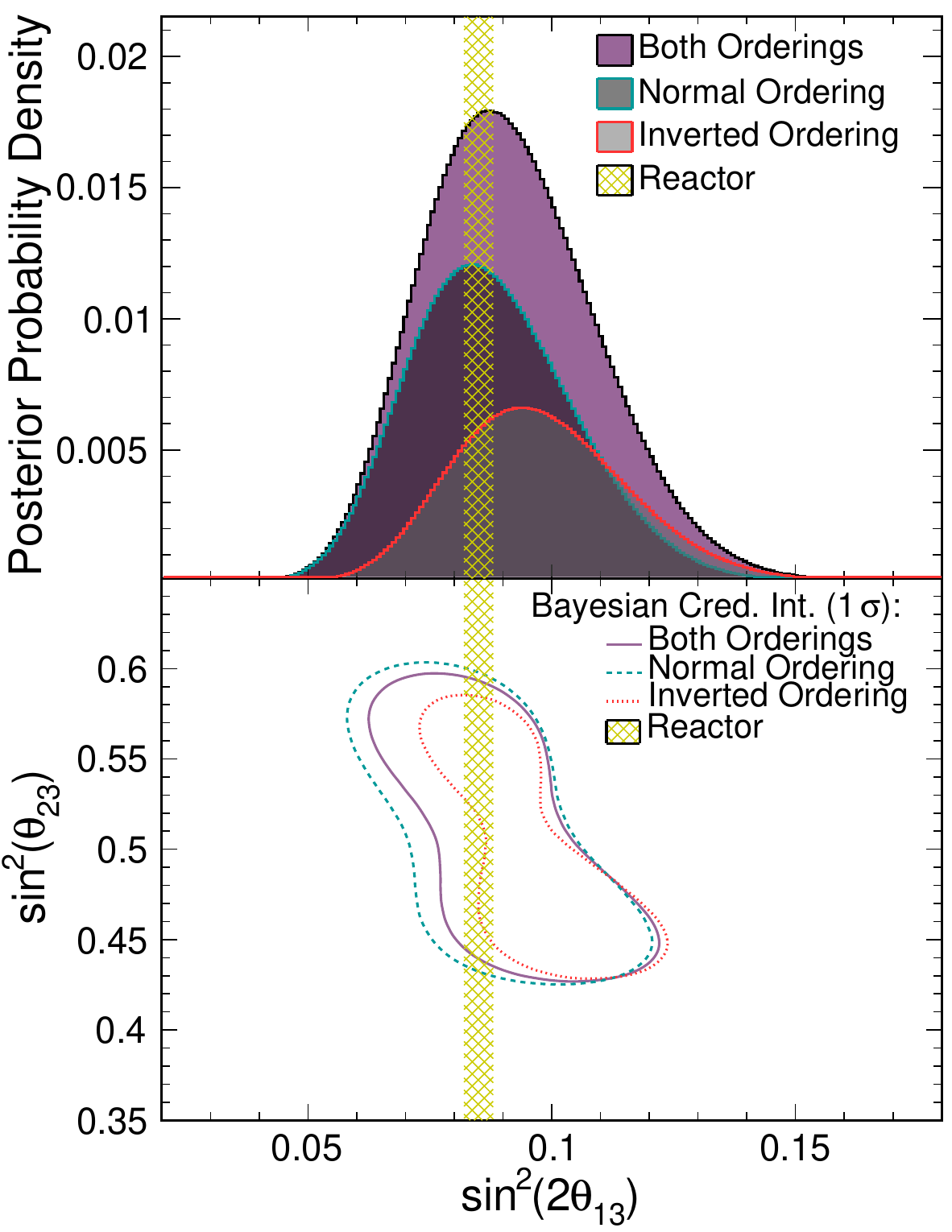}
	
	\caption{Comparison of the 2D \sstth{13}--\ssth{23} credible intervals
    (bottom) and 1D \sstth{13} posterior probability densities (top) between both
    (purple), normal (light-blue) and inverted (red) mass orderings with a
    prior uniform in \sstth{13} and \ssth{23}. The reactor experiments'
    1\,$\sigma$ interval in $\sstth{13}$ from the PDG
    2019~\cite{ParticleDataGroup:2018ovx} is shown in yellow hatched bar.}
	\label{fig:dataworcth13th23mo}
\end{figure}

In Fig.~\ref{fig:dataworcth13th23mo} we study the impact of applying the reactor constraint on our mass-ordering inference.
We first observe that the NOvA and reactor measurements for \sstth{13} are in good agreement, as evidenced by the overlap of the yellow bar (indicating the reactor 1\,$\sigma$ range) with the peak of the ``both orderings'' posterior in the upper plot.
While we know of no widely accepted measure for quantifying such consistency, we combined this ``both orderings'' posterior with a Gaussian probability distribution centered at $\sstth{13} = 0.085$ and of width~0.003 (representing the reactor constraint) using the technique of conflation \cite{Hill:2011, Fox:2010id}.
The result is identical to the reactor distribution we began with, which (as the examples in Ref.~\cite{Fox:2010id} show) does not occur if the probability densities being conflated agree poorly.
This accords with the fact that the PPP values did not meaningfully improve when removing the reactor constraint, as noted previously.
When we examine the orderings separately, we find that the normal ordering contains more posterior probability, as expected from Fig.~\ref{fig:ppp bievent} and as reflected in the larger NO contour in the lower panel here.
Moreover, as we see in the marginal posterior distribution shown in the top panel, the posterior restricted to IO prefers generally larger values of \sstth{13}.
The associated correlation with \ssth{23} also pushes \thatm into the lower octant, as seen in the lower panel; this point will be developed further momentarily.
Because the value of \sstth{13} measured by the reactor experiments is more consistent with the NOvA NO posterior, this results in the slightly stronger preference for the NO in Table~\ref{tab:bayesfactorwrc} as compared to \ref{tab:bayesfactorworc}.
However, the difference is small, which indicates that the (mild) NO preference observed in the data is largely independent of the reactor constraint.

\begin{figure*}
  \centering
  \includegraphics[width=0.75\textwidth]{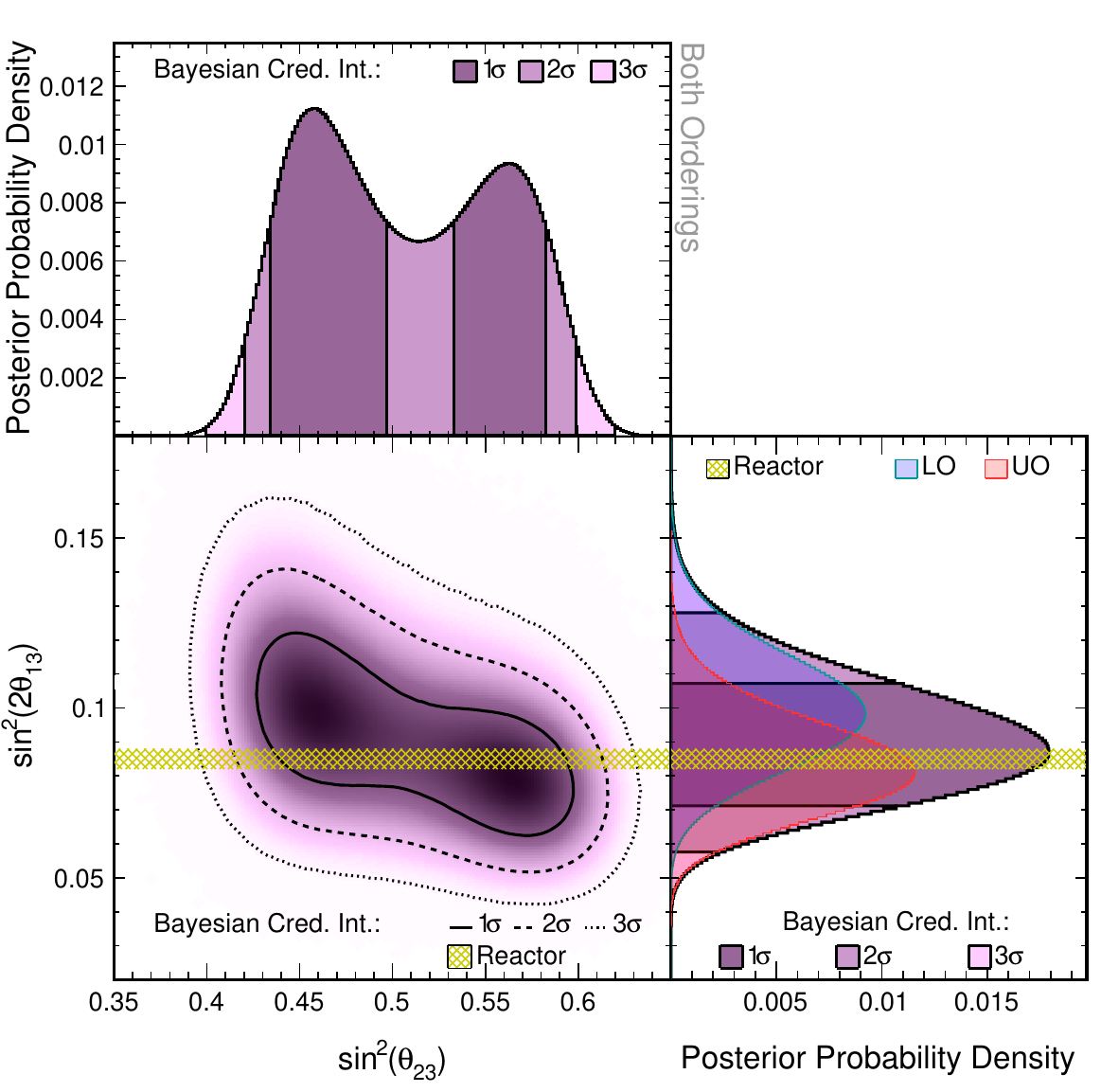}
  \caption{\ssth{23}--\sstth{13} posterior probability densities
           with 1, 2, and 3\,$\sigma$ credible intervals, marginalized over both
           mass orderings. The posterior density was extracted from a fit with
           a prior uniform in \sstth{13}. The reactor experiments' 1\,$\sigma$
           interval in \sstth{13} from the PDG
           2019~\cite{ParticleDataGroup:2018ovx} is shown in the yellow hatched
           bar. The right panel shows the posterior probability for \sstth{13},
           with its contributions from the upper octant (UO, transparent, red
           outline) and lower octant (LO, transparent, blue outline) of
           \thatm{}.}
  \label{fig:dataworcth13th23}
\end{figure*}

The degeneracy in the measurement between \ssth{23} and \sstth{13} we noted in Fig.~\ref{fig:ppp bievent} can be studied directly by examining their joint posterior probability distribution, marginalized over all other parameters and the mass ordering.
We show this in Fig.~\ref{fig:dataworcth13th23}. 
The central panel exhibits a clear anticorrelation between the octant of \thatm{} and the value of \sstth{13}, which is expected since both parameters enter at leading order in the \numutonue{} and \numutonuebar{} oscillation probabilities.
Here the overlap of the reactor measurements (again indicated by the yellow hatched bar) with our marginal posterior for \sstth{13} (right panel) favors the upper octant over the lower octant when we constrain the results to specifically the upper or lower octant of \thatm{}.
Thus, we see that the preference for the upper octant of \thatm{} in Table~\ref{tab:bayesfactorwrc} is an emergent behavior that arises from the application of the reactor constraint.
Similar changes in the strength of the octant preference when the reactor constraint is applied have been noted in results from T2K~\cite{T2K:2023smv} and simulation studies for DUNE~\cite{DUNE:2020jqi}, though to our knowledge this is the first time the underlying $\theta_{13}-\theta_{23}$ anticorrelation has been examined in detail with a data result.
We also note that though the marginal posterior for \ssth{23}, in the top panel, shows a higher posterior density in the lower octant, the total posterior probability integrated across the upper octant is slightly larger than the corresponding lower octant probability; but as Table~\ref{tab:bayesfactorworc} makes clear, this preference is entirely insignificant.

We emphasize that reactor neutrino experiments and accelerator neutrino
experiments measure the PMNS \sstth{13} by examining different sectors of
neutrino oscillations, over a wide range of baselines. Reactor neutrino
experiments measure the \nuebar survival probability \pnuenuebar with low-energy
(few-\unit{MeV}) \nuebar{}s over a short (few-\unit{km}) baseline.
Conversely, accelerator neutrino experiments
simultaneously measure \nue appearance in a few-\unit{GeV} \numu beam, and
\nuebar appearance in a \numubar beam, both over a long (hundreds-of-\unit{km})
baseline. In long-baseline measurements \pnumunumu, \pnumunumubar, \pnumunue,
and \pnumunuebar are all exploited to constrain the PMNS oscillation
parameters, including \sstth{13}. 
Thus, the consistency observed between long-
and short-baseline measurements lends support to the PMNS interpretation
of neutrino oscillations.

\section{Conclusions}
We have refit the NOvA dataset of $13.6 \times 10^{20}$\,POT in neutrino beam mode and $12.5 \times 10^{20}$\,POT in antineutrino mode using a Bayesian statistical approach.
NOvA data continues to be consistent with maximal mixing for \ssth{23} and the regions of the \ssth{23}--\dmsq{32} and \dcp--\ssth{23} spaces preferred by this analysis are consistent with those of the previous analysis done using a frequentist fit.

With the introduction of the Bayesian analysis, we also expand the neutrino oscillation parameters measured by NOvA.
We report for the first time NOvA measurements that do not require constraints on \sstth{13} from reactor antineutrino oscillations.
Moreover, we are also able to include new results for \sstth{13} and the Jarlskog invariant \J{}; these were impractical to produce under the preceding frequentist method due to the necessity of Feldman--Cousins corrections.
The inferences on \J{} provide a parametrization-independent measurement of CP violation and indicate that NOvA data has a weak preference for CP violation, which becomes slightly more pronounced when assuming the inverted mass ordering.

As NOvA measures a convolution of mixing angles, \dcp, and mass ordering in the oscillation probabilities, removing the external constraint on \threac{} reduces our constraining power to determine the octant of \thatm{}, mass ordering, and \dcp.
While our sensitivity is reduced without the external constraint (particularly for the octant), we note that the conclusions arising from the analysis remain unchanged.

This measurement of \sstth{13} using electron neutrinos and antineutrinos with energies in the GeV range and propagating for hundreds of kilometers is fully consistent with measurements performed using few-MeV electron antineutrinos from nuclear reactors propagating for a few kilometers.
The consistency of results using the PMNS framework across a broad regime of conditions bolsters its applicability.
More stringent tests of CP violation and the consistency of our \sstth{13} measurement with those of reactors will be possible with increased statistics in upcoming NOvA measurements.

\section{Acknowledgments}

This document was prepared by the NOvA collaboration using the resources of the Fermi National Accelerator Laboratory (Fermilab), a U.S. Department of Energy, Office of Science, HEP User Facility. Fermilab is managed by Fermi Research Alliance, LLC (FRA), acting under Contract No. DE-AC02-07CH11359. This work was supported by the U.S. Department of Energy; the U.S. National Science Foundation; the Department of Science and Technology, India; the European Research Council; the MSMT CR, GA UK, Czech Republic; the RAS, MSHE, and RFBR, Russia; CNPq and FAPEG, Brazil; UKRI, STFC and the Royal Society, United Kingdom; and the state and University of Minnesota.  We are grateful for the contributions of the staffs of the University of Minnesota at the Ash River Laboratory, and of Fermilab. For the purpose of open access, the author has applied a Creative Commons Attribution (CC BY) license to any Author Accepted Manuscript version arising.

\appendix 
\section{Determining step sizes for ARIA}
\label{app:step sizes}

As noted in Sec. \ref{subsubsec: ARIA}, an \mrrtt{} chain is derived sample-by-sample using a repeated two-step procedure:
\begin{enumerate}
	\item \label{item:proposal} \textbf{Proposal}: The coordinates of a potential new sample are selected from a probability distribution centered on the current sample (or initial seed).
	\item \label{item:acceptance} \textbf{Acceptance}: The proposal selected above is either accepted or rejected according to the rule of detailed balance, i.e., that every step in the chain be exactly reversible.  
\end{enumerate}
If accepted, the proposed coordinates become the next sample.
If rejected, the previous sample is repeated to become the next sample.

The \mrrtt{} algorithm does not specify the distribution to be used in step~\ref{item:proposal} above, however.
In our implementation, we use the most common choice, a multivariate Gaussian:
\begin{widetext}
\begin{equation}
	\label{eq:proposal}
	g(\vec{x}\,' | \vec{x}) = (2\pi)^{-\frac{N}{2}} (\det \Sigma)^{-\frac{1}{2}} \exp\left(- \frac{1}{2} \left( \vec{x}\,' - \vec{x}\, \right)^T \Sigma^{-1}  \left( \vec{x}\,' - \vec{x}\, \right) \right),
\end{equation}
\end{widetext}
where $\vec{x}$ represents the current sample coordinates, $\vec{x}\,'$ the proposed next coordinates, and $N$ the dimensionality of the coordinate space.
The matrix $\Sigma$ imposes a length scale on the ``distance'' between successive samples, and (especially when it is diagonal) its elements are usually called the ``step sizes'' of the sampling for each degree of freedom.
The ideal asymptotic fraction of samples accepted in step~\ref{item:acceptance}, $\alpha$, is 23.4\% under a wide range of circumstances~\cite{Gelman:1997a,Roberts:1998a}.
Though this figure is strictly true only for $N \rightarrow \infty$, it has been shown to hold approximately even for parameter counts as low as $N=5$~\cite{Roberts:2001a}.
Because the outcome of step~\ref{item:acceptance} is related to the proposals generated in step~\ref{item:proposal}, we tuned the values of $\Sigma$ to arrive at $\alpha = 23.4\%$.

\begin{figure}[htpb]
	\centering
	\includegraphics[width=\columnwidth]{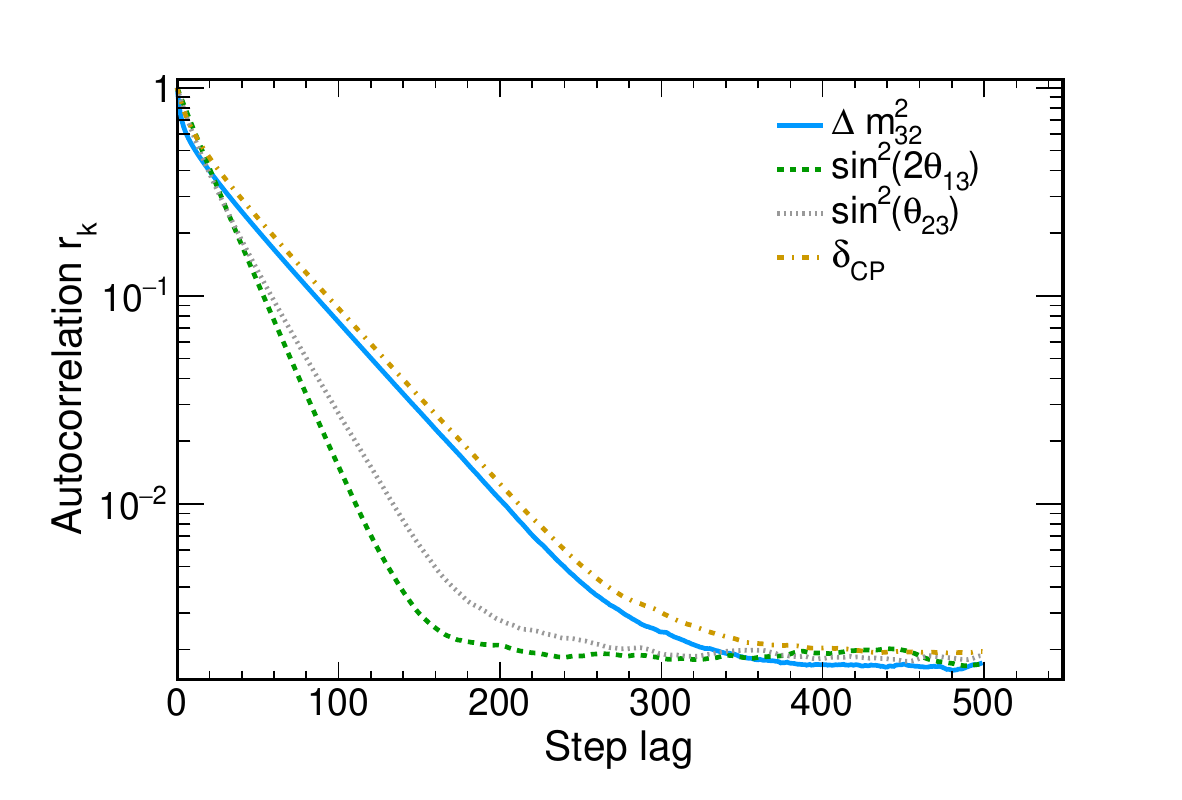}
	\caption{Lag autocorrelation (see Eq.~\ref{eq:autocorrelation}) computed using a \mrrtt{} chain sampling only oscillation parameters.}
	\label{fig:autocorr stats-only}
\end{figure}

Our overall heuristic in the tuning procedure is to maintain step sizes that yield similar autocorrelations (defined rigorously below) across all the parameters.
This results in the most efficient exploration of the parameter space~\cite{Brooks:2011a}.
We first optimized the step sizes for the parameters of interest, $\theta_{13}$, $\theta_{23}$, $|\Delta m_{32}^2|$, and \dcp{}.
We constructed a chain that sampled only those parameters using a unit matrix for $\Sigma$.
We computed $\alpha$ for this chain and scaled the relevant elements of $\Sigma$ in order to arrive at a tolerable preliminary acceptance rate of about 20\%. 
We then computed the $k$-lag autocorrelation for each parameter $\theta$, which measures the average correlation between MCMC sample $n$ and sample $n+k$ across all $n$~\cite{Croarkin:2013yxl}:
\begin{equation}
	\label{eq:autocorrelation}
	r_k = \frac{ \sum_{n=1}^{N-k}{\left( \theta_{n} - \bar{\theta} \right)\left(\theta_{n+k} - \bar{\theta} \right)}}{\sum_{n=1}^{N}{\left( \theta_{n} - \bar{\theta} \right)^2}},
\end{equation}
where $\theta_{n}$ refers to the value of parameter $\theta$ at step $n$, and $\bar{\theta}$ is its mean value.
These autocorrelations are shown in Fig.~\ref{fig:autocorr stats-only}.
Using these $r_k$, we further adjusted the elements of $\Sigma$ so the oscillation parameters would have similar autocorrelations.

To optimize the step sizes for the nuisance parameters (systematic uncertainties), we constructed a chain sampling only those parameters, again beginning with $\Sigma$ entries of unity for them.
As with the oscillation parameters, we adjusted $\Sigma$ to ensure none of the nuisance parameters had significantly different autocorrelations from the others.
We then constructed a new, much longer chain, and subsequently computed a covariance matrix over the nuisance parameters from it.
The Cholesky decomposition of this matrix, $L$, was used in the next step.

\begin{figure}[htpb]
	\centering
	\includegraphics[width=\columnwidth]{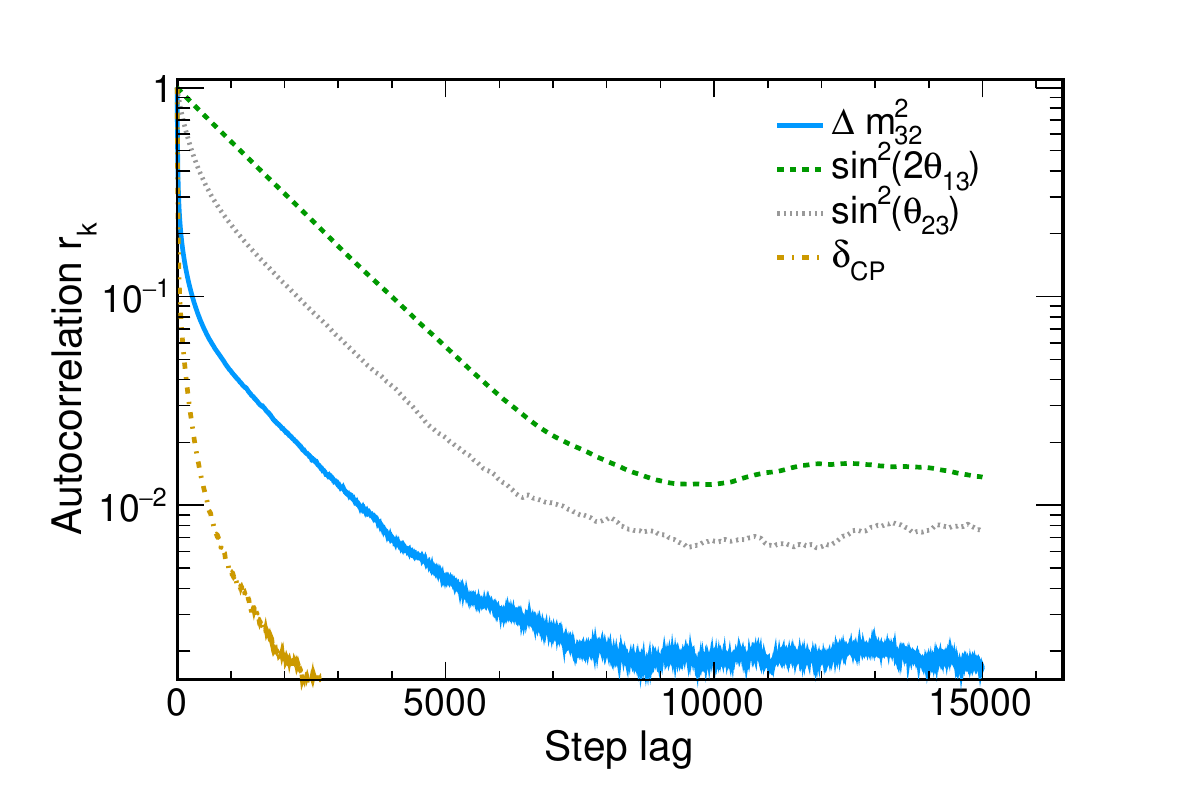}
	\caption{Lag autocorrelation (see Eq.~\ref{eq:autocorrelation}) computed using a \mrrtt{} chain after step size tuning with all parameters included.}
	\label{fig:autocorr final}
\end{figure}

A final chain, this time sampling both oscillation and nuisance parameters, was constructed using the adjusted $\Sigma$ entries for both.
While sampling, the proposed values for the nuisance parameters were multiplied by the decomposed covariance scaled by a tunable factor, $\beta L$. 
Using this chain, we recomputed the autocorrelations for all the parameters.
The elements of $\Sigma$ were readjusted to obtain similar autocorrelations from the oscillation parameters, now in the presence of the nuisance parameters.
We also adjusted $\beta$ to yield similar autocorrelations to those of the oscillation parameters.
A final global scale was applied to $\Sigma$ and $\beta$ to finally arrive at $\alpha = 23.4\%$.
The autocorrelations for the oscillation parameters at the end of the tuning procedure are shown in Fig.~\ref{fig:autocorr final}.

\section{Determining warmup fraction and effective sample size for ARIA}
\label{app:warmup and thinning}

The property of sample proportionality to posterior density in the \mrrtt{} method is only guaranteed as asymptotic behavior.
Therefore, it is usually necessary to discard some number of samples $N_0$ at the beginning of the Markov chain while the chain ``burns in'' or ``warms up.''
Moreover, it is usually impossible to find a choice of step sizes $\Sigma$ in Eq.~\ref{eq:proposal} that proposes samples that are both fully uncorrelated with the previous one and whose acceptance probabilities are high enough to not impose severe computing requirements.
Chains can, in principle, be ``thinned'' by discarding all but the $k$th sample to reduce autocorrelations.
Though we do not actually thin the chains, the resulting fraction $N_{\mathrm{eff}} = \frac{1}{k}(N - N_0)$, the ``number of effective samples,'' still corresponds to the statistical power of the chain.
The plateau in Fig.~\ref{fig:autocorr final} of around or less than 1\% represents the asymptotic behavior of the autocorrelations with the step size tuning procedure we use.
Therefore we interpret our effective sample size as being computed as above with $k=10^4$.

Because autocorrelations in our analysis are relatively long (see Fig.~\ref{fig:autocorr final}) and ARIA runs fairly quickly, we produced very long chains of $5 \times 10^6$ samples each.
Thus, when we compared our posterior densities using $N_0 = 0$, $N_0 = \{1, 3, 5\} \times 10^3$, $N_0 = \{1, 3, 5\} \times 10^4$, and $N_0 = \{1, 3\} \times 10^5$, we found them all to be indistinguishable.
Thus, we did not find any need to discard warmup samples from our chains.

\begin{figure*}%[htpb]
	\centering
	\begin{subfigure}[b]{0.9\columnwidth}
		\begin{center}
      \includegraphics[width=0.9\textwidth]{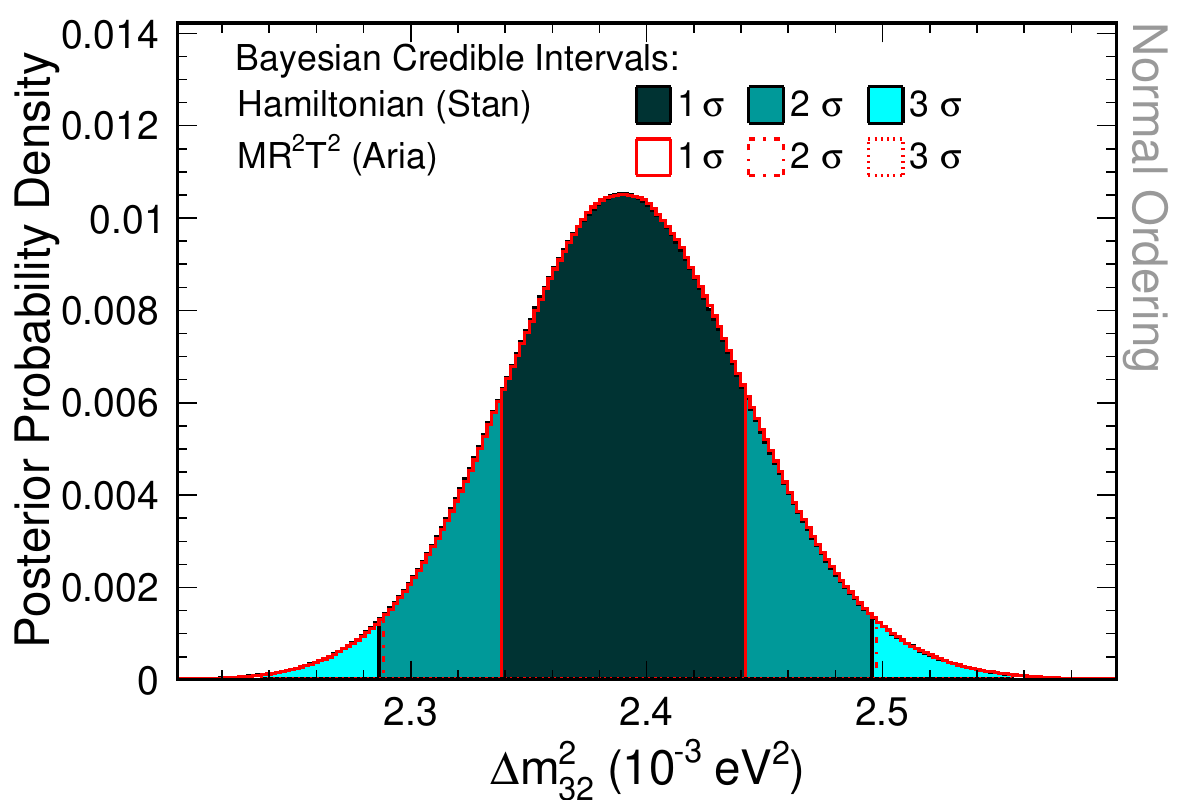}
		\end{center}
	\end{subfigure}
	\begin{subfigure}[b]{0.9\columnwidth}
		\begin{center}
      \includegraphics[width=0.9\textwidth]{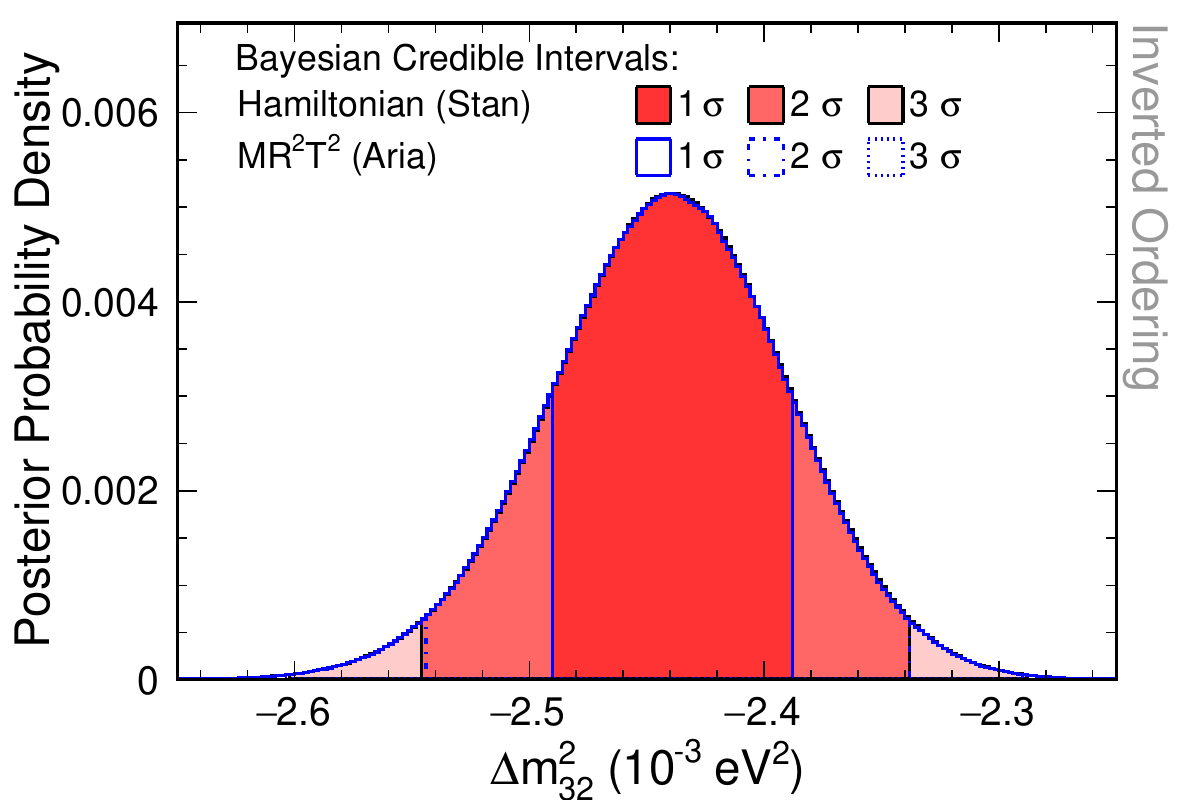}
		\end{center}
	\end{subfigure}
  \caption{Posterior probability density in \dmsq{32} for normal ordering (left) and
           inverted ordering (right). Samples from Stan are shown as the shaded
           distributions (1\,$\sigma$, 2\,$\sigma$, 3\,$\sigma$ regions
           indicated by darkest to lightest fill). Samples from ARIA are shown
           as the red (normal ordering) or blue (inverted ordering) overlaid
           lines (solid, dashed, dotted lines represent the boundaries of the
           1\,$\sigma$, 2\,$\sigma$, 3\,$\sigma$ regions, respectively).}
	\label{fig:aria-stan dm32}
\end{figure*}
\begin{figure*}
	\centering
	\begin{subfigure}[b]{0.9\columnwidth}
		\begin{center}
      \includegraphics[width=0.9\textwidth]{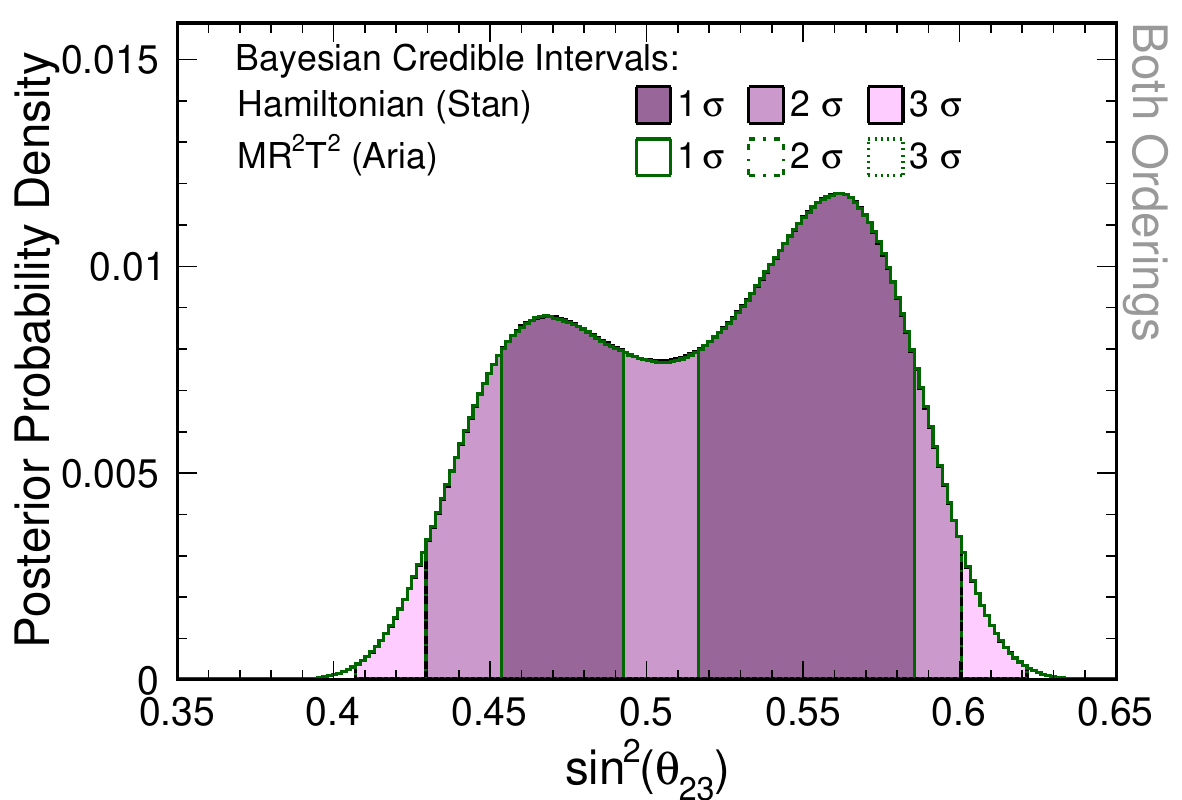}
		\end{center}
	\end{subfigure}
	\begin{subfigure}[b]{0.9\columnwidth}
		\begin{center}
      \includegraphics[width=0.9\textwidth]{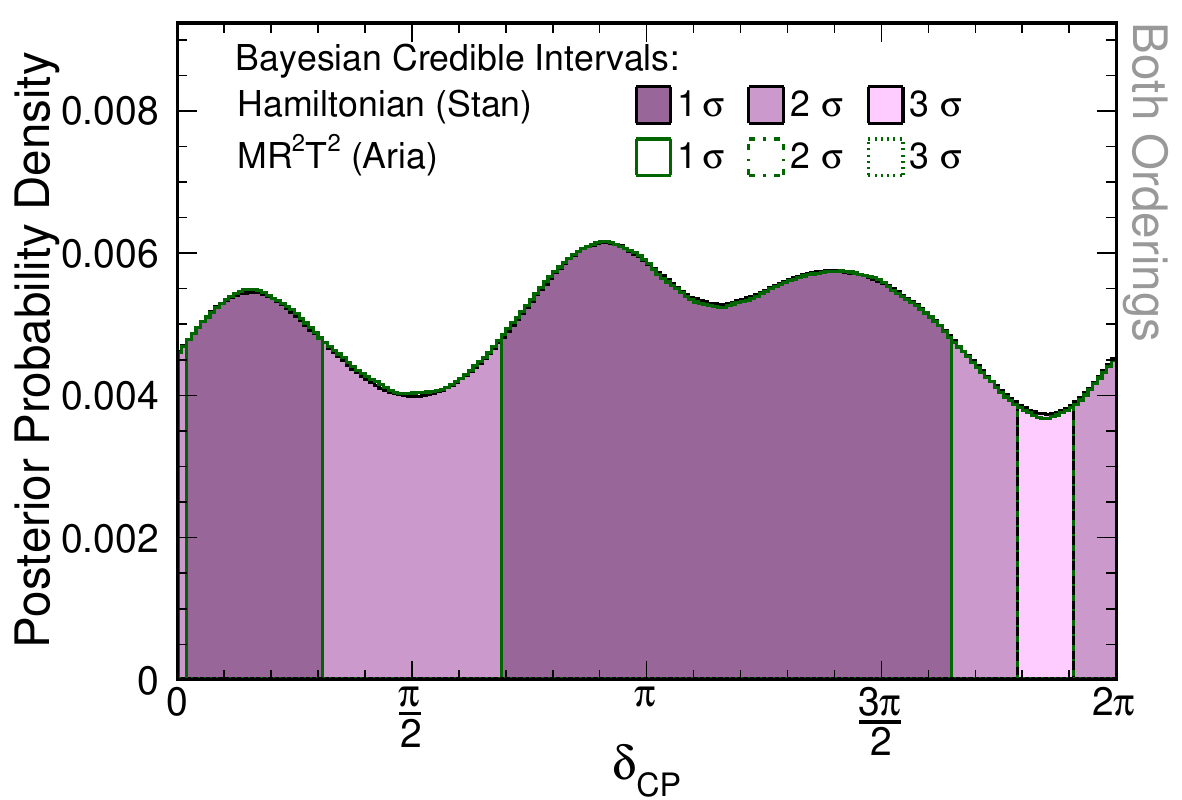}
		\end{center}
	\end{subfigure}
	\caption{Posterior probability density in \ssth{23} (left) and \dcp (right), marginalized over both mass orderings.  Line styles as in Fig.~\ref{fig:aria-stan dm32}.}
	\label{fig:aria-stan ssth23-dCP}
\end{figure*}

\section{Algorithm choices for \hmcmc{}}
\label{app:Stan defaults}

As noted in Sec.~\ref{sec:MCMC details}, \hmcmc{} generates proposals by numerically integrating a Hamiltonian for a fictitious particle, whose potential arises from treating the log-posterior in analogy to gravity:
\begin{alignat}{3}
	\label{eq:Hamilton's eq}
	\frac{d\vec{q}}{dt} & = &\frac{\partial H}{\partial \vec{p}} & = \frac{\partial T}{\partial \vec{p}} \\
	\frac{d\vec{p}}{dt} & = -&\frac{\partial H}{\partial \vec{q}}  & = - \frac{\partial T}{\partial \vec{q}} - \frac{\partial V}{\partial \vec{q}} \nonumber
\end{alignat}
where $T$ and $V$ are the kinetic and potential energies of the system, respectively.

There are two ingredients of \hmcmc{} left unspecified by the method.
In both cases Stan's default choices were found to be suitable for our needs.
The first is the distribution of kinetic energies from which $T$ in Eq.~\ref{eq:Hamilton's eq} is chosen.
Stan's default is the Euclidean-Gaussian kinetic energy distribution:
\begin{equation}
	\label{eq:Euclidean-Gaussian KE}
	T(\vec{q}, \vec{p}) = \frac{1}{2}\vec{p}\,^{\text{T}} M^{-1} \vec{p} + \log |M| + \text{const.}
\end{equation}
Here the mass matrix $M$ (analogous to the effect of mass in gravitation) is a parameter that is automatically inferred by Stan during its warm-up sampling by iterative adjustments based on a running covariance over the samples.
The second implementation choice is how long the integrator is allowed to run for each particular trajectory.
Stan uses an algorithm called No-U-Turns (NUTS)~\cite{Hoffman:2014a}, which is a heuristic method that halts integration when two trajectories, extending in each direction from the starting point along the initial momentum, begin to converge towards one another.
An upper limit of integrator steps is also supplied as a parameter to Stan; in this analysis, we find that all our trajectories end within $2^{11}$ steps.

\section{Equivalence of ARIA and Stan results}
\label{app:ARIA-Stan equivalence}

We extracted the posterior distributions for all of the results shown in this paper using both the ARIA and Stan samplers described in Sec.~\ref{sec:MCMC details}.
The posteriors were in all cases nearly indistinguishable, with tiny differences that occasionally caused the boundaries of credible intervals to shift by single bins.
This is illustrated in Figs.~\ref{fig:aria-stan dm32}-\ref{fig:aria-stan ssth23-dCP}.

\bibliography{cites}

\begin{thebibliography}{103}%
\makeatletter
\providecommand \@ifxundefined [1]{%
 \@ifx{#1\undefined}
}%
\providecommand \@ifnum [1]{%
 \ifnum #1\expandafter \@firstoftwo
 \else \expandafter \@secondoftwo
 \fi
}%
\providecommand \@ifx [1]{%
 \ifx #1\expandafter \@firstoftwo
 \else \expandafter \@secondoftwo
 \fi
}%
\providecommand \natexlab [1]{#1}%
\providecommand \enquote  [1]{``#1''}%
\providecommand \bibnamefont  [1]{#1}%
\providecommand \bibfnamefont [1]{#1}%
\providecommand \citenamefont [1]{#1}%
\providecommand \href@noop [0]{\@secondoftwo}%
\providecommand \href [0]{\begingroup \@sanitize@url \@href}%
\providecommand \@href[1]{\@@startlink{#1}\@@href}%
\providecommand \@@href[1]{\endgroup#1\@@endlink}%
\providecommand \@sanitize@url [0]{\catcode `\\12\catcode `\$12\catcode
  `\&12\catcode `\#12\catcode `\^12\catcode `\_12\catcode `\%12\relax}%
\providecommand \@@startlink[1]{}%
\providecommand \@@endlink[0]{}%
\providecommand \url  [0]{\begingroup\@sanitize@url \@url }%
\providecommand \@url [1]{\endgroup\@href {#1}{\urlprefix }}%
\providecommand \urlprefix  [0]{URL }%
\providecommand \Eprint [0]{\href }%
\providecommand \doibase [0]{https://doi.org/}%
\providecommand \selectlanguage [0]{\@gobble}%
\providecommand \bibinfo  [0]{\@secondoftwo}%
\providecommand \bibfield  [0]{\@secondoftwo}%
\providecommand \translation [1]{[#1]}%
\providecommand \BibitemOpen [0]{}%
\providecommand \bibitemStop [0]{}%
\providecommand \bibitemNoStop [0]{.\EOS\space}%
\providecommand \EOS [0]{\spacefactor3000\relax}%
\providecommand \BibitemShut  [1]{\csname bibitem#1\endcsname}%
\let\auto@bib@innerbib\@empty
%</preamble>
\bibitem [{\citenamefont {Fukuda}\ \emph {et~al.}(1998)\citenamefont {Fukuda}
  \emph {et~al.}}]{Fukuda:1998mi}%
  \BibitemOpen
  \bibfield  {author} {\bibinfo {author} {\bibfnamefont {Y.}~\bibnamefont
  {Fukuda}} \emph {et~al.} (\bibinfo {collaboration} {Super-Kamiokande}),\
  }\href {https://doi.org/10.1103/PhysRevLett.81.1562} {\bibfield  {journal}
  {\bibinfo  {journal} {Phys. Rev. Lett.}\ }\textbf {\bibinfo {volume} {81}},\
  \bibinfo {pages} {1562} (\bibinfo {year} {1998})},\ \Eprint
  {https://arxiv.org/abs/hep-ex/9807003} {arXiv:hep-ex/9807003 [hep-ex]}
  \BibitemShut {NoStop}%
%%CITATION = HEP-EX/9807003;%%
\bibitem [{\citenamefont {Fukuda}\ \emph {et~al.}(2002)\citenamefont {Fukuda}
  \emph {et~al.}}]{Fukuda:2002pe}%
  \BibitemOpen
  \bibfield  {author} {\bibinfo {author} {\bibfnamefont {S.}~\bibnamefont
  {Fukuda}} \emph {et~al.} (\bibinfo {collaboration} {Super-Kamiokande}),\
  }\href {https://doi.org/10.1016/S0370-2693(02)02090-7} {\bibfield  {journal}
  {\bibinfo  {journal} {Phys. Lett. B}\ }\textbf {\bibinfo {volume} {539}},\
  \bibinfo {pages} {179} (\bibinfo {year} {2002})},\ \Eprint
  {https://arxiv.org/abs/hep-ex/0205075} {arXiv:hep-ex/0205075 [hep-ex]}
  \BibitemShut {NoStop}%
%%CITATION = HEP-EX/0205075;%%
\bibitem [{\citenamefont {Ahmad}\ \emph {et~al.}(2002)\citenamefont {Ahmad}
  \emph {et~al.}}]{Ahmad:2002jz}%
  \BibitemOpen
  \bibfield  {author} {\bibinfo {author} {\bibfnamefont {Q.~R.}\ \bibnamefont
  {Ahmad}} \emph {et~al.} (\bibinfo {collaboration} {SNO}),\ }\href
  {https://doi.org/10.1103/PhysRevLett.89.011301} {\bibfield  {journal}
  {\bibinfo  {journal} {Phys. Rev. Lett.}\ }\textbf {\bibinfo {volume} {89}},\
  \bibinfo {pages} {011301} (\bibinfo {year} {2002})},\ \Eprint
  {https://arxiv.org/abs/nucl-ex/0204008} {arXiv:nucl-ex/0204008 [nucl-ex]}
  \BibitemShut {NoStop}%
%%CITATION = NUCL-EX/0204008;%%
\bibitem [{\citenamefont {Eguchi}\ \emph {et~al.}(2003)\citenamefont {Eguchi}
  \emph {et~al.}}]{Eguchi:2002dm}%
  \BibitemOpen
  \bibfield  {author} {\bibinfo {author} {\bibfnamefont {K.}~\bibnamefont
  {Eguchi}} \emph {et~al.} (\bibinfo {collaboration} {KamLAND}),\ }\href
  {https://doi.org/10.1103/PhysRevLett.90.021802} {\bibfield  {journal}
  {\bibinfo  {journal} {Phys. Rev. Lett.}\ }\textbf {\bibinfo {volume} {90}},\
  \bibinfo {pages} {021802} (\bibinfo {year} {2003})},\ \Eprint
  {https://arxiv.org/abs/hep-ex/0212021} {arXiv:hep-ex/0212021 [hep-ex]}
  \BibitemShut {NoStop}%
%%CITATION = HEP-EX/0212021;%%
\bibitem [{\citenamefont {Michael}\ \emph {et~al.}(2006)\citenamefont {Michael}
  \emph {et~al.}}]{Michael:2006rx}%
  \BibitemOpen
  \bibfield  {author} {\bibinfo {author} {\bibfnamefont {D.~G.}\ \bibnamefont
  {Michael}} \emph {et~al.} (\bibinfo {collaboration} {MINOS}),\ }\href
  {https://doi.org/10.1103/PhysRevLett.97.191801} {\bibfield  {journal}
  {\bibinfo  {journal} {Phys. Rev. Lett.}\ }\textbf {\bibinfo {volume} {97}},\
  \bibinfo {pages} {191801} (\bibinfo {year} {2006})},\ \Eprint
  {https://arxiv.org/abs/hep-ex/0607088} {arXiv:hep-ex/0607088 [hep-ex]}
  \BibitemShut {NoStop}%
%%CITATION = HEP-EX/0607088;%%
\bibitem [{\citenamefont {Abe}\ \emph {et~al.}(2011)\citenamefont {Abe} \emph
  {et~al.}}]{Abe:2011sj}%
  \BibitemOpen
  \bibfield  {author} {\bibinfo {author} {\bibfnamefont {K.}~\bibnamefont
  {Abe}} \emph {et~al.} (\bibinfo {collaboration} {T2K}),\ }\href
  {https://doi.org/10.1103/PhysRevLett.107.041801} {\bibfield  {journal}
  {\bibinfo  {journal} {Phys. Rev. Lett.}\ }\textbf {\bibinfo {volume} {107}},\
  \bibinfo {pages} {041801} (\bibinfo {year} {2011})},\ \Eprint
  {https://arxiv.org/abs/1106.2822} {arXiv:1106.2822 [hep-ex]} \BibitemShut
  {NoStop}%
%%CITATION = ARXIV:1106.2822;%%
\bibitem [{\citenamefont {Abe}\ \emph {et~al.}(2012)\citenamefont {Abe} \emph
  {et~al.}}]{Abe:2011fz}%
  \BibitemOpen
  \bibfield  {author} {\bibinfo {author} {\bibfnamefont {Y.}~\bibnamefont
  {Abe}} \emph {et~al.} (\bibinfo {collaboration} {Double Chooz}),\ }\href
  {https://doi.org/10.1103/PhysRevLett.108.131801} {\bibfield  {journal}
  {\bibinfo  {journal} {Phys. Rev. Lett.}\ }\textbf {\bibinfo {volume} {108}},\
  \bibinfo {pages} {131801} (\bibinfo {year} {2012})},\ \Eprint
  {https://arxiv.org/abs/1112.6353} {arXiv:1112.6353 [hep-ex]} \BibitemShut
  {NoStop}%
%%CITATION = ARXIV:1112.6353;%%
\bibitem [{\citenamefont {An}\ \emph {et~al.}(2012)\citenamefont {An} \emph
  {et~al.}}]{An:2012eh}%
  \BibitemOpen
  \bibfield  {author} {\bibinfo {author} {\bibfnamefont {F.~P.}\ \bibnamefont
  {An}} \emph {et~al.} (\bibinfo {collaboration} {Daya Bay}),\ }\href
  {https://doi.org/10.1103/PhysRevLett.108.171803} {\bibfield  {journal}
  {\bibinfo  {journal} {Phys. Rev. Lett.}\ }\textbf {\bibinfo {volume} {108}},\
  \bibinfo {pages} {171803} (\bibinfo {year} {2012})},\ \Eprint
  {https://arxiv.org/abs/1203.1669} {arXiv:1203.1669 [hep-ex]} \BibitemShut
  {NoStop}%
%%CITATION = ARXIV:1203.1669;%%
\bibitem [{\citenamefont {Ahn}\ \emph {et~al.}(2012)\citenamefont {Ahn} \emph
  {et~al.}}]{Ahn:2012nd}%
  \BibitemOpen
  \bibfield  {author} {\bibinfo {author} {\bibfnamefont {J.~K.}\ \bibnamefont
  {Ahn}} \emph {et~al.} (\bibinfo {collaboration} {RENO}),\ }\href
  {https://doi.org/10.1103/PhysRevLett.108.191802} {\bibfield  {journal}
  {\bibinfo  {journal} {Phys. Rev. Lett.}\ }\textbf {\bibinfo {volume} {108}},\
  \bibinfo {pages} {191802} (\bibinfo {year} {2012})},\ \Eprint
  {https://arxiv.org/abs/1204.0626} {arXiv:1204.0626 [hep-ex]} \BibitemShut
  {NoStop}%
%%CITATION = ARXIV:1204.0626;%%
\bibitem [{\citenamefont {Mohapatra}\ and\ \citenamefont
  {Smirnov}(2006)}]{Mohapatra:2006gs}%
  \BibitemOpen
  \bibfield  {author} {\bibinfo {author} {\bibfnamefont {R.~N.}\ \bibnamefont
  {Mohapatra}}\ and\ \bibinfo {author} {\bibfnamefont {A.~Y.}\ \bibnamefont
  {Smirnov}},\ }\bibfield  {booktitle} {\emph {\bibinfo {booktitle}
  {{Elementary particle physics. Proceedings, Corfu Summer Institute,
  CORFU2005, Corfu, Greece, September 4-26, 2005}}},\ }\href
  {https://doi.org/10.1146/annurev.nucl.56.080805.140534} {\bibfield  {journal}
  {\bibinfo  {journal} {Ann. Rev. Nucl. Part. Sci.}\ }\textbf {\bibinfo
  {volume} {56}},\ \bibinfo {pages} {569} (\bibinfo {year} {2006})},\ \Eprint
  {https://arxiv.org/abs/hep-ph/0603118} {arXiv:hep-ph/0603118 [hep-ph]}
  \BibitemShut {NoStop}%
%%CITATION = HEP-PH/0603118;%%
\bibitem [{\citenamefont {Nunokawa}\ \emph {et~al.}(2008)\citenamefont
  {Nunokawa}, \citenamefont {Parke},\ and\ \citenamefont
  {Valle}}]{Nunokawa:2007qh}%
  \BibitemOpen
  \bibfield  {author} {\bibinfo {author} {\bibfnamefont {H.}~\bibnamefont
  {Nunokawa}}, \bibinfo {author} {\bibfnamefont {S.~J.}\ \bibnamefont
  {Parke}},\ and\ \bibinfo {author} {\bibfnamefont {J.~W.~F.}\ \bibnamefont
  {Valle}},\ }\href {https://doi.org/10.1016/j.ppnp.2007.10.001} {\bibfield
  {journal} {\bibinfo  {journal} {Prog. Part. Nucl. Phys.}\ }\textbf {\bibinfo
  {volume} {60}},\ \bibinfo {pages} {338} (\bibinfo {year} {2008})},\ \Eprint
  {https://arxiv.org/abs/0710.0554} {arXiv:0710.0554 [hep-ph]} \BibitemShut
  {NoStop}%
%%CITATION = ARXIV:0710.0554;%%
\bibitem [{\citenamefont {Altarelli}\ and\ \citenamefont
  {Feruglio}(2010)}]{Altarelli:2010gt}%
  \BibitemOpen
  \bibfield  {author} {\bibinfo {author} {\bibfnamefont {G.}~\bibnamefont
  {Altarelli}}\ and\ \bibinfo {author} {\bibfnamefont {F.}~\bibnamefont
  {Feruglio}},\ }\href {https://doi.org/10.1103/RevModPhys.82.2701} {\bibfield
  {journal} {\bibinfo  {journal} {Rev. Mod. Phys.}\ }\textbf {\bibinfo {volume}
  {82}},\ \bibinfo {pages} {2701} (\bibinfo {year} {2010})},\ \Eprint
  {https://arxiv.org/abs/1002.0211} {arXiv:1002.0211 [hep-ph]} \BibitemShut
  {NoStop}%
%%CITATION = ARXIV:1002.0211;%%
\bibitem [{\citenamefont {King}(2015)}]{King:2015aea}%
  \BibitemOpen
  \bibfield  {author} {\bibinfo {author} {\bibfnamefont {S.~F.}\ \bibnamefont
  {King}},\ }\href {https://doi.org/10.1088/0954-3899/42/12/123001} {\bibfield
  {journal} {\bibinfo  {journal} {J. Phys. G}\ }\textbf {\bibinfo {volume}
  {42}},\ \bibinfo {pages} {123001} (\bibinfo {year} {2015})},\ \Eprint
  {https://arxiv.org/abs/1510.02091} {arXiv:1510.02091 [hep-ph]} \BibitemShut
  {NoStop}%
%%CITATION = ARXIV:1510.02091;%%
\bibitem [{\citenamefont {Petcov}(2018)}]{Petcov:2017ggy}%
  \BibitemOpen
  \bibfield  {author} {\bibinfo {author} {\bibfnamefont {S.~T.}\ \bibnamefont
  {Petcov}},\ }\href {https://doi.org/10.1140/epjc/s10052-018-6158-5}
  {\bibfield  {journal} {\bibinfo  {journal} {Eur. Phys. J. C}\ }\textbf
  {\bibinfo {volume} {78}},\ \bibinfo {pages} {709} (\bibinfo {year} {2018})},\
  \Eprint {https://arxiv.org/abs/1711.10806} {arXiv:1711.10806 [hep-ph]}
  \BibitemShut {NoStop}%
%%CITATION = ARXIV:1711.10806;%%
\bibitem [{\citenamefont {Fukugita}\ and\ \citenamefont
  {Yanagida}(1986)}]{Fukugita:1986hr}%
  \BibitemOpen
  \bibfield  {author} {\bibinfo {author} {\bibfnamefont {M.}~\bibnamefont
  {Fukugita}}\ and\ \bibinfo {author} {\bibfnamefont {T.}~\bibnamefont
  {Yanagida}},\ }\href {https://doi.org/10.1016/0370-2693(86)91126-3}
  {\bibfield  {journal} {\bibinfo  {journal} {Phys. Lett. B}\ }\textbf
  {\bibinfo {volume} {174}},\ \bibinfo {pages} {45} (\bibinfo {year}
  {1986})}\BibitemShut {NoStop}%
\bibitem [{\citenamefont {Buchmuller}\ and\ \citenamefont
  {Plumacher}(1996)}]{Buchmuller:1996pa}%
  \BibitemOpen
  \bibfield  {author} {\bibinfo {author} {\bibfnamefont {W.}~\bibnamefont
  {Buchmuller}}\ and\ \bibinfo {author} {\bibfnamefont {M.}~\bibnamefont
  {Plumacher}},\ }\href {https://doi.org/10.1016/S0370-2693(96)01232-4}
  {\bibfield  {journal} {\bibinfo  {journal} {Phys. Lett. B}\ }\textbf
  {\bibinfo {volume} {389}},\ \bibinfo {pages} {73} (\bibinfo {year} {1996})},\
  \Eprint {https://arxiv.org/abs/hep-ph/9608308} {arXiv:hep-ph/9608308}
  \BibitemShut {NoStop}%
\bibitem [{\citenamefont {Buchmuller}\ \emph
  {et~al.}(2005{\natexlab{a}})\citenamefont {Buchmuller}, \citenamefont
  {Di~Bari},\ and\ \citenamefont {Plumacher}}]{Buchmuller:2004nz}%
  \BibitemOpen
  \bibfield  {author} {\bibinfo {author} {\bibfnamefont {W.}~\bibnamefont
  {Buchmuller}}, \bibinfo {author} {\bibfnamefont {P.}~\bibnamefont
  {Di~Bari}},\ and\ \bibinfo {author} {\bibfnamefont {M.}~\bibnamefont
  {Plumacher}},\ }\href {https://doi.org/10.1016/j.aop.2004.02.003} {\bibfield
  {journal} {\bibinfo  {journal} {Annals Phys.}\ }\textbf {\bibinfo {volume}
  {315}},\ \bibinfo {pages} {305} (\bibinfo {year} {2005}{\natexlab{a}})},\
  \Eprint {https://arxiv.org/abs/hep-ph/0401240} {arXiv:hep-ph/0401240}
  \BibitemShut {NoStop}%
\bibitem [{\citenamefont {Buchmuller}\ \emph
  {et~al.}(2005{\natexlab{b}})\citenamefont {Buchmuller}, \citenamefont
  {Peccei},\ and\ \citenamefont {Yanagida}}]{Buchmuller:2005eh}%
  \BibitemOpen
  \bibfield  {author} {\bibinfo {author} {\bibfnamefont {W.}~\bibnamefont
  {Buchmuller}}, \bibinfo {author} {\bibfnamefont {R.~D.}\ \bibnamefont
  {Peccei}},\ and\ \bibinfo {author} {\bibfnamefont {T.}~\bibnamefont
  {Yanagida}},\ }\href {https://doi.org/10.1146/annurev.nucl.55.090704.151558}
  {\bibfield  {journal} {\bibinfo  {journal} {Ann. Rev. Nucl. Part. Sci.}\
  }\textbf {\bibinfo {volume} {55}},\ \bibinfo {pages} {311} (\bibinfo {year}
  {2005}{\natexlab{b}})},\ \Eprint {https://arxiv.org/abs/hep-ph/0502169}
  {arXiv:hep-ph/0502169} \BibitemShut {NoStop}%
\bibitem [{\citenamefont {Pilaftsis}(1997)}]{Pilaftsis:1997jf}%
  \BibitemOpen
  \bibfield  {author} {\bibinfo {author} {\bibfnamefont {A.}~\bibnamefont
  {Pilaftsis}},\ }\href {https://doi.org/10.1103/PhysRevD.56.5431} {\bibfield
  {journal} {\bibinfo  {journal} {Phys. Rev. D}\ }\textbf {\bibinfo {volume}
  {56}},\ \bibinfo {pages} {5431} (\bibinfo {year} {1997})},\ \Eprint
  {https://arxiv.org/abs/hep-ph/9707235} {arXiv:hep-ph/9707235} \BibitemShut
  {NoStop}%
\bibitem [{\citenamefont {Acero}\ \emph
  {et~al.}(2022{\natexlab{a}})\citenamefont {Acero} \emph
  {et~al.}}]{NOvA:2022wnj}%
  \BibitemOpen
  \bibfield  {author} {\bibinfo {author} {\bibfnamefont {M.~A.}\ \bibnamefont
  {Acero}} \emph {et~al.} (\bibinfo {collaboration} {NOvA}),\ }\href@noop {}
  {\bibfield  {journal} {\bibinfo  {journal} {Submitted to \textit{Phys. Rev.
  D}}\ } (\bibinfo {year} {2022}{\natexlab{a}})},\ \Eprint
  {https://arxiv.org/abs/2207.14353} {arXiv:2207.14353 [hep-ex]} \BibitemShut
  {NoStop}%
\bibitem [{\citenamefont {Abe}\ \emph {et~al.}(2023{\natexlab{a}})\citenamefont
  {Abe} \emph {et~al.}}]{T2K:2023mcm}%
  \BibitemOpen
  \bibfield  {author} {\bibinfo {author} {\bibfnamefont {K.}~\bibnamefont
  {Abe}} \emph {et~al.} (\bibinfo {collaboration} {T2K}),\ }\href
  {https://doi.org/10.1103/PhysRevD.108.072011} {\bibfield  {journal} {\bibinfo
   {journal} {Phys. Rev. D}\ }\textbf {\bibinfo {volume} {108}},\ \bibinfo
  {pages} {072011} (\bibinfo {year} {2023}{\natexlab{a}})},\ \Eprint
  {https://arxiv.org/abs/2305.09916} {arXiv:2305.09916 [hep-ex]} \BibitemShut
  {NoStop}%
\bibitem [{\citenamefont {Jiang}\ \emph {et~al.}(2019)\citenamefont {Jiang}
  \emph {et~al.}}]{Super-Kamiokande:2019gzr}%
  \BibitemOpen
  \bibfield  {author} {\bibinfo {author} {\bibfnamefont {M.}~\bibnamefont
  {Jiang}} \emph {et~al.} (\bibinfo {collaboration} {Super-Kamiokande}),\
  }\href {https://doi.org/10.1093/ptep/ptz015} {\bibfield  {journal} {\bibinfo
  {journal} {PTEP}\ }\textbf {\bibinfo {volume} {2019}},\ \bibinfo {pages}
  {053F01} (\bibinfo {year} {2019})},\ \Eprint
  {https://arxiv.org/abs/1901.03230} {arXiv:1901.03230 [hep-ex]} \BibitemShut
  {NoStop}%
\bibitem [{\citenamefont {Abbasi}\ \emph {et~al.}(2023)\citenamefont {Abbasi}
  \emph {et~al.}}]{IceCubeCollaboration:2023wtb}%
  \BibitemOpen
  \bibfield  {author} {\bibinfo {author} {\bibfnamefont {R.}~\bibnamefont
  {Abbasi}} \emph {et~al.} (\bibinfo {collaboration} {(IceCube Collaboration)*,
  IceCube}),\ }\href {https://doi.org/10.1103/PhysRevD.108.012014} {\bibfield
  {journal} {\bibinfo  {journal} {Phys. Rev. D}\ }\textbf {\bibinfo {volume}
  {108}},\ \bibinfo {pages} {012014} (\bibinfo {year} {2023})},\ \Eprint
  {https://arxiv.org/abs/2304.12236} {arXiv:2304.12236 [hep-ex]} \BibitemShut
  {NoStop}%
\bibitem [{\citenamefont {Workman}\ \emph {et~al.}(2022)\citenamefont {Workman}
  \emph {et~al.}}]{ParticleDataGroup:2022pth}%
  \BibitemOpen
  \bibfield  {author} {\bibinfo {author} {\bibfnamefont {R.~L.}\ \bibnamefont
  {Workman}} \emph {et~al.} (\bibinfo {collaboration} {Particle Data Group}),\
  }\href {https://doi.org/10.1093/ptep/ptac097} {\bibfield  {journal} {\bibinfo
   {journal} {PTEP}\ }\textbf {\bibinfo {volume} {2022}},\ \bibinfo {pages}
  {083C01} (\bibinfo {year} {2022})}\BibitemShut {NoStop}%
\bibitem [{\citenamefont {Wolfenstein}(1978)}]{Wolfenstein:1977ue}%
  \BibitemOpen
  \bibfield  {author} {\bibinfo {author} {\bibfnamefont {L.}~\bibnamefont
  {Wolfenstein}},\ }\href {https://doi.org/10.1103/PhysRevD.17.2369} {\bibfield
   {journal} {\bibinfo  {journal} {Phys. Rev. D}\ }\textbf {\bibinfo {volume}
  {17}},\ \bibinfo {pages} {2369} (\bibinfo {year} {1978})}\BibitemShut
  {NoStop}%
\bibitem [{\citenamefont {Abe}\ \emph {et~al.}(2023{\natexlab{b}})\citenamefont
  {Abe} \emph {et~al.}}]{T2K:2023smv}%
  \BibitemOpen
  \bibfield  {author} {\bibinfo {author} {\bibfnamefont {K.}~\bibnamefont
  {Abe}} \emph {et~al.} (\bibinfo {collaboration} {T2K}),\ }\href
  {https://doi.org/10.1140/epjc/s10052-023-11819-x} {\bibfield  {journal}
  {\bibinfo  {journal} {Eur. Phys. J. C}\ }\textbf {\bibinfo {volume} {83}},\
  \bibinfo {pages} {782} (\bibinfo {year} {2023}{\natexlab{b}})},\ \Eprint
  {https://arxiv.org/abs/2303.03222} {arXiv:2303.03222 [hep-ex]} \BibitemShut
  {NoStop}%
\bibitem [{\citenamefont {Adamson}\ \emph {et~al.}(2013)\citenamefont {Adamson}
  \emph {et~al.}}]{MINOS:2013xrl}%
  \BibitemOpen
  \bibfield  {author} {\bibinfo {author} {\bibfnamefont {P.}~\bibnamefont
  {Adamson}} \emph {et~al.} (\bibinfo {collaboration} {MINOS}),\ }\href
  {https://doi.org/10.1103/PhysRevLett.110.171801} {\bibfield  {journal}
  {\bibinfo  {journal} {Phys. Rev. Lett.}\ }\textbf {\bibinfo {volume} {110}},\
  \bibinfo {pages} {171801} (\bibinfo {year} {2013})},\ \Eprint
  {https://arxiv.org/abs/1301.4581} {arXiv:1301.4581 [hep-ex]} \BibitemShut
  {NoStop}%
\bibitem [{\citenamefont {Esteban}\ \emph {et~al.}(2020)\citenamefont
  {Esteban}, \citenamefont {Gonzalez-Garcia}, \citenamefont {Maltoni},
  \citenamefont {Schwetz},\ and\ \citenamefont {Zhou}}]{Esteban:2020cvm}%
  \BibitemOpen
  \bibfield  {author} {\bibinfo {author} {\bibfnamefont {I.}~\bibnamefont
  {Esteban}}, \bibinfo {author} {\bibfnamefont {M.~C.}\ \bibnamefont
  {Gonzalez-Garcia}}, \bibinfo {author} {\bibfnamefont {M.}~\bibnamefont
  {Maltoni}}, \bibinfo {author} {\bibfnamefont {T.}~\bibnamefont {Schwetz}},\
  and\ \bibinfo {author} {\bibfnamefont {A.}~\bibnamefont {Zhou}},\ }\href
  {https://doi.org/10.1007/JHEP09(2020)178} {\bibfield  {journal} {\bibinfo
  {journal} {JHEP}\ }\textbf {\bibinfo {volume} {09}},\ \bibinfo {pages}
  {178}},\ \Eprint {https://arxiv.org/abs/2007.14792} {arXiv:2007.14792
  [hep-ph]} \BibitemShut {NoStop}%
\bibitem [{\citenamefont {Adamson}\ \emph
  {et~al.}(2016{\natexlab{a}})\citenamefont {Adamson} \emph
  {et~al.}}]{NOvA:2016vij}%
  \BibitemOpen
  \bibfield  {author} {\bibinfo {author} {\bibfnamefont {P.}~\bibnamefont
  {Adamson}} \emph {et~al.} (\bibinfo {collaboration} {NOvA}),\ }\href
  {https://doi.org/10.1103/PhysRevD.93.051104} {\bibfield  {journal} {\bibinfo
  {journal} {Phys. Rev. D}\ }\textbf {\bibinfo {volume} {93}},\ \bibinfo
  {pages} {051104} (\bibinfo {year} {2016}{\natexlab{a}})},\ \Eprint
  {https://arxiv.org/abs/1601.05037} {arXiv:1601.05037 [hep-ex]} \BibitemShut
  {NoStop}%
\bibitem [{\citenamefont {Adamson}\ \emph
  {et~al.}(2016{\natexlab{b}})\citenamefont {Adamson} \emph
  {et~al.}}]{NOvA:2016kwd}%
  \BibitemOpen
  \bibfield  {author} {\bibinfo {author} {\bibfnamefont {P.}~\bibnamefont
  {Adamson}} \emph {et~al.} (\bibinfo {collaboration} {NOvA}),\ }\href
  {https://doi.org/10.1103/PhysRevLett.116.151806} {\bibfield  {journal}
  {\bibinfo  {journal} {Phys. Rev. Lett.}\ }\textbf {\bibinfo {volume} {116}},\
  \bibinfo {pages} {151806} (\bibinfo {year} {2016}{\natexlab{b}})},\ \Eprint
  {https://arxiv.org/abs/1601.05022} {arXiv:1601.05022 [hep-ex]} \BibitemShut
  {NoStop}%
\bibitem [{\citenamefont {Adamson}\ \emph
  {et~al.}(2017{\natexlab{a}})\citenamefont {Adamson} \emph
  {et~al.}}]{NOvA:2017ohq}%
  \BibitemOpen
  \bibfield  {author} {\bibinfo {author} {\bibfnamefont {P.}~\bibnamefont
  {Adamson}} \emph {et~al.} (\bibinfo {collaboration} {NOvA}),\ }\href
  {https://doi.org/10.1103/PhysRevLett.118.151802} {\bibfield  {journal}
  {\bibinfo  {journal} {Phys. Rev. Lett.}\ }\textbf {\bibinfo {volume} {118}},\
  \bibinfo {pages} {151802} (\bibinfo {year} {2017}{\natexlab{a}})},\ \Eprint
  {https://arxiv.org/abs/1701.05891} {arXiv:1701.05891 [hep-ex]} \BibitemShut
  {NoStop}%
\bibitem [{\citenamefont {Adamson}\ \emph
  {et~al.}(2017{\natexlab{b}})\citenamefont {Adamson} \emph
  {et~al.}}]{NOvA:2017abs}%
  \BibitemOpen
  \bibfield  {author} {\bibinfo {author} {\bibfnamefont {P.}~\bibnamefont
  {Adamson}} \emph {et~al.} (\bibinfo {collaboration} {NOvA}),\ }\href
  {https://doi.org/10.1103/PhysRevLett.118.231801} {\bibfield  {journal}
  {\bibinfo  {journal} {Phys. Rev. Lett.}\ }\textbf {\bibinfo {volume} {118}},\
  \bibinfo {pages} {231801} (\bibinfo {year} {2017}{\natexlab{b}})},\ \Eprint
  {https://arxiv.org/abs/1703.03328} {arXiv:1703.03328 [hep-ex]} \BibitemShut
  {NoStop}%
\bibitem [{\citenamefont {Acero}\ \emph {et~al.}(2018)\citenamefont {Acero}
  \emph {et~al.}}]{NOvA:2018gge}%
  \BibitemOpen
  \bibfield  {author} {\bibinfo {author} {\bibfnamefont {M.~A.}\ \bibnamefont
  {Acero}} \emph {et~al.} (\bibinfo {collaboration} {NOvA}),\ }\href
  {https://doi.org/10.1103/PhysRevD.98.032012} {\bibfield  {journal} {\bibinfo
  {journal} {Phys. Rev. D}\ }\textbf {\bibinfo {volume} {98}},\ \bibinfo
  {pages} {032012} (\bibinfo {year} {2018})},\ \Eprint
  {https://arxiv.org/abs/1806.00096} {arXiv:1806.00096 [hep-ex]} \BibitemShut
  {NoStop}%
\bibitem [{\citenamefont {Acero}\ \emph {et~al.}(2019)\citenamefont {Acero}
  \emph {et~al.}}]{NOvA:2019cyt}%
  \BibitemOpen
  \bibfield  {author} {\bibinfo {author} {\bibfnamefont {M.~A.}\ \bibnamefont
  {Acero}} \emph {et~al.} (\bibinfo {collaboration} {NOvA}),\ }\href
  {https://doi.org/10.1103/PhysRevLett.123.151803} {\bibfield  {journal}
  {\bibinfo  {journal} {Phys. Rev. Lett.}\ }\textbf {\bibinfo {volume} {123}},\
  \bibinfo {pages} {151803} (\bibinfo {year} {2019})},\ \Eprint
  {https://arxiv.org/abs/1906.04907} {arXiv:1906.04907 [hep-ex]} \BibitemShut
  {NoStop}%
\bibitem [{\citenamefont {Acero}\ \emph
  {et~al.}(2022{\natexlab{b}})\citenamefont {Acero} \emph
  {et~al.}}]{NOvA:2021nfi}%
  \BibitemOpen
  \bibfield  {author} {\bibinfo {author} {\bibfnamefont {M.~A.}\ \bibnamefont
  {Acero}} \emph {et~al.} (\bibinfo {collaboration} {NOvA}),\ }\href
  {https://doi.org/10.1103/PhysRevD.106.032004} {\bibfield  {journal} {\bibinfo
   {journal} {Phys. Rev. D}\ }\textbf {\bibinfo {volume} {106}},\ \bibinfo
  {pages} {032004} (\bibinfo {year} {2022}{\natexlab{b}})},\ \Eprint
  {https://arxiv.org/abs/2108.08219} {arXiv:2108.08219 [hep-ex]} \BibitemShut
  {NoStop}%
\bibitem [{\citenamefont {Feldman}\ and\ \citenamefont
  {Cousins}(1998)}]{Feldman:1997qc}%
  \BibitemOpen
  \bibfield  {author} {\bibinfo {author} {\bibfnamefont {G.~J.}\ \bibnamefont
  {Feldman}}\ and\ \bibinfo {author} {\bibfnamefont {R.~D.}\ \bibnamefont
  {Cousins}},\ }\href {https://doi.org/10.1103/PhysRevD.57.3873} {\bibfield
  {journal} {\bibinfo  {journal} {Phys. Rev. D}\ }\textbf {\bibinfo {volume}
  {57}},\ \bibinfo {pages} {3873} (\bibinfo {year} {1998})},\ \Eprint
  {https://arxiv.org/abs/physics/9711021} {arXiv:physics/9711021} \BibitemShut
  {NoStop}%
\bibitem [{\citenamefont {Adamson}\ \emph
  {et~al.}(2016{\natexlab{c}})\citenamefont {Adamson} \emph
  {et~al.}}]{NuMI:Adamson2016}%
  \BibitemOpen
  \bibfield  {author} {\bibinfo {author} {\bibfnamefont {P.}~\bibnamefont
  {Adamson}} \emph {et~al.},\ }\href
  {https://doi.org/https://doi.org/10.1016/j.nima.2015.08.063} {\bibfield
  {journal} {\bibinfo  {journal} {Nucl. Instrum. Methods Phys Res., Sect. A}\
  }\textbf {\bibinfo {volume} {806}},\ \bibinfo {pages} {279} (\bibinfo {year}
  {2016}{\natexlab{c}})},\ \Eprint {https://arxiv.org/abs/1507.06690}
  {arXiv:1507.06690 [hep-ex]} \BibitemShut {NoStop}%
\bibitem [{\citenamefont {Agostinelli}\ \emph {et~al.}(2003)\citenamefont
  {Agostinelli} \emph {et~al.}}]{Agostinelli:2002hh}%
  \BibitemOpen
  \bibfield  {author} {\bibinfo {author} {\bibfnamefont {S.}~\bibnamefont
  {Agostinelli}} \emph {et~al.} (\bibinfo {collaboration} {GEANT4}),\ }\href
  {https://doi.org/10.1016/S0168-9002(03)01368-8} {\bibfield  {journal}
  {\bibinfo  {journal} {Nucl. Instrum. Meth. A}\ }\textbf {\bibinfo {volume}
  {506}},\ \bibinfo {pages} {250} (\bibinfo {year} {2003})}\BibitemShut
  {NoStop}%
%%CITATION = NUIMA,A506,250;%%
\bibitem [{\citenamefont {{Geant4 Collaboration}}(2017)}]{Geant:2017ats}%
  \BibitemOpen
  \bibfield  {author} {\bibinfo {author} {\bibnamefont {{Geant4
  Collaboration}}},\ }\href
  {https://geant4-data.web.cern.ch/ReleaseNotes/ReleaseNotes4.10.4.html}
  {\bibfield  {journal} {\bibinfo  {journal} {geant4-data.web.cern.ch,
  https://geant4-data.web.cern.ch/ ReleaseNotes/ReleaseNotes4.10.4.html}\ }
  (\bibinfo {year} {2017})}\BibitemShut {NoStop}%
\bibitem [{\citenamefont {Aliaga}\ \emph {et~al.}(2016)\citenamefont {Aliaga}
  \emph {et~al.}}]{Aliaga:2016oaz}%
  \BibitemOpen
  \bibfield  {author} {\bibinfo {author} {\bibfnamefont {L.}~\bibnamefont
  {Aliaga}} \emph {et~al.} (\bibinfo {collaboration} {MINERvA}),\ }\href
  {https://doi.org/10.1103/PhysRevD.94.092005, 10.1103/PhysRevD.95.039903}
  {\bibfield  {journal} {\bibinfo  {journal} {Phys. Rev.}\ }\textbf {\bibinfo
  {volume} {D94}},\ \bibinfo {pages} {092005} (\bibinfo {year} {2016})},\
  \bibinfo {note} {[Addendum: Phys. Rev. {\bf D95}, no.3, 039903 (2017)]},\
  \Eprint {https://arxiv.org/abs/1607.00704} {arXiv:1607.00704 [hep-ex]}
  \BibitemShut {NoStop}%
%%CITATION = ARXIV:1607.00704;%%
\bibitem [{\citenamefont {Paley}\ \emph {et~al.}(2014)\citenamefont {Paley}
  \emph {et~al.}}]{Paley:2014rpb}%
  \BibitemOpen
  \bibfield  {author} {\bibinfo {author} {\bibfnamefont {J.~M.}\ \bibnamefont
  {Paley}} \emph {et~al.} (\bibinfo {collaboration} {MIPP}),\ }\href
  {https://doi.org/10.1103/PhysRevD.90.032001} {\bibfield  {journal} {\bibinfo
  {journal} {Phys. Rev.}\ }\textbf {\bibinfo {volume} {D90}},\ \bibinfo {pages}
  {032001} (\bibinfo {year} {2014})},\ \Eprint
  {https://arxiv.org/abs/1404.5882} {arXiv:1404.5882 [hep-ex]} \BibitemShut
  {NoStop}%
%%CITATION = ARXIV:1404.5882;%%
\bibitem [{\citenamefont {Alt}\ \emph {et~al.}(2007)\citenamefont {Alt} \emph
  {et~al.}}]{Alt:2006fr}%
  \BibitemOpen
  \bibfield  {author} {\bibinfo {author} {\bibfnamefont {C.}~\bibnamefont
  {Alt}} \emph {et~al.} (\bibinfo {collaboration} {NA49}),\ }\href
  {https://doi.org/10.1140/epjc/s10052-006-0165-7} {\bibfield  {journal}
  {\bibinfo  {journal} {Eur. Phys. J.}\ }\textbf {\bibinfo {volume} {C49}},\
  \bibinfo {pages} {897} (\bibinfo {year} {2007})},\ \Eprint
  {https://arxiv.org/abs/hep-ex/0606028} {arXiv:hep-ex/0606028 [hep-ex]}
  \BibitemShut {NoStop}%
%%CITATION = HEP-EX/0606028;%%
\bibitem [{\citenamefont {Abgrall}\ \emph {et~al.}(2011)\citenamefont {Abgrall}
  \emph {et~al.}}]{Abgrall:2011ae}%
  \BibitemOpen
  \bibfield  {author} {\bibinfo {author} {\bibfnamefont {N.}~\bibnamefont
  {Abgrall}} \emph {et~al.} (\bibinfo {collaboration} {NA61/SHINE}),\ }\href
  {https://doi.org/10.1103/PhysRevC.84.034604} {\bibfield  {journal} {\bibinfo
  {journal} {Phys. Rev.}\ }\textbf {\bibinfo {volume} {C84}},\ \bibinfo {pages}
  {034604} (\bibinfo {year} {2011})},\ \Eprint
  {https://arxiv.org/abs/1102.0983} {arXiv:1102.0983 [hep-ex]} \BibitemShut
  {NoStop}%
%%CITATION = ARXIV:1102.0983;%%
\bibitem [{\citenamefont {Barton}\ \emph {et~al.}(1983)\citenamefont {Barton}
  \emph {et~al.}}]{Barton:1982dg}%
  \BibitemOpen
  \bibfield  {author} {\bibinfo {author} {\bibfnamefont {D.~S.}\ \bibnamefont
  {Barton}} \emph {et~al.},\ }\href {https://doi.org/10.1103/PhysRevD.27.2580}
  {\bibfield  {journal} {\bibinfo  {journal} {Phys. Rev.}\ }\textbf {\bibinfo
  {volume} {D27}},\ \bibinfo {pages} {2580} (\bibinfo {year}
  {1983})}\BibitemShut {NoStop}%
%%CITATION = PHRVA,D27,2580;%%
\bibitem [{\citenamefont {Seun}(2007)}]{Seun:2007zz}%
  \BibitemOpen
  \bibfield  {author} {\bibinfo {author} {\bibfnamefont {S.~M.}\ \bibnamefont
  {Seun}},\ }\emph {\bibinfo {title} {{Measurement of $\pi-K$ ratios from the
  NuMI target}}},\ \href {https://doi.org/10.2172/935004} {Ph.D. thesis},\
  \bibinfo  {school} {Harvard U.} (\bibinfo {year} {2007})\BibitemShut
  {NoStop}%
%%CITATION = FERMILAB-THESIS-2007-61;%%
\bibitem [{\citenamefont {Tinti}(2010)}]{Tinti:2010zz}%
  \BibitemOpen
  \bibfield  {author} {\bibinfo {author} {\bibfnamefont {G.~M.}\ \bibnamefont
  {Tinti}},\ }\emph {\bibinfo {title} {{Sterile neutrino oscillations in MINOS
  and hadron production in pC collisions}}},\ \href
  {https://doi.org/10.2172/992263} {Ph.D. thesis},\ \bibinfo  {school} {Oxford
  U.} (\bibinfo {year} {2010})\BibitemShut {NoStop}%
%%CITATION = FERMILAB-THESIS-2010-44;%%
\bibitem [{\citenamefont {Lebedev}(2007)}]{Lebedev:2007zz}%
  \BibitemOpen
  \bibfield  {author} {\bibinfo {author} {\bibfnamefont {A.~V.}\ \bibnamefont
  {Lebedev}},\ }\emph {\bibinfo {title} {{Ratio of pion kaon production in
  proton carbon interactions}}},\ \href {https://doi.org/10.2172/948174} {Ph.D.
  thesis},\ \bibinfo  {school} {Harvard U.} (\bibinfo {year}
  {2007})\BibitemShut {NoStop}%
%%CITATION = FERMILAB-THESIS-2007-76;%%
\bibitem [{\citenamefont {Baatar}\ \emph {et~al.}(2013)\citenamefont {Baatar}
  \emph {et~al.}}]{Baatar:2012fua}%
  \BibitemOpen
  \bibfield  {author} {\bibinfo {author} {\bibfnamefont {B.}~\bibnamefont
  {Baatar}} \emph {et~al.} (\bibinfo {collaboration} {NA49}),\ }\href
  {https://doi.org/10.1140/epjc/s10052-013-2364-3} {\bibfield  {journal}
  {\bibinfo  {journal} {Eur. Phys. J.}\ }\textbf {\bibinfo {volume} {C73}},\
  \bibinfo {pages} {2364} (\bibinfo {year} {2013})},\ \Eprint
  {https://arxiv.org/abs/1207.6520} {arXiv:1207.6520 [hep-ex]} \BibitemShut
  {NoStop}%
%%CITATION = ARXIV:1207.6520;%%
\bibitem [{\citenamefont {Skubic}\ \emph {et~al.}(1978)\citenamefont {Skubic}
  \emph {et~al.}}]{Skubic:1978fi}%
  \BibitemOpen
  \bibfield  {author} {\bibinfo {author} {\bibfnamefont {P.}~\bibnamefont
  {Skubic}} \emph {et~al.},\ }\href {https://doi.org/10.1103/PhysRevD.18.3115}
  {\bibfield  {journal} {\bibinfo  {journal} {Phys. Rev.}\ }\textbf {\bibinfo
  {volume} {D18}},\ \bibinfo {pages} {3115} (\bibinfo {year}
  {1978})}\BibitemShut {NoStop}%
%%CITATION = PHRVA,D18,3115;%%
\bibitem [{\citenamefont {Denisov}\ \emph {et~al.}(1973)\citenamefont
  {Denisov}, \citenamefont {Donskov}, \citenamefont {Gorin}, \citenamefont
  {Krasnokutsky}, \citenamefont {Petrukhin}, \citenamefont {Prokoshkin},\ and\
  \citenamefont {Stoyanova}}]{Denisov:1973zv}%
  \BibitemOpen
  \bibfield  {author} {\bibinfo {author} {\bibfnamefont {S.~P.}\ \bibnamefont
  {Denisov}}, \bibinfo {author} {\bibfnamefont {S.~V.}\ \bibnamefont
  {Donskov}}, \bibinfo {author} {\bibfnamefont {{\relax Yu}.~P.}\ \bibnamefont
  {Gorin}}, \bibinfo {author} {\bibfnamefont {R.~N.}\ \bibnamefont
  {Krasnokutsky}}, \bibinfo {author} {\bibfnamefont {A.~I.}\ \bibnamefont
  {Petrukhin}}, \bibinfo {author} {\bibfnamefont {{\relax Yu}.~D.}\
  \bibnamefont {Prokoshkin}},\ and\ \bibinfo {author} {\bibfnamefont {D.~A.}\
  \bibnamefont {Stoyanova}},\ }\href
  {https://doi.org/10.1016/0550-3213(73)90351-9} {\bibfield  {journal}
  {\bibinfo  {journal} {Nucl. Phys.}\ }\textbf {\bibinfo {volume} {B61}},\
  \bibinfo {pages} {62} (\bibinfo {year} {1973})}\BibitemShut {NoStop}%
%%CITATION = NUPHA,B61,62;%%
\bibitem [{\citenamefont {Carroll}\ \emph {et~al.}(1979)\citenamefont {Carroll}
  \emph {et~al.}}]{Carroll:1978hc}%
  \BibitemOpen
  \bibfield  {author} {\bibinfo {author} {\bibfnamefont {A.~S.}\ \bibnamefont
  {Carroll}} \emph {et~al.},\ }\href
  {https://doi.org/10.1016/0370-2693(79)90226-0} {\bibfield  {journal}
  {\bibinfo  {journal} {Phys. Lett.}\ }\textbf {\bibinfo {volume} {80B}},\
  \bibinfo {pages} {319} (\bibinfo {year} {1979})}\BibitemShut {NoStop}%
%%CITATION = PHLTA,80B,319;%%
\bibitem [{\citenamefont {Abe}\ \emph {et~al.}(2013)\citenamefont {Abe} \emph
  {et~al.}}]{Abe:2012av}%
  \BibitemOpen
  \bibfield  {author} {\bibinfo {author} {\bibfnamefont {K.}~\bibnamefont
  {Abe}} \emph {et~al.} (\bibinfo {collaboration} {T2K}),\ }\href
  {https://doi.org/10.1103/PhysRevD.87.012001, 10.1103/PhysRevD.87.019902}
  {\bibfield  {journal} {\bibinfo  {journal} {Phys. Rev.}\ }\textbf {\bibinfo
  {volume} {D87}},\ \bibinfo {pages} {012001} (\bibinfo {year} {2013})},\
  \bibinfo {note} {[Addendum: Phys. Rev. {\bf D87}, no.1, 019902 (2013)]},\
  \Eprint {https://arxiv.org/abs/1211.0469} {arXiv:1211.0469 [hep-ex]}
  \BibitemShut {NoStop}%
%%CITATION = ARXIV:1211.0469;%%
\bibitem [{\citenamefont {Gaisser}\ \emph {et~al.}(1975)\citenamefont
  {Gaisser}, \citenamefont {Yodh}, \citenamefont {Barger},\ and\ \citenamefont
  {Halzen}}]{Gaisser:1975et}%
  \BibitemOpen
  \bibfield  {author} {\bibinfo {author} {\bibfnamefont {T.~K.}\ \bibnamefont
  {Gaisser}}, \bibinfo {author} {\bibfnamefont {G.~B.}\ \bibnamefont {Yodh}},
  \bibinfo {author} {\bibfnamefont {V.~D.}\ \bibnamefont {Barger}},\ and\
  \bibinfo {author} {\bibfnamefont {F.}~\bibnamefont {Halzen}},\ }in\
  \href@noop {} {\emph {\bibinfo {booktitle} {{14th International Cosmic Ray
  Conference (ICRC 1975) Munich, Germany, August 15-29, 1975}}}}\ (\bibinfo
  {year} {1975})\ pp.\ \bibinfo {pages} {2161--2166}\BibitemShut {NoStop}%
%%CITATION = INSPIRE-105858;%%
\bibitem [{\citenamefont {Cronin}\ \emph {et~al.}(1957)\citenamefont {Cronin},
  \citenamefont {Cool},\ and\ \citenamefont {Abashian}}]{Cronin:1957zz}%
  \BibitemOpen
  \bibfield  {author} {\bibinfo {author} {\bibfnamefont {J.~W.}\ \bibnamefont
  {Cronin}}, \bibinfo {author} {\bibfnamefont {R.}~\bibnamefont {Cool}},\ and\
  \bibinfo {author} {\bibfnamefont {A.}~\bibnamefont {Abashian}},\ }\href
  {https://doi.org/10.1103/PhysRev.107.1121} {\bibfield  {journal} {\bibinfo
  {journal} {Phys. Rev.}\ }\textbf {\bibinfo {volume} {107}},\ \bibinfo {pages}
  {1121} (\bibinfo {year} {1957})}\BibitemShut {NoStop}%
%%CITATION = PHRVA,107,1121;%%
\bibitem [{\citenamefont {Allaby}\ \emph {et~al.}(1969)\citenamefont {Allaby}
  \emph {et~al.}}]{Allaby:1969de}%
  \BibitemOpen
  \bibfield  {author} {\bibinfo {author} {\bibfnamefont {J.~V.}\ \bibnamefont
  {Allaby}} \emph {et~al.} (\bibinfo {collaboration} {IHEP-CERN}),\ }\href
  {https://doi.org/10.1016/0370-2693(69)90184-1} {\bibfield  {journal}
  {\bibinfo  {journal} {Phys. Lett.}\ }\textbf {\bibinfo {volume} {30B}},\
  \bibinfo {pages} {500} (\bibinfo {year} {1969})}\BibitemShut {NoStop}%
%%CITATION = PHLTA,30B,500;%%
\bibitem [{\citenamefont {Longo}\ and\ \citenamefont
  {Moyer}(1962)}]{Longo:1962zz}%
  \BibitemOpen
  \bibfield  {author} {\bibinfo {author} {\bibfnamefont {M.~J.}\ \bibnamefont
  {Longo}}\ and\ \bibinfo {author} {\bibfnamefont {B.~J.}\ \bibnamefont
  {Moyer}},\ }\href {https://doi.org/10.1103/PhysRev.125.701} {\bibfield
  {journal} {\bibinfo  {journal} {Phys. Rev.}\ }\textbf {\bibinfo {volume}
  {125}},\ \bibinfo {pages} {701} (\bibinfo {year} {1962})}\BibitemShut
  {NoStop}%
%%CITATION = PHRVA,125,701;%%
\bibitem [{\citenamefont {Bobchenko}\ \emph {et~al.}(1979)\citenamefont
  {Bobchenko} \emph {et~al.}}]{Bobchenko:1979hp}%
  \BibitemOpen
  \bibfield  {author} {\bibinfo {author} {\bibfnamefont {B.~M.}\ \bibnamefont
  {Bobchenko}} \emph {et~al.},\ }\href@noop {} {\bibfield  {journal} {\bibinfo
  {journal} {Sov. J. Nucl. Phys.}\ }\textbf {\bibinfo {volume} {30}},\ \bibinfo
  {pages} {805} (\bibinfo {year} {1979})},\ \bibinfo {note} {[Yad. Fiz. 30,
  1553 (1979)]}\BibitemShut {NoStop}%
%%CITATION = SJNCA,30,805;%%
\bibitem [{\citenamefont {Fedorov}\ \emph {et~al.}(1978)\citenamefont
  {Fedorov}, \citenamefont {Grishuk}, \citenamefont {Kosov}, \citenamefont
  {Leksin}, \citenamefont {Pivnyuk}, \citenamefont {Shevchenko}, \citenamefont
  {Stolin}, \citenamefont {Vlasov},\ and\ \citenamefont
  {Vorobev}}]{Fedorov:1977an}%
  \BibitemOpen
  \bibfield  {author} {\bibinfo {author} {\bibfnamefont {V.~B.}\ \bibnamefont
  {Fedorov}}, \bibinfo {author} {\bibfnamefont {{\relax Yu}.~G.}\ \bibnamefont
  {Grishuk}}, \bibinfo {author} {\bibfnamefont {M.~V.}\ \bibnamefont {Kosov}},
  \bibinfo {author} {\bibfnamefont {G.~A.}\ \bibnamefont {Leksin}}, \bibinfo
  {author} {\bibfnamefont {N.~A.}\ \bibnamefont {Pivnyuk}}, \bibinfo {author}
  {\bibfnamefont {S.~V.}\ \bibnamefont {Shevchenko}}, \bibinfo {author}
  {\bibfnamefont {V.~L.}\ \bibnamefont {Stolin}}, \bibinfo {author}
  {\bibfnamefont {A.~V.}\ \bibnamefont {Vlasov}},\ and\ \bibinfo {author}
  {\bibfnamefont {L.~S.}\ \bibnamefont {Vorobev}},\ }\href@noop {} {\bibfield
  {journal} {\bibinfo  {journal} {Sov. J. Nucl. Phys.}\ }\textbf {\bibinfo
  {volume} {27}},\ \bibinfo {pages} {222} (\bibinfo {year} {1978})},\ \bibinfo
  {note} {[Yad. Fiz.27,413(1978)]}\BibitemShut {NoStop}%
%%CITATION = SJNCA,27,222;%%
\bibitem [{\citenamefont {Abrams}\ \emph {et~al.}(1970)\citenamefont {Abrams},
  \citenamefont {Cool}, \citenamefont {Giacomelli}, \citenamefont {Kycia},
  \citenamefont {Leontic}, \citenamefont {Li},\ and\ \citenamefont
  {Michael}}]{Abrams:1969jm}%
  \BibitemOpen
  \bibfield  {author} {\bibinfo {author} {\bibfnamefont {R.~J.}\ \bibnamefont
  {Abrams}}, \bibinfo {author} {\bibfnamefont {R.~L.}\ \bibnamefont {Cool}},
  \bibinfo {author} {\bibfnamefont {G.}~\bibnamefont {Giacomelli}}, \bibinfo
  {author} {\bibfnamefont {T.~F.}\ \bibnamefont {Kycia}}, \bibinfo {author}
  {\bibfnamefont {B.~A.}\ \bibnamefont {Leontic}}, \bibinfo {author}
  {\bibfnamefont {K.~K.}\ \bibnamefont {Li}},\ and\ \bibinfo {author}
  {\bibfnamefont {D.~N.}\ \bibnamefont {Michael}},\ }\href
  {https://doi.org/10.1103/PhysRevD.1.1917} {\bibfield  {journal} {\bibinfo
  {journal} {Phys. Rev.}\ }\textbf {\bibinfo {volume} {D1}},\ \bibinfo {pages}
  {1917} (\bibinfo {year} {1970})}\BibitemShut {NoStop}%
%%CITATION = PHRVA,D1,1917;%%
\bibitem [{\citenamefont {Andreopoulos}\ \emph {et~al.}(2010)\citenamefont
  {Andreopoulos} \emph {et~al.}}]{Andreopoulos:2009rq}%
  \BibitemOpen
  \bibfield  {author} {\bibinfo {author} {\bibfnamefont {C.}~\bibnamefont
  {Andreopoulos}} \emph {et~al.},\ }\href
  {https://doi.org/10.1016/j.nima.2009.12.009} {\bibfield  {journal} {\bibinfo
  {journal} {Nucl. Instrum. Meth. A}\ }\textbf {\bibinfo {volume} {614}},\
  \bibinfo {pages} {87} (\bibinfo {year} {2010})},\ \Eprint
  {https://arxiv.org/abs/0905.2517} {arXiv:0905.2517 [hep-ph]} \BibitemShut
  {NoStop}%
\bibitem [{\citenamefont {Andreopoulos}\ \emph {et~al.}(2015)\citenamefont
  {Andreopoulos}, \citenamefont {Barry}, \citenamefont {Dytman}, \citenamefont
  {Gallagher}, \citenamefont {Golan}, \citenamefont {Hatcher}, \citenamefont
  {Perdue},\ and\ \citenamefont {Yarba}}]{Andreopoulos:2015wxa}%
  \BibitemOpen
  \bibfield  {author} {\bibinfo {author} {\bibfnamefont {C.}~\bibnamefont
  {Andreopoulos}}, \bibinfo {author} {\bibfnamefont {C.}~\bibnamefont {Barry}},
  \bibinfo {author} {\bibfnamefont {S.}~\bibnamefont {Dytman}}, \bibinfo
  {author} {\bibfnamefont {H.}~\bibnamefont {Gallagher}}, \bibinfo {author}
  {\bibfnamefont {T.}~\bibnamefont {Golan}}, \bibinfo {author} {\bibfnamefont
  {R.}~\bibnamefont {Hatcher}}, \bibinfo {author} {\bibfnamefont
  {G.}~\bibnamefont {Perdue}},\ and\ \bibinfo {author} {\bibfnamefont
  {J.}~\bibnamefont {Yarba}},\ }\href@noop {} {\bibinfo {title} {{The GENIE
  Neutrino Monte Carlo Generator: Physics and User Manual}}} (\bibinfo {year}
  {2015}),\ \Eprint {https://arxiv.org/abs/1510.05494} {arXiv:1510.05494
  [hep-ph]} \BibitemShut {NoStop}%
\bibitem [{\citenamefont {Nieves}\ \emph {et~al.}(2011)\citenamefont {Nieves},
  \citenamefont {Ruiz~Simo},\ and\ \citenamefont
  {Vicente~Vacas}}]{Nieves:2011pp}%
  \BibitemOpen
  \bibfield  {author} {\bibinfo {author} {\bibfnamefont {J.}~\bibnamefont
  {Nieves}}, \bibinfo {author} {\bibfnamefont {I.}~\bibnamefont {Ruiz~Simo}},\
  and\ \bibinfo {author} {\bibfnamefont {M.~J.}\ \bibnamefont
  {Vicente~Vacas}},\ }\href {https://doi.org/10.1103/PhysRevC.83.045501}
  {\bibfield  {journal} {\bibinfo  {journal} {Phys. Rev. C}\ }\textbf {\bibinfo
  {volume} {83}},\ \bibinfo {pages} {045501} (\bibinfo {year} {2011})},\
  \Eprint {https://arxiv.org/abs/1102.2777} {arXiv:1102.2777 [hep-ph]}
  \BibitemShut {NoStop}%
\bibitem [{\citenamefont {Gran}\ \emph {et~al.}(2013)\citenamefont {Gran},
  \citenamefont {Nieves}, \citenamefont {Sanchez},\ and\ \citenamefont
  {Vicente~Vacas}}]{Gran:2013kda}%
  \BibitemOpen
  \bibfield  {author} {\bibinfo {author} {\bibfnamefont {R.}~\bibnamefont
  {Gran}}, \bibinfo {author} {\bibfnamefont {J.}~\bibnamefont {Nieves}},
  \bibinfo {author} {\bibfnamefont {F.}~\bibnamefont {Sanchez}},\ and\ \bibinfo
  {author} {\bibfnamefont {M.~J.}\ \bibnamefont {Vicente~Vacas}},\ }\href
  {https://doi.org/10.1103/PhysRevD.88.113007} {\bibfield  {journal} {\bibinfo
  {journal} {Phys. Rev. D}\ }\textbf {\bibinfo {volume} {88}},\ \bibinfo
  {pages} {113007} (\bibinfo {year} {2013})},\ \Eprint
  {https://arxiv.org/abs/1307.8105} {arXiv:1307.8105 [hep-ph]} \BibitemShut
  {NoStop}%
%%CITATION = ARXIV:1307.8105;%%
\bibitem [{\citenamefont {Allardyce}\ \emph {et~al.}(1973)\citenamefont
  {Allardyce} \emph {et~al.}}]{Allardyce:1973ce}%
  \BibitemOpen
  \bibfield  {author} {\bibinfo {author} {\bibfnamefont {B.~W.}\ \bibnamefont
  {Allardyce}} \emph {et~al.},\ }\href
  {https://doi.org/10.1016/0375-9474(73)90049-3} {\bibfield  {journal}
  {\bibinfo  {journal} {Nucl. Phys. A}\ }\textbf {\bibinfo {volume} {209}},\
  \bibinfo {pages} {1} (\bibinfo {year} {1973})}\BibitemShut {NoStop}%
%%CITATION = NUPHA,A209,1;%%
\bibitem [{\citenamefont {Saunders}\ \emph {et~al.}(1996)\citenamefont
  {Saunders}, \citenamefont {Hoeibraten}, \citenamefont {Kraushaar},
  \citenamefont {Kriss}, \citenamefont {Peterson}, \citenamefont {Ristinen},
  \citenamefont {Brack}, \citenamefont {Hofman}, \citenamefont {Gibson},\ and\
  \citenamefont {Morris}}]{Saunders:1996ic}%
  \BibitemOpen
  \bibfield  {author} {\bibinfo {author} {\bibfnamefont {A.}~\bibnamefont
  {Saunders}}, \bibinfo {author} {\bibfnamefont {S.}~\bibnamefont
  {Hoeibraten}}, \bibinfo {author} {\bibfnamefont {J.~J.}\ \bibnamefont
  {Kraushaar}}, \bibinfo {author} {\bibfnamefont {B.~J.}\ \bibnamefont
  {Kriss}}, \bibinfo {author} {\bibfnamefont {R.~J.}\ \bibnamefont {Peterson}},
  \bibinfo {author} {\bibfnamefont {R.~A.}\ \bibnamefont {Ristinen}}, \bibinfo
  {author} {\bibfnamefont {J.~T.}\ \bibnamefont {Brack}}, \bibinfo {author}
  {\bibfnamefont {G.}~\bibnamefont {Hofman}}, \bibinfo {author} {\bibfnamefont
  {E.~F.}\ \bibnamefont {Gibson}},\ and\ \bibinfo {author} {\bibfnamefont
  {C.~L.}\ \bibnamefont {Morris}},\ }\href
  {https://doi.org/10.1103/PhysRevC.53.1745} {\bibfield  {journal} {\bibinfo
  {journal} {Phys. Rev. C}\ }\textbf {\bibinfo {volume} {53}},\ \bibinfo
  {pages} {1745} (\bibinfo {year} {1996})}\BibitemShut {NoStop}%
%%CITATION = PHRVA,C53,1745;%%
\bibitem [{\citenamefont {Meirav}\ \emph {et~al.}(1989)\citenamefont {Meirav},
  \citenamefont {Friedman}, \citenamefont {Johnson}, \citenamefont
  {Olszewski},\ and\ \citenamefont {Weber}}]{Meirav:1988pn}%
  \BibitemOpen
  \bibfield  {author} {\bibinfo {author} {\bibfnamefont {O.}~\bibnamefont
  {Meirav}}, \bibinfo {author} {\bibfnamefont {E.}~\bibnamefont {Friedman}},
  \bibinfo {author} {\bibfnamefont {R.~R.}\ \bibnamefont {Johnson}}, \bibinfo
  {author} {\bibfnamefont {R.}~\bibnamefont {Olszewski}},\ and\ \bibinfo
  {author} {\bibfnamefont {P.}~\bibnamefont {Weber}},\ }\href
  {https://doi.org/10.1103/PhysRevC.40.843} {\bibfield  {journal} {\bibinfo
  {journal} {Phys. Rev. C}\ }\textbf {\bibinfo {volume} {40}},\ \bibinfo
  {pages} {843} (\bibinfo {year} {1989})}\BibitemShut {NoStop}%
%%CITATION = PHRVA,C40,843;%%
\bibitem [{\citenamefont {Levenson}\ \emph {et~al.}(1983)\citenamefont
  {Levenson} \emph {et~al.}}]{Levenson:1983xu}%
  \BibitemOpen
  \bibfield  {author} {\bibinfo {author} {\bibfnamefont {S.~M.}\ \bibnamefont
  {Levenson}} \emph {et~al.},\ }\href {https://doi.org/10.1103/PhysRevC.28.326}
  {\bibfield  {journal} {\bibinfo  {journal} {Phys. Rev. C}\ }\textbf {\bibinfo
  {volume} {28}},\ \bibinfo {pages} {326} (\bibinfo {year} {1983})}\BibitemShut
  {NoStop}%
%%CITATION = PHRVA,C28,326;%%
\bibitem [{\citenamefont {Ashery}\ \emph {et~al.}(1981)\citenamefont {Ashery},
  \citenamefont {Navon}, \citenamefont {Azuelos}, \citenamefont {Walter},
  \citenamefont {Pfeiffer},\ and\ \citenamefont {Schleputz}}]{Ashery:1981tq}%
  \BibitemOpen
  \bibfield  {author} {\bibinfo {author} {\bibfnamefont {D.}~\bibnamefont
  {Ashery}}, \bibinfo {author} {\bibfnamefont {I.}~\bibnamefont {Navon}},
  \bibinfo {author} {\bibfnamefont {G.}~\bibnamefont {Azuelos}}, \bibinfo
  {author} {\bibfnamefont {H.~K.}\ \bibnamefont {Walter}}, \bibinfo {author}
  {\bibfnamefont {H.~J.}\ \bibnamefont {Pfeiffer}},\ and\ \bibinfo {author}
  {\bibfnamefont {F.~W.}\ \bibnamefont {Schleputz}},\ }\href
  {https://doi.org/10.1103/PhysRevC.23.2173} {\bibfield  {journal} {\bibinfo
  {journal} {Phys. Rev. C}\ }\textbf {\bibinfo {volume} {23}},\ \bibinfo
  {pages} {2173} (\bibinfo {year} {1981})}\BibitemShut {NoStop}%
%%CITATION = PHRVA,C23,2173;%%
\bibitem [{\citenamefont {Ashery}\ \emph {et~al.}(1984)\citenamefont {Ashery}
  \emph {et~al.}}]{Ashery:1984ne}%
  \BibitemOpen
  \bibfield  {author} {\bibinfo {author} {\bibfnamefont {D.}~\bibnamefont
  {Ashery}} \emph {et~al.},\ }\href {https://doi.org/10.1103/PhysRevC.30.946}
  {\bibfield  {journal} {\bibinfo  {journal} {Phys. Rev. C}\ }\textbf {\bibinfo
  {volume} {30}},\ \bibinfo {pages} {946} (\bibinfo {year} {1984})}\BibitemShut
  {NoStop}%
%%CITATION = PHRVA,C30,946;%%
\bibitem [{\citenamefont {Pinzon~Guerra}\ \emph {et~al.}(2017)\citenamefont
  {Pinzon~Guerra} \emph {et~al.}}]{PinzonGuerra:2016uae}%
  \BibitemOpen
  \bibfield  {author} {\bibinfo {author} {\bibfnamefont {E.~S.}\ \bibnamefont
  {Pinzon~Guerra}} \emph {et~al.} (\bibinfo {collaboration} {DUET}),\ }\href
  {https://doi.org/10.1103/PhysRevC.95.045203} {\bibfield  {journal} {\bibinfo
  {journal} {Phys. Rev. C}\ }\textbf {\bibinfo {volume} {95}},\ \bibinfo
  {pages} {045203} (\bibinfo {year} {2017})},\ \Eprint
  {https://arxiv.org/abs/1611.05612} {arXiv:1611.05612 [hep-ex]} \BibitemShut
  {NoStop}%
%%CITATION = ARXIV:1611.05612;%%
\bibitem [{\citenamefont {Aurisano}\ \emph {et~al.}(2015)\citenamefont
  {Aurisano}, \citenamefont {Backhouse}, \citenamefont {Hatcher}, \citenamefont
  {Mayer}, \citenamefont {Musser}, \citenamefont {Patterson}, \citenamefont
  {Schroeter},\ and\ \citenamefont {Sousa}}]{Aurisano:2015oxj}%
  \BibitemOpen
  \bibfield  {author} {\bibinfo {author} {\bibfnamefont {A.}~\bibnamefont
  {Aurisano}}, \bibinfo {author} {\bibfnamefont {C.}~\bibnamefont {Backhouse}},
  \bibinfo {author} {\bibfnamefont {R.}~\bibnamefont {Hatcher}}, \bibinfo
  {author} {\bibfnamefont {N.}~\bibnamefont {Mayer}}, \bibinfo {author}
  {\bibfnamefont {J.}~\bibnamefont {Musser}}, \bibinfo {author} {\bibfnamefont
  {R.}~\bibnamefont {Patterson}}, \bibinfo {author} {\bibfnamefont
  {R.}~\bibnamefont {Schroeter}},\ and\ \bibinfo {author} {\bibfnamefont
  {A.}~\bibnamefont {Sousa}} (\bibinfo {collaboration} {NOvA}),\ }\bibfield
  {booktitle} {\emph {\bibinfo {booktitle} {{Proceedings, 21st International
  Conference on Computing in High Energy and Nuclear Physics (CHEP 2015):
  Okinawa, Japan, April 13-17, 2015}}},\ }\href
  {https://doi.org/10.1088/1742-6596/664/7/072002} {\bibfield  {journal}
  {\bibinfo  {journal} {J. Phys. Conf. Ser.}\ }\textbf {\bibinfo {volume}
  {664}},\ \bibinfo {pages} {072002} (\bibinfo {year} {2015})}\BibitemShut
  {NoStop}%
%%CITATION = 00462,664,072002;%%
\bibitem [{\citenamefont {Aurisano}\ \emph {et~al.}(2016)\citenamefont
  {Aurisano}, \citenamefont {Radovic}, \citenamefont {Rocco}, \citenamefont
  {Himmel}, \citenamefont {Messier}, \citenamefont {Niner}, \citenamefont
  {Pawloski}, \citenamefont {Psihas}, \citenamefont {Sousa},\ and\
  \citenamefont {Vahle}}]{Aurisano:2016jvx}%
  \BibitemOpen
  \bibfield  {author} {\bibinfo {author} {\bibfnamefont {A.}~\bibnamefont
  {Aurisano}}, \bibinfo {author} {\bibfnamefont {A.}~\bibnamefont {Radovic}},
  \bibinfo {author} {\bibfnamefont {D.}~\bibnamefont {Rocco}}, \bibinfo
  {author} {\bibfnamefont {A.}~\bibnamefont {Himmel}}, \bibinfo {author}
  {\bibfnamefont {M.~D.}\ \bibnamefont {Messier}}, \bibinfo {author}
  {\bibfnamefont {E.}~\bibnamefont {Niner}}, \bibinfo {author} {\bibfnamefont
  {G.}~\bibnamefont {Pawloski}}, \bibinfo {author} {\bibfnamefont
  {F.}~\bibnamefont {Psihas}}, \bibinfo {author} {\bibfnamefont
  {A.}~\bibnamefont {Sousa}},\ and\ \bibinfo {author} {\bibfnamefont
  {P.}~\bibnamefont {Vahle}},\ }\href
  {https://doi.org/10.1088/1748-0221/11/09/P09001} {\bibfield  {journal}
  {\bibinfo  {journal} {JINST}\ }\textbf {\bibinfo {volume} {11}}\bibfield
  {number} {\bibinfo  {number} { (09)},\ \bibinfo {pages} {P09001}},\ }\Eprint
  {https://arxiv.org/abs/1604.01444} {arXiv:1604.01444 [hep-ex]} \BibitemShut
  {NoStop}%
%%CITATION = ARXIV:1604.01444;%%
\bibitem [{\citenamefont {Psihas}\ \emph {et~al.}(2019)\citenamefont {Psihas},
  \citenamefont {Niner}, \citenamefont {Groh}, \citenamefont {Murphy},
  \citenamefont {Aurisano}, \citenamefont {Himmel}, \citenamefont {Lang},
  \citenamefont {Messier}, \citenamefont {Radovic},\ and\ \citenamefont
  {Sousa}}]{Psihas:2019ksa}%
  \BibitemOpen
  \bibfield  {author} {\bibinfo {author} {\bibfnamefont {F.}~\bibnamefont
  {Psihas}}, \bibinfo {author} {\bibfnamefont {E.}~\bibnamefont {Niner}},
  \bibinfo {author} {\bibfnamefont {M.}~\bibnamefont {Groh}}, \bibinfo {author}
  {\bibfnamefont {R.}~\bibnamefont {Murphy}}, \bibinfo {author} {\bibfnamefont
  {A.}~\bibnamefont {Aurisano}}, \bibinfo {author} {\bibfnamefont
  {A.}~\bibnamefont {Himmel}}, \bibinfo {author} {\bibfnamefont
  {K.}~\bibnamefont {Lang}}, \bibinfo {author} {\bibfnamefont {M.~D.}\
  \bibnamefont {Messier}}, \bibinfo {author} {\bibfnamefont {A.}~\bibnamefont
  {Radovic}},\ and\ \bibinfo {author} {\bibfnamefont {A.}~\bibnamefont
  {Sousa}},\ }\href {https://doi.org/10.1103/PhysRevD.100.073005} {\bibfield
  {journal} {\bibinfo  {journal} {Phys. Rev. D}\ }\textbf {\bibinfo {volume}
  {100}},\ \bibinfo {pages} {073005} (\bibinfo {year} {2019})},\ \Eprint
  {https://arxiv.org/abs/1906.00713} {arXiv:1906.00713 [physics.ins-det]}
  \BibitemShut {NoStop}%
\bibitem [{\citenamefont {{Mr. Bayes}}\ and\ \citenamefont {{Mr.
  Price}}(1763)}]{Bayes:1763}%
  \BibitemOpen
  \bibfield  {author} {\bibinfo {author} {\bibnamefont {{Mr. Bayes}}}\ and\
  \bibinfo {author} {\bibnamefont {{Mr. Price}}},\ }\href
  {http://www.jstor.org/stable/105741} {\bibfield  {journal} {\bibinfo
  {journal} {Philosophical Transactions (1683-1775)}\ }\textbf {\bibinfo
  {volume} {53}},\ \bibinfo {pages} {370} (\bibinfo {year} {1763})}\BibitemShut
  {NoStop}%
\bibitem [{\citenamefont {Speagle}(2020)}]{Speagle:2020}%
  \BibitemOpen
  \bibfield  {author} {\bibinfo {author} {\bibfnamefont {J.~S.}\ \bibnamefont
  {Speagle}},\ }\href@noop {} {\bibinfo {title} {{A Conceptual Introduction to
  Markov Chain Monte Carlo Methods}}} (\bibinfo {year} {2020}),\ \Eprint
  {https://arxiv.org/abs/1909.12313} {arXiv:1909.12313 [stat.OT]} \BibitemShut
  {NoStop}%
\bibitem [{\citenamefont {Brooks}\ \emph {et~al.}(2011)\citenamefont {Brooks},
  \citenamefont {Gelman}, \citenamefont {Jones},\ and\ \citenamefont
  {Meng}}]{Brooks:2011a}%
  \BibitemOpen
  \bibfield  {author} {\bibinfo {author} {\bibfnamefont {S.}~\bibnamefont
  {Brooks}}, \bibinfo {author} {\bibfnamefont {A.}~\bibnamefont {Gelman}},
  \bibinfo {author} {\bibfnamefont {G.}~\bibnamefont {Jones}},\ and\ \bibinfo
  {author} {\bibfnamefont {X.-L.}\ \bibnamefont {Meng}},\ }\href@noop {} {\emph
  {\bibinfo {title} {Handbook of Markov Chain Monte Carlo}}}\ (\bibinfo {year}
  {2011})\ pp.\ \bibinfo {pages} {1--592}\BibitemShut {NoStop}%
\bibitem [{\citenamefont {Metropolis}\ \emph {et~al.}(1953)\citenamefont
  {Metropolis}, \citenamefont {Rosenbluth}, \citenamefont {Rosenbluth},
  \citenamefont {Teller},\ and\ \citenamefont {Teller}}]{Metropolis:1953am}%
  \BibitemOpen
  \bibfield  {author} {\bibinfo {author} {\bibfnamefont {N.}~\bibnamefont
  {Metropolis}}, \bibinfo {author} {\bibfnamefont {A.~W.}\ \bibnamefont
  {Rosenbluth}}, \bibinfo {author} {\bibfnamefont {M.~N.}\ \bibnamefont
  {Rosenbluth}}, \bibinfo {author} {\bibfnamefont {A.~H.}\ \bibnamefont
  {Teller}},\ and\ \bibinfo {author} {\bibfnamefont {E.}~\bibnamefont
  {Teller}},\ }\href {https://doi.org/10.1063/1.1699114} {\bibfield  {journal}
  {\bibinfo  {journal} {J. Chem. Phys.}\ }\textbf {\bibinfo {volume} {21}},\
  \bibinfo {pages} {1087} (\bibinfo {year} {1953})}\BibitemShut {NoStop}%
\bibitem [{\citenamefont {Hastings}(1970)}]{Hastings:1970aa}%
  \BibitemOpen
  \bibfield  {author} {\bibinfo {author} {\bibfnamefont {W.~K.}\ \bibnamefont
  {Hastings}},\ }\href {https://doi.org/10.1093/biomet/57.1.97} {\bibfield
  {journal} {\bibinfo  {journal} {Biometrika}\ }\textbf {\bibinfo {volume}
  {57}},\ \bibinfo {pages} {97} (\bibinfo {year} {1970})}\BibitemShut {NoStop}%
\bibitem [{\citenamefont {Gubernatis}(2005)}]{Gubernatis:2005zz}%
  \BibitemOpen
  \bibfield  {author} {\bibinfo {author} {\bibfnamefont {J.~E.}\ \bibnamefont
  {Gubernatis}},\ }\href {https://doi.org/10.1063/1.1887186} {\bibfield
  {journal} {\bibinfo  {journal} {Phys. Plasmas}\ }\textbf {\bibinfo {volume}
  {12}},\ \bibinfo {pages} {057303} (\bibinfo {year} {2005})}\BibitemShut
  {NoStop}%
\bibitem [{\citenamefont {Rosenbluth}()}]{Rosenbluth:2003}%
  \BibitemOpen
  \bibfield  {author} {\bibinfo {author} {\bibfnamefont {M.}~\bibnamefont
  {Rosenbluth}},\ }\href {https://doi.org/10.1063/1.1632112} {\bibfield
  {journal} {\bibinfo  {journal} {AIP Conference Proceedings}\ }\textbf
  {\bibinfo {volume} {690}},\ \bibinfo {pages} {22}}\BibitemShut {NoStop}%
\bibitem [{\citenamefont {{Stan Development Team}}(2021)}]{Stan-manual}%
  \BibitemOpen
  \bibfield  {author} {\bibinfo {author} {\bibnamefont {{Stan Development
  Team}}},\ }\href@noop {} {\emph {\bibinfo {title} {Stan Modeling Language
  Users Guide and Reference Manual, version 2.26.1}}},\ \bibinfo {organization}
  {{Stan Development Team}} (\bibinfo {year} {2021})\BibitemShut {NoStop}%
\bibitem [{\citenamefont {Betancourt}()}]{Betancourt:2017}%
  \BibitemOpen
  \bibfield  {author} {\bibinfo {author} {\bibfnamefont {M.}~\bibnamefont
  {Betancourt}},\ }\href@noop {} {\ }\Eprint {https://arxiv.org/abs/1701.02434}
  {arXiv:1701.02434 [stat.ME]} \BibitemShut {NoStop}%
\bibitem [{\citenamefont {Tanabashi}\ \emph {et~al.}(date)\citenamefont
  {Tanabashi} \emph {et~al.}}]{ParticleDataGroup:2018ovx}%
  \BibitemOpen
  \bibfield  {author} {\bibinfo {author} {\bibfnamefont {M.}~\bibnamefont
  {Tanabashi}} \emph {et~al.} (\bibinfo {collaboration} {Particle Data
  Group}),\ }\href {https://doi.org/10.1103/PhysRevD.98.030001} {\bibfield
  {journal} {\bibinfo  {journal} {Phys. Rev. D}\ }\textbf {\bibinfo {volume}
  {98}},\ \bibinfo {pages} {030001} (\bibinfo {year} {2018 and 2019
  update})}\BibitemShut {NoStop}%
\bibitem [{\citenamefont {Nayak}(2021)}]{Nayak:2021eal}%
  \BibitemOpen
  \bibfield  {author} {\bibinfo {author} {\bibfnamefont {N.}~\bibnamefont
  {Nayak}},\ }\emph {\bibinfo {title} {{A Joint Measurement of
  $\nu_{\mu}$-Disappearance and $\nu_{e}$-Appearance in the NuMI beam using the
  NOvA Experiment}}},\ \href@noop {} {Ph.D. thesis},\ \bibinfo  {school} {UC,
  Irvine (main), UC, Irvine} (\bibinfo {year} {2021})\BibitemShut {NoStop}%
\bibitem [{\citenamefont {Gelman}\ \emph {et~al.}(1996)\citenamefont {Gelman},
  \citenamefont {Meng},\ and\ \citenamefont {Stern}}]{Gelman:1996}%
  \BibitemOpen
  \bibfield  {author} {\bibinfo {author} {\bibfnamefont {A.}~\bibnamefont
  {Gelman}}, \bibinfo {author} {\bibfnamefont {X.-L.}\ \bibnamefont {Meng}},\
  and\ \bibinfo {author} {\bibfnamefont {H.}~\bibnamefont {Stern}},\ }\href
  {http://www.jstor.org/stable/24306036} {\bibfield  {journal} {\bibinfo
  {journal} {Statistica Sinica}\ }\textbf {\bibinfo {volume} {6}},\ \bibinfo
  {pages} {733} (\bibinfo {year} {1996})}\BibitemShut {NoStop}%
\bibitem [{\citenamefont {Gelman}\ \emph {et~al.}(2002)\citenamefont {Gelman},
  \citenamefont {Goegebeur}, \citenamefont {Tuerlinckx},\ and\ \citenamefont
  {Van~Mechelen}}]{Gelman:2002}%
  \BibitemOpen
  \bibfield  {author} {\bibinfo {author} {\bibfnamefont {A.}~\bibnamefont
  {Gelman}}, \bibinfo {author} {\bibfnamefont {Y.}~\bibnamefont {Goegebeur}},
  \bibinfo {author} {\bibfnamefont {F.}~\bibnamefont {Tuerlinckx}},\ and\
  \bibinfo {author} {\bibfnamefont {I.~v.}\ \bibnamefont {Van~Mechelen}},\
  }\href {https://doi.org/10.1111/1467-9876.00190} {\bibfield  {journal}
  {\bibinfo  {journal} {Journal of the Royal Statistical Society Series C:
  Applied Statistics}\ }\textbf {\bibinfo {volume} {49}},\ \bibinfo {pages}
  {247} (\bibinfo {year} {2002})}\BibitemShut {NoStop}%
\bibitem [{\citenamefont {Gelman}(2013)}]{Gelman:2013}%
  \BibitemOpen
  \bibfield  {author} {\bibinfo {author} {\bibfnamefont {A.}~\bibnamefont
  {Gelman}},\ }\href {https://doi.org/10.1214/13-EJS854} {\bibfield  {journal}
  {\bibinfo  {journal} {Electronic Journal of Statistics}\ }\textbf {\bibinfo
  {volume} {7}},\ \bibinfo {pages} {2595} (\bibinfo {year} {2013})}\BibitemShut
  {NoStop}%
\bibitem [{sup()}]{supp}%
  \BibitemOpen
  \href@noop {} {}\bibinfo {note} {See Ancillary Material 
  for one- and two-dimensional posterior density
  distributions under all combinations of usage of reactor constraint and
  marginalization scheme.}\BibitemShut {Stop}%
\bibitem [{\citenamefont {Jeffreys}(1961)}]{Jeffreys:1961}%
  \BibitemOpen
  \bibfield  {author} {\bibinfo {author} {\bibfnamefont {H.}~\bibnamefont
  {Jeffreys}},\ }\href@noop {} {\emph {\bibinfo {title} {Theory of
  Probability}}},\ \bibinfo {edition} {3rd}\ ed.\ (\bibinfo  {publisher}
  {Oxford},\ \bibinfo {address} {Oxford, England},\ \bibinfo {year}
  {1961})\BibitemShut {NoStop}%
\bibitem [{\citenamefont {Kass}\ and\ \citenamefont
  {Raftery}(1995)}]{KassAndRaftery:1995}%
  \BibitemOpen
  \bibfield  {author} {\bibinfo {author} {\bibfnamefont {R.~E.}\ \bibnamefont
  {Kass}}\ and\ \bibinfo {author} {\bibfnamefont {A.~E.}\ \bibnamefont
  {Raftery}},\ }\href {https://doi.org/10.1080/01621459.1995.10476572}
  {\bibfield  {journal} {\bibinfo  {journal} {Journal of the American
  Statistical Association}\ }\textbf {\bibinfo {volume} {90}},\ \bibinfo
  {pages} {773} (\bibinfo {year} {1995})}\BibitemShut {NoStop}%
\bibitem [{\citenamefont {Denton}\ and\ \citenamefont
  {Parke}(2019)}]{Denton:2019}%
  \BibitemOpen
  \bibfield  {author} {\bibinfo {author} {\bibfnamefont {P.~B.}\ \bibnamefont
  {Denton}}\ and\ \bibinfo {author} {\bibfnamefont {S.~J.}\ \bibnamefont
  {Parke}},\ }\href {https://doi.org/10.1103/PhysRevD.100.053004} {\bibfield
  {journal} {\bibinfo  {journal} {Phys. Rev. D}\ }\textbf {\bibinfo {volume}
  {100}},\ \bibinfo {pages} {053004} (\bibinfo {year} {2019})}\BibitemShut
  {NoStop}%
\bibitem [{\citenamefont {Jarlskog}(1985)}]{Jarlskog:1985}%
  \BibitemOpen
  \bibfield  {author} {\bibinfo {author} {\bibfnamefont {C.}~\bibnamefont
  {Jarlskog}},\ }\href {https://doi.org/10.1103/PhysRevLett.55.1039} {\bibfield
   {journal} {\bibinfo  {journal} {Phys. Rev. Lett.}\ }\textbf {\bibinfo
  {volume} {55}},\ \bibinfo {pages} {1039} (\bibinfo {year}
  {1985})}\BibitemShut {NoStop}%
\bibitem [{\citenamefont {Denton}\ and\ \citenamefont
  {Pestes}(2021)}]{Denton:2020igp}%
  \BibitemOpen
  \bibfield  {author} {\bibinfo {author} {\bibfnamefont {P.~B.}\ \bibnamefont
  {Denton}}\ and\ \bibinfo {author} {\bibfnamefont {R.}~\bibnamefont
  {Pestes}},\ }\href {https://doi.org/10.1007/JHEP05(2021)139} {\bibfield
  {journal} {\bibinfo  {journal} {JHEP}\ }\textbf {\bibinfo {volume} {05}},\
  \bibinfo {pages} {139}},\ \Eprint {https://arxiv.org/abs/2006.09384}
  {arXiv:2006.09384 [hep-ph]} \BibitemShut {NoStop}%
\bibitem [{\citenamefont {Dickey}(1971)}]{Dickey:1971}%
  \BibitemOpen
  \bibfield  {author} {\bibinfo {author} {\bibfnamefont {J.~M.}\ \bibnamefont
  {Dickey}},\ }\href {https://doi.org/10.1214/aoms/1177693507} {\bibfield
  {journal} {\bibinfo  {journal} {The Annals of Mathematical Statistics}\
  }\textbf {\bibinfo {volume} {42}},\ \bibinfo {pages} {204} (\bibinfo {year}
  {1971})}\BibitemShut {NoStop}%
\bibitem [{\citenamefont {Mulder}\ \emph {et~al.}(2020)\citenamefont {Mulder},
  \citenamefont {Wagenmakers},\ and\ \citenamefont {Marsman}}]{Mulder:2020}%
  \BibitemOpen
  \bibfield  {author} {\bibinfo {author} {\bibfnamefont {J.}~\bibnamefont
  {Mulder}}, \bibinfo {author} {\bibfnamefont {E.~J.}\ \bibnamefont
  {Wagenmakers}},\ and\ \bibinfo {author} {\bibfnamefont {M.}~\bibnamefont
  {Marsman}}\ }\href {https://doi.org/10.48550/arxiv.2004.09899}
  {10.48550/arxiv.2004.09899} (\bibinfo {year} {2020})\BibitemShut {NoStop}%
\bibitem [{\citenamefont {Hill}(2011)}]{Hill:2011}%
  \BibitemOpen
  \bibfield  {author} {\bibinfo {author} {\bibfnamefont {T.~P.}\ \bibnamefont
  {Hill}},\ }\href {https://doi.org/10.1090/S0002-9947-2011-05340-7} {\bibfield
   {journal} {\bibinfo  {journal} {Trans. Amer. Math. Soc.}\ }\textbf {\bibinfo
  {volume} {363}},\ \bibinfo {pages} {3351} (\bibinfo {year} {2011})},\ \Eprint
  {https://arxiv.org/abs/0808.1808} {0808.1808} \BibitemShut {NoStop}%
\bibitem [{\citenamefont {Fox}\ \emph {et~al.}(2011)\citenamefont {Fox},
  \citenamefont {Hill},\ and\ \citenamefont {Miller}}]{Fox:2010id}%
  \BibitemOpen
  \bibfield  {author} {\bibinfo {author} {\bibfnamefont {R.~F.}\ \bibnamefont
  {Fox}}, \bibinfo {author} {\bibfnamefont {T.~P.}\ \bibnamefont {Hill}},\ and\
  \bibinfo {author} {\bibfnamefont {J.}~\bibnamefont {Miller}},\ }\href
  {https://doi.org/10.1063/1.3593373} {\bibfield  {journal} {\bibinfo
  {journal} {Chaos}\ }\textbf {\bibinfo {volume} {21}},\ \bibinfo {pages}
  {033102} (\bibinfo {year} {2011})},\ \Eprint
  {https://arxiv.org/abs/1005.4978} {arXiv:1005.4978 [physics.data-an]}
  \BibitemShut {NoStop}%
\bibitem [{\citenamefont {Abi}\ \emph {et~al.}(2020)\citenamefont {Abi} \emph
  {et~al.}}]{DUNE:2020jqi}%
  \BibitemOpen
  \bibfield  {author} {\bibinfo {author} {\bibfnamefont {B.}~\bibnamefont
  {Abi}} \emph {et~al.} (\bibinfo {collaboration} {DUNE}),\ }\href
  {https://doi.org/10.1140/epjc/s10052-020-08456-z} {\bibfield  {journal}
  {\bibinfo  {journal} {Eur. Phys. J. C}\ }\textbf {\bibinfo {volume} {80}},\
  \bibinfo {pages} {978} (\bibinfo {year} {2020})},\ \Eprint
  {https://arxiv.org/abs/2006.16043} {arXiv:2006.16043 [hep-ex]} \BibitemShut
  {NoStop}%
\bibitem [{\citenamefont {Gelman}\ \emph {et~al.}(1997)\citenamefont {Gelman},
  \citenamefont {Gilks},\ and\ \citenamefont {Roberts}}]{Gelman:1997a}%
  \BibitemOpen
  \bibfield  {author} {\bibinfo {author} {\bibfnamefont {A.}~\bibnamefont
  {Gelman}}, \bibinfo {author} {\bibfnamefont {W.~R.}\ \bibnamefont {Gilks}},\
  and\ \bibinfo {author} {\bibfnamefont {G.~O.}\ \bibnamefont {Roberts}},\
  }\href {https://doi.org/10.1214/aoap/1034625254} {\bibfield  {journal}
  {\bibinfo  {journal} {The Annals of Applied Probability}\ }\textbf {\bibinfo
  {volume} {7}},\ \bibinfo {pages} {110 } (\bibinfo {year} {1997})}\BibitemShut
  {NoStop}%
\bibitem [{\citenamefont {Roberts}(1998)}]{Roberts:1998a}%
  \BibitemOpen
  \bibfield  {author} {\bibinfo {author} {\bibfnamefont {G.~O.}\ \bibnamefont
  {Roberts}},\ }\href {https://doi.org/10.1080/17442509808834136} {\bibfield
  {journal} {\bibinfo  {journal} {Stochastics and Stochastic Reports}\ }\textbf
  {\bibinfo {volume} {62}},\ \bibinfo {pages} {275} (\bibinfo {year}
  {1998})}\BibitemShut {NoStop}%
\bibitem [{\citenamefont {Roberts}\ and\ \citenamefont
  {Rosenthal}(2001)}]{Roberts:2001a}%
  \BibitemOpen
  \bibfield  {author} {\bibinfo {author} {\bibfnamefont {G.~O.}\ \bibnamefont
  {Roberts}}\ and\ \bibinfo {author} {\bibfnamefont {J.~S.}\ \bibnamefont
  {Rosenthal}},\ }\href {https://doi.org/10.1214/ss/1015346320} {\bibfield
  {journal} {\bibinfo  {journal} {Statistical Science}\ }\textbf {\bibinfo
  {volume} {16}},\ \bibinfo {pages} {351 } (\bibinfo {year}
  {2001})}\BibitemShut {NoStop}%
\bibitem [{\citenamefont {Croarkin}\ and\ \citenamefont
  {Tobias}(2013)}]{Croarkin:2013yxl}%
  \BibitemOpen
  \bibinfo {editor} {\bibfnamefont {C.}~\bibnamefont {Croarkin}}\ and\ \bibinfo
  {editor} {\bibfnamefont {P.}~\bibnamefont {Tobias}},\ eds.,\ \href
  {https://doi.org/10.18434/M32189} {\emph {\bibinfo {title} {NIST/SEMATECH
  e-Handbook of Statistical Methods}}}\ (\bibinfo {year} {2013})\BibitemShut
  {NoStop}%
\bibitem [{\citenamefont {Hoffman}\ and\ \citenamefont
  {Gelman}(2014)}]{Hoffman:2014a}%
  \BibitemOpen
  \bibfield  {author} {\bibinfo {author} {\bibfnamefont {M.~D.}\ \bibnamefont
  {Hoffman}}\ and\ \bibinfo {author} {\bibfnamefont {A.}~\bibnamefont
  {Gelman}},\ }\href {http://jmlr.org/papers/v15/hoffman14a.html} {\bibfield
  {journal} {\bibinfo  {journal} {Journal of Machine Learning Research}\
  }\textbf {\bibinfo {volume} {15}},\ \bibinfo {pages} {1593} (\bibinfo {year}
  {2014})}\BibitemShut {NoStop}%
\end{thebibliography}%

\end{document}